\documentclass[xcolor=dvipsnames]{desyproc}

{
}
\usepackage[backend=biber,style=numeric-comp,firstinits=true,sorting=none,doi=false,mcite=true,subentry]{biblatex}
\usepackage{ILD_biblatex}

\DeclareFieldFormat[article]{title}{}
\usepackage[unicode=true]{hyperref}
\usepackage{cancel}
\usepackage{wrapfig,rotating}

\usepackage{pgfpages}
\usepackage{subfig}
\usepackage{pgfpages}
\usepackage{fancybox,graphicx}
\usepackage{epstopdf}
\AppendGraphicsExtensions{.eps.gz}
\usepackage[english]{babel}
\usepackage[latin1]{inputenc}
\usepackage{times}
\usepackage[T1]{fontenc}
\def\leqsim{\mathbin{\;\raise1pt\hbox{$<$}\kern-8pt\lower3pt\hbox{$\sim$}\;}}
\def\geqsim{\mathbin{\;\raise1pt\hbox{$>$}\kern-8pt\lower3pt\hbox{$\sim$}\;}}
\def\MXN#1{\mbox{$ M_{\tilde{\chi}^0_#1}                                $}}
\def\MXC#1{\mbox{$ M_{\tilde{\chi}^{\pm}_#1}                            $}}

\def\XPM#1{\mbox{$ \tilde{\chi}^{\pm}_#1                                $}}

\def\XN#1{\mbox{$ \tilde{\chi}^0_#1                                     $}}

\def\p#1{\mbox{$ \mbox{\bf p}_1                                         $}}

\newcommand{\stau}    {\mbox{$ \tilde{\tau}                                $}}
\newcommand{\stone}   {\mbox{$ \tilde{\tau}_1                              $}}

\newcommand{\ba}{\begin{array}}
\newcommand{\ea}{\end{array}}
\newcommand{\bc}{\begin{center}}
\newcommand{\ec}{\end{center}}
\newcommand{\be}{\begin{eqnarray}}
\newcommand{\eeq}{\end{eqnarray}}
\newcommand{\bes}{\begin{eqnarray*}}
\newcommand{\ees}{\end{eqnarray*}}
\newcommand{\Kz}{\ifmmode {\rm K^0_s} \else ${\rm K^0_s} $ \fi}
\newcommand{\Zz}{\ifmmode {\rm Z^0} \else ${\rm Z^0 } $ \fi}
\newcommand{\xxbar}{\ifmmode {\rm x\bar{x}} \else ${\rm x\bar{x}} $ \fi}
\newcommand{\rphi}{\ifmmode {\rm R\phi} \else ${\rm R\phi} $ \fi}
\def    \missEt      {\ifmmode{/\mkern-11mu E_t}\else{${/\mkern-11mu E_t}$}\fi}
\def    \missE       {\ifmmode{/\mkern-11mu E}\else{${/\mkern-11mu E}$}\fi}
\def    \missp       {\ifmmode{/\mkern-11mu p}\else{${/\mkern-11mu p}$}\fi}
\def    \misspt      {\ifmmode{/\mkern-11mu p_t}\else{${/\mkern-11mu p_t}$}\fi}

\begin{document}
%\tikzstyle{every picture}+=[remember picture]
%\tikzstyle{na} = [baseline=-.5ex]
\title{
  What pp SUSY limits mean for future e$^+$e$^-$ colliders\footnote{
    %Contribution to LCWS19, Sendai, Japan}}
Contribution to the International Workshop on Future Linear Colliders (LCWS2019), Sendai, Japan,28 October-1 November, 2019. C19-10-28}}
%***********************************************************************
% AUTHORS INFORMATION AREA
%***********************************************************************
\author{{\slshape Mikael Berggren}
% Optional short acknowledgment: remove next line if non-needed
% DO NOT MODIFY THE FOLLOWING '\vspace' ARGUMENT
\vspace{.3cm}\\
% Addresses and institutions (remove "1- " in case of a single institution)
DESY,
Notkestra\ss e 85, D-22607 Hamburg -Germany
}
%%***********************************************************************
% END OF AUTHORS INFORMATION AREA
%***********************************************************************
\acronym{LCWS2019}
\maketitle
%%%   \titlepage
\begin{abstract}
  It is well-known that e$^+$e$^-$ colliders have the power to with certainty exclude or
  discover any SUSY model that predicts a  Next to
  lightest SUSY particle (an NLSP) that has a mass up to slightly below the half the centre-of-mass energy
  of the collider. 
  
  Here, we present an estimation of the power of present and future hadron colliders to
  extend the reach of searches for SUSY, with particular emphasis
  %on the level of
  %coverage, i.e. the
  %with what certainty
  whether
  it can be claimed that either discovery or
  exclusion is {\it guaranteed} in a region of  LSP and NLSP masses -
  no set of values of the other SUSY
  could change the conclusion.
  We study this by, reasonably, assuming that the most challenging
  scenario would be one where the lightest SUSY particles are the electroweak bosinos,
  and that sfermions are out of reach.
  A scan over SUSY parameter space was done, only requiring that the NLSP
   was a bosino with mass not larger than a few TeV.
  The mass-spectrum, cross-sections and decay branching ratios found in this region were
  confronted with projections of sensitivity at future hadron colliders. In our conclusions
  we weigh in the maturity of the analysis the projections are based upon.
  The conclusion is that although future hadron colliders have a large discovery-reach,
  i.e. potential to discover {\it some} SUSY model,
  hardly any models with low-to-medium LSP-NLSP mass-differences can be excluded
  with certainty.
  The models that are expected to be excluded/discovered are, on one hand, those with mass-differences
  larger than those allowed by models with GUT-scale $M_1$-$M_2$ unification,
  and on the other hand, a tiny region where the mass-difference is so small that
  the NLSP decays in the tracking volume of the detectors. Excluding the latter
  possibility does not, however, allow to exclude the possibility of a Wino or
  Higgsino LSP: at any value of the LSP mass, we could identify models where the
  NLSP lifetime would be too short for a signal to be seen.
  
\end{abstract}

\section{Introduction: SUSY and future colliders}
If Supersymmetry (SUSY) \mcite{susy,*Wess:1974tw,*Nilles:1983ge,*Haber:1984rc,*Barbieri:1982eh} is to explain the
current problems of the Standard Model for
particles physics, such as
the naturalness of the theory, the hierarchy problem, the nature of Dark Matter,
or the possible discrepancy between the observed and predicted value on the
muon magnetic moment (g-2),  a light electroweak SUSY sector is preferred.
From LEP II, it is known that an electroweak sector with masses below $\sim$ 100 GeV
is excluded, except for some very special cases.
From LHC, we know that
a coloured sector with masses below  $\sim$ 1 TeV is also excluded.
However, except for the third generation squarks, the coloured sector
  does not contribute significantly to clarify the mentioned issues with the SM.

 The model-space of the electroweak sector of SUSY can conveniently be sub-divided by the nature of
the Lightest SUSY Particle (the LSP) as the Bino-, Higgsino- or Wino-region, defined by whether $M_1$, $\mu$, or $M_2$ is
the smallest of the three, and thus which field is the largest contributor to the mass-eigenstate (not to be
confused with {\it pure} Wino, Bino or Higgsino models, where the respective contributions are close to 100 \%).
Alternatively, one can classify by the size of the mass-difference, $\Delta(M)$, between
the LSP and the next-to-lightest SUSY particle (the NLSP),
as high $\Delta(M)$ or low $\Delta(M)$. The first case coincides with the Bino-region, the second
contains the  Higgsino- and Wino-regions, which differ in important experimental consequences.
In other words: In the  Higgsino- and Wino-regions,
the electroweak
  SUSY sector 
  is ``compressed'', i.e. the masses of some of the  other electroweak bosinos tend to be close to the LSP mass.
  In this situation, most decays of massive sparticles are via cascades,
  and at the end of these cascades, the mass difference is small, in turn
  meaning that the final decay into the invisible LSP releases little energy.
  While such events show large missing energy, this is of no help at hadron colliders - contrary to
  the case at lepton colliders - since the initial energy is unknown.
  Therefore, to address such cases at hadron colliders, one must resort to missing transverse momentum, a much more delicate signal.
  Consequently, for such topologies, current limits from LHC are  for specific models, and the results from LEP II
  \mcite{lepsusywg,aleph,*Heister:2001nk,*Heister:2003zk,*Heister:2002mn,Abdallah:2003xe,Achard:2003ge,Abbiendi:2003ji}
  are those that yield the model-independent exclusions.
  The same observations are also valid if the NLSP is a slepton in general,
  and the $\stau$ in particular.

  The organisation of the note is as follows. We first discuss how to compare different options on an equal footing in section 2,
  and present our scan-range, tools and general observations about the mass-spectra in section 3.
  In section 4, we discuss the interpretation of the electroweak SUSY chapter of the physics Briefing-book \cite{Strategy:2019vxc}
  to the
  update of the European strategy for particle physics (the ESU) in view of
  our observations. In section 5, for reference, we summarise the ILC projections, before concluding in section 6.
  
   \begin{wrapfigure}{R}{8.5cm}
    \begin{center}
      \subfloat[][$\mu$ vs. $M_1$]{ \includegraphics [scale=0.4]{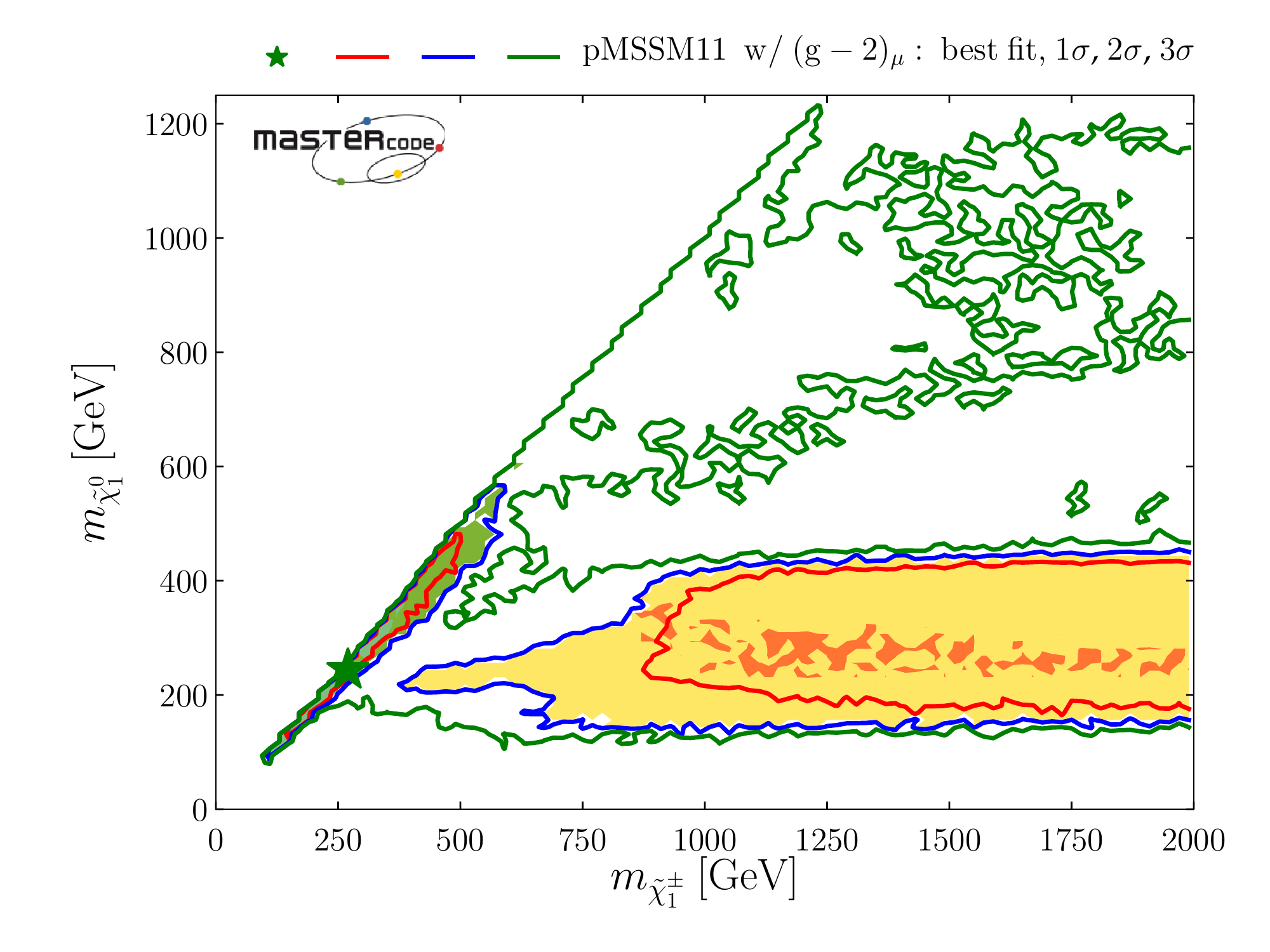}}
     \caption{ pMSSM11 fit by MasterCode to LHC13/LEP/g-2/DM(=100\%~LSP)/precision observables  in the $\MXC{1}$ - $\MXN{1}$ plane. From \cite{Bagnaschi:2017tru}. \label{fig:mastercode}}
     \end{center}
     \end{wrapfigure}
     \section{Comparing options}
  For asserting the capabilities of future facilities to explore SUSY, it is important to make the
  distinction between discovery potential and exclusion potential.
The former is power to  discover {\it some} model,
while the latter is the power to  exclude {\it all} models
compatible with the shown parameters, i.e. marginalising over all non-shown
parameters.
The methodologies needed in the two cases are different. In the first case,
one would concentrate on specific models yielding signatures that are
observable as far into uncharted territory as possible,
while in the latter case one needs to determine which model is the
most difficult, and evaluate whether that worst-case would be observed, if it is realised in nature.
The latter was indeed the focus at LEP II. The limits from there have
been marginalised over all other parameters, and can
be considered definite.

A further consideration that must be made in weighing different future projects
against each other is the level of understanding.
This includes the level of maturity of the project,
ranging from existing results (e.g. from LEP or LHC), over
existing/in construction new detectors and machines (HL-LHC),
TDR-level new facilities, such as ILC or CLIC,
to conceptual extensions to existing facilities (HE-LHC, LHeC).
It further extends to new conceptual ideas, such as the different
options for FCC, and continues to emerging technologies, e.g. plasma acceleration or $\mu$-colliders.
One must also consider the level of detail of studies done, which range
from fully simulated, well defined detectors and accelerators (LHC, HL-LHC, ILC, and CLIC),
fully simulated  evolving concepts, e.g. CepC, over detailed fast simulation
(i.e. with more detail than purely parametric simulations), to
parametric simulations with parametric input from full simulation of the proposed detector,
or simply using parameters from an existing detector at a new facility.
Also pure four-vector smearing of generated objects and simple cross-section level estimates
can be used as initial estimates.
In the case of cross-section and luminosity scaling estimates, one should also
consider whether they were done at the level of the final published exclusion reach,
or had access to more basic information of the extrapolated experiment
(background event count, efficiency tables etc.)
Finally, when it comes to interpreting the results of such studies it is also important to
consider whether they were done by detector experts or not.
This is of particular importance for systematics-limited experiments,
and cases where detailed knowledge of object-finding and reconstruction is 
essential.

\section{Estimating SUSY reach}

  \begin{figure}[t]
    \begin{center}
      \subfloat[][$\mu$ vs. $M_1$]{\includegraphics [scale=0.28]{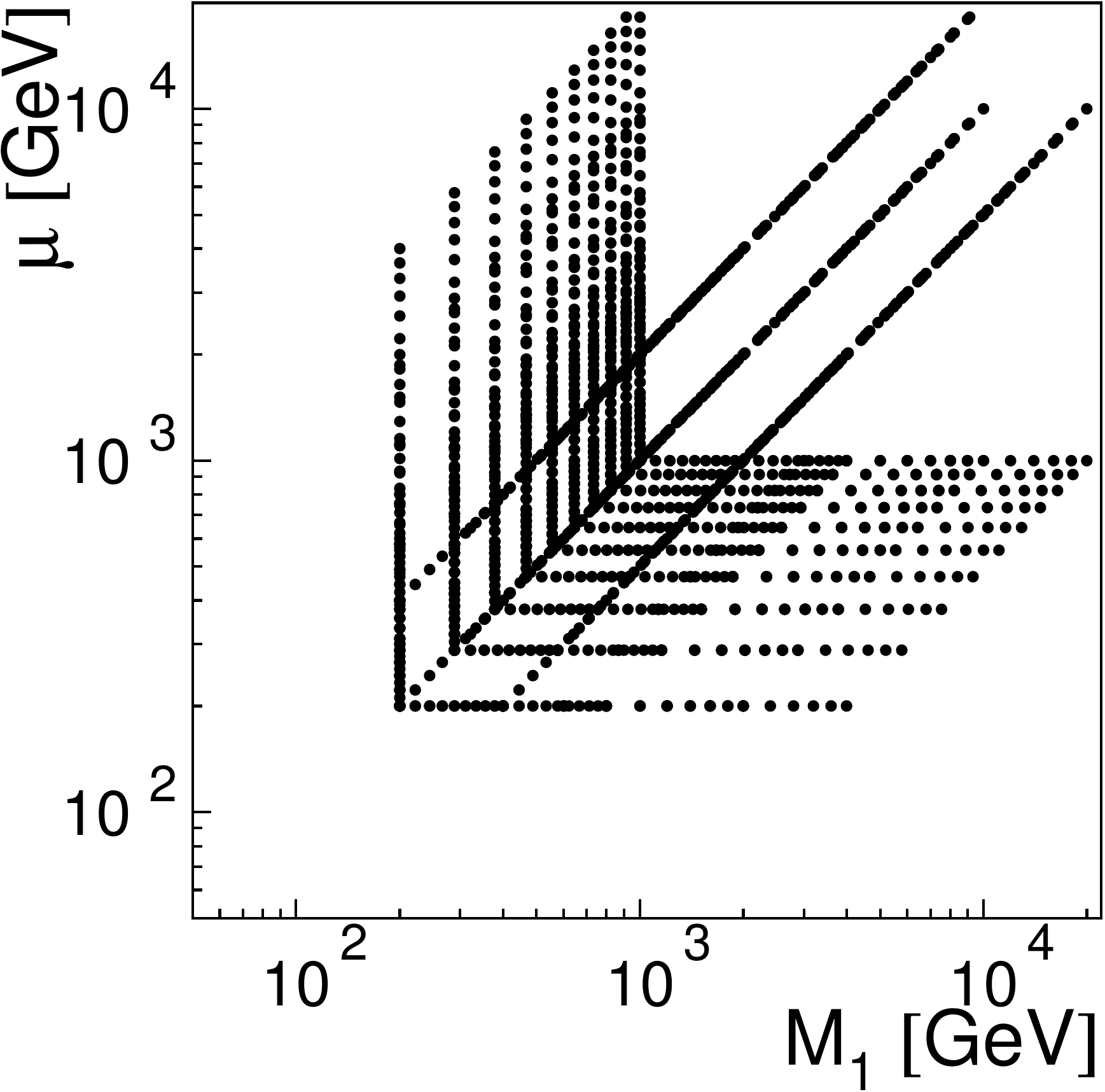}}
      \subfloat[][$\mu$ vs. $M_2$]{\includegraphics [scale=0.28]{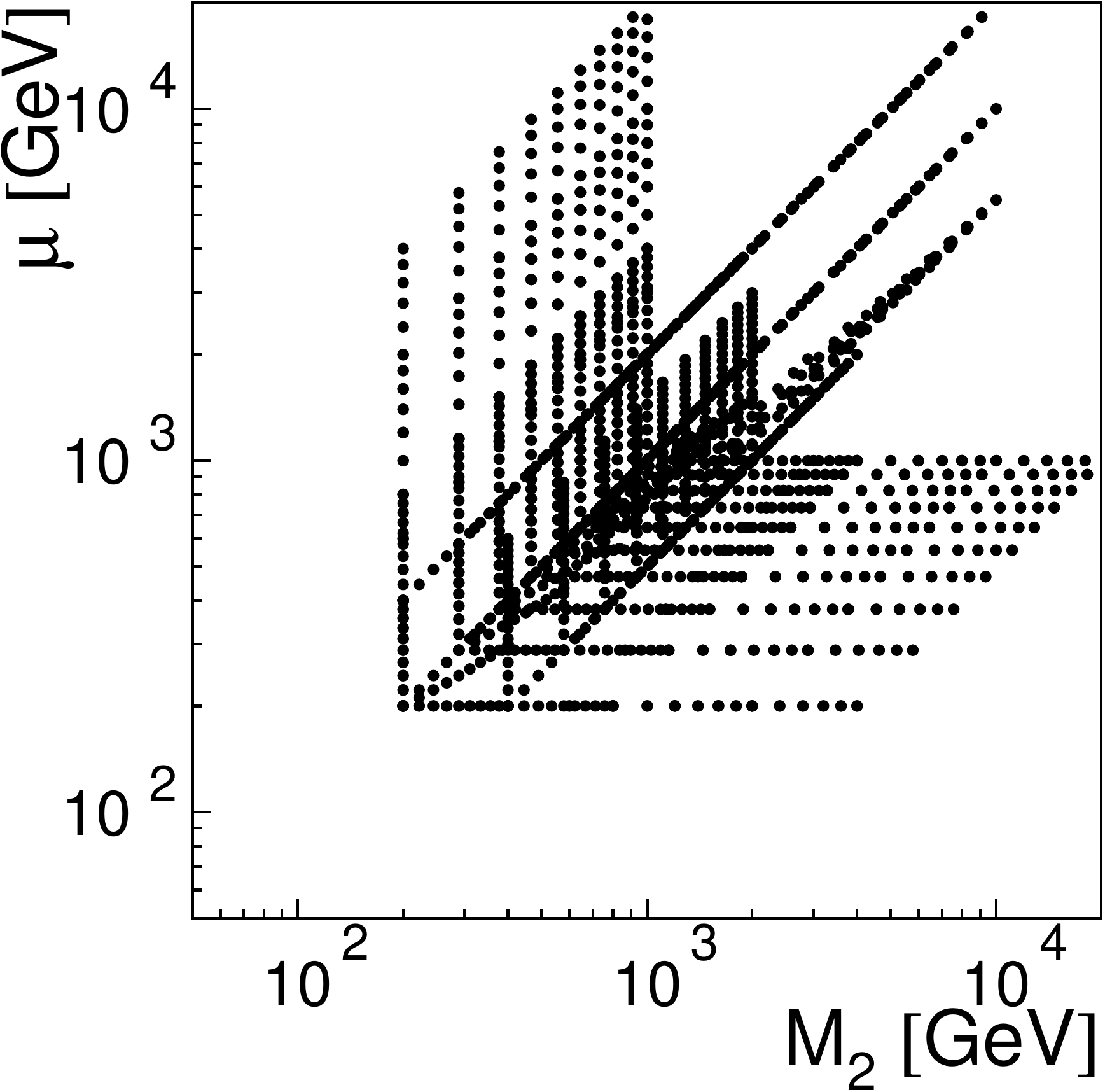}}
      \subfloat[][$M_1$ vs. $M_2$]{\includegraphics [scale=0.28]{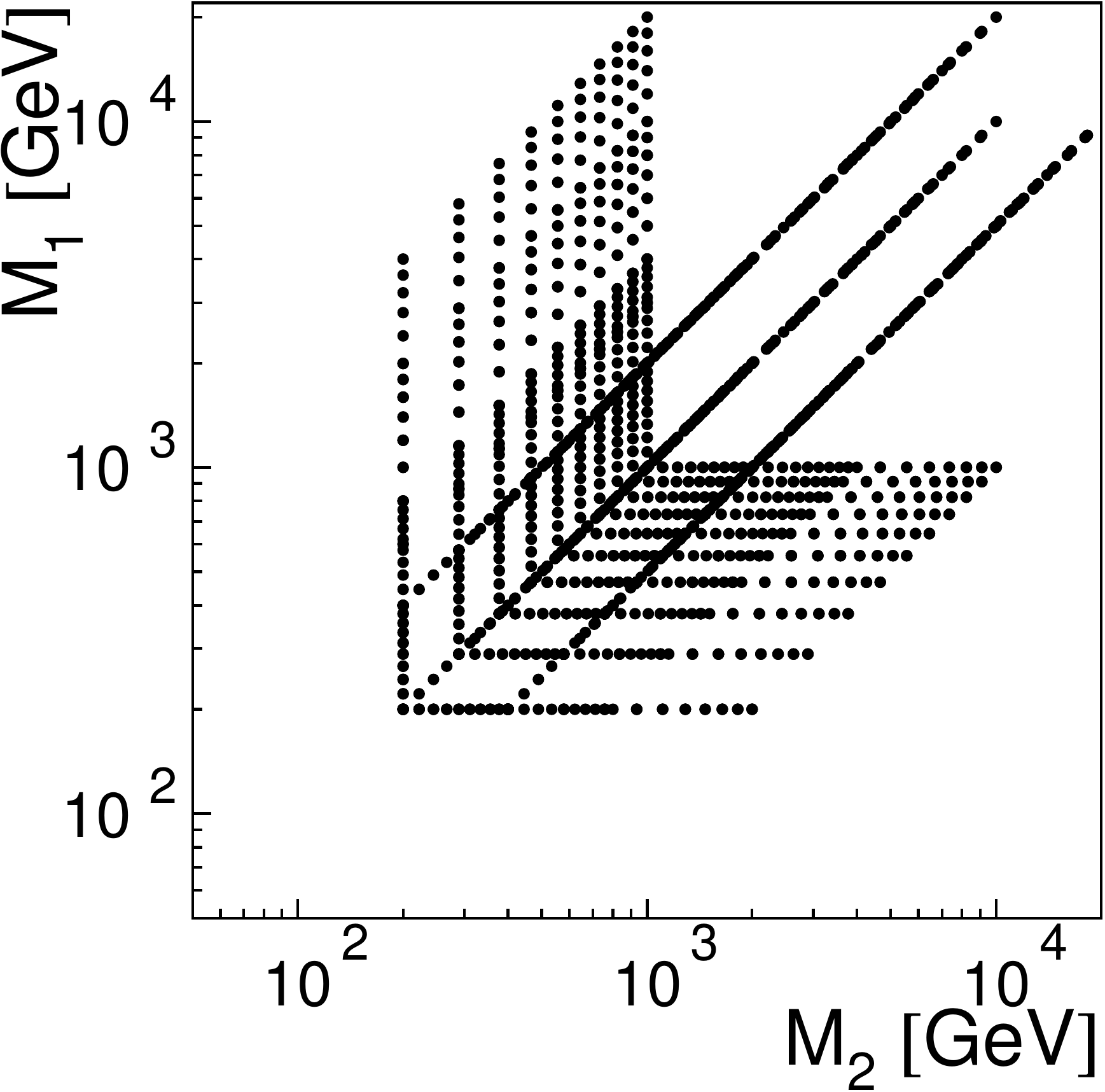}}
      \caption{The scanned points in $\mu$, $M_1$ and $M_2$.\label{fig:cube}}
     \end{center}
     \end{figure}
 Several groups \mcite{susyfits,*Bagnaschi:2017tru,*Bagnaschi:2018zwg,*Caron:2016hib,*Aad:2015baa} have
  combined current experimental observations with SUSY theory to estimate whereto in the
  parameter space observations point, and to estimate what regions actually are excluded at present.
  One example, from the MasterCode group \cite{Bagnaschi:2017tru}, is show as Figure \ref{fig:mastercode}.
  It, interestingly, indicates that current results point to the aforementioned ``compressed region''.
  This type of studies
  aims at answering the question
about where SUSY is most likely to be found, and to compare this with the
estimated capabilities of present or future facilities and techniques.
It should be noted that to arrive at definite conclusions,
the analyses typically include non-HEP observations.
Whether, and how, these can be put in to a HEP context already contains
assumptions.
E.g.\ it is often assumed that SUSY is the sole source of WIMP dark matter,
and that direct or indirect searches for WIMPs can be translated into HEP
observations.
However, other well motivated candidates for dark matter exists,  the prime example
being the QCD axion \cite{Marsh:2015xka}, so it might well be that SUSY is realised in nature, but
is not the (full) explanation for Dark Matter.
In addition, the estimates sometimes include results that hint to physics beyond the standard model,
but that are not yet solidly established, g-2 of the muon being one example. 

 \begin{figure}[b]
    \begin{center}
      \subfloat[][Higgsino-like LSP ($\mu < M_1, M_2$)]{\includegraphics [scale=0.27]{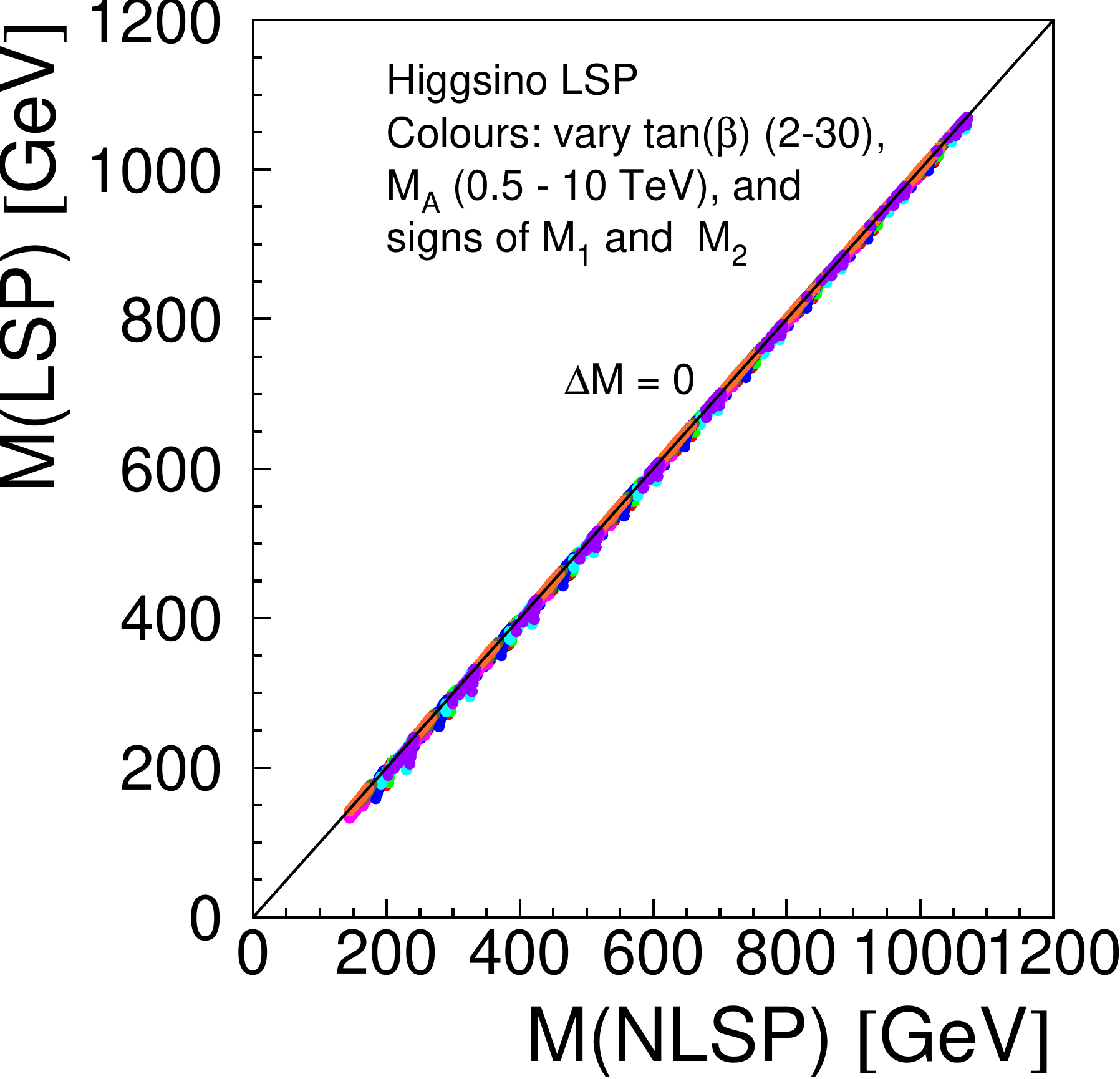}}
      \subfloat[][Wino-like LSP ($M_2 < M_1, \mu $)]{\includegraphics [scale=0.27]{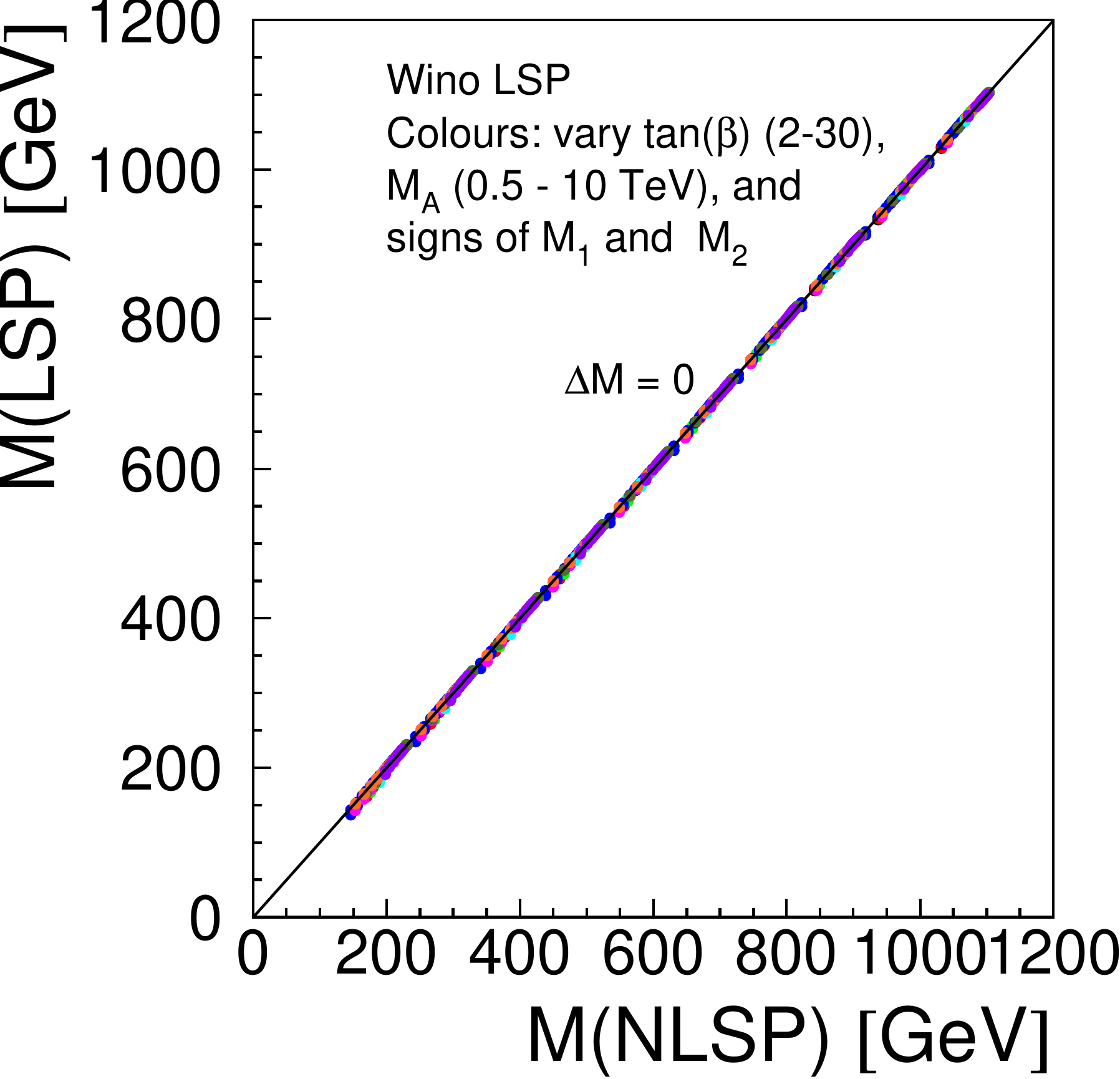}}
      \subfloat[][Bino-like LSP($M_1 < M_2, \mu $)]{\includegraphics [scale=0.27]{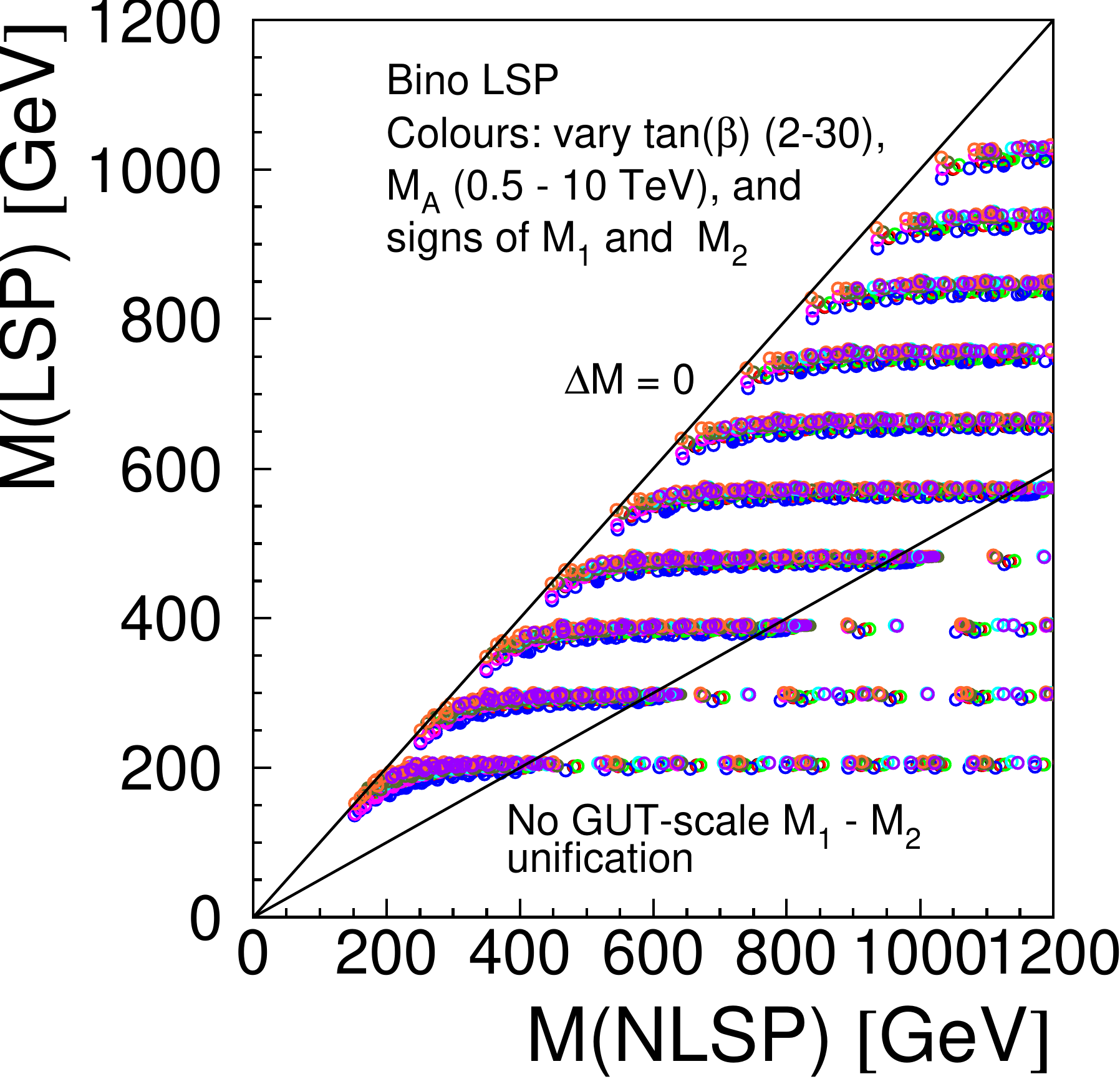}}
      \caption{The LSP mass vs. the NLSP mass for the three cases. The colour coding is the following:
         Points with the same colour have varying values of the 
         bosino parameters, as per Figure \ref{fig:cube}, while the colours are: All point have $\tan{\beta}$=10, except
         light green ($\tan{\beta}$=3) and  blue ($\tan{\beta}$=30). All have $M_A$= 5 TeV except black ($M_A$= 0.5 TeV)
         and magenta ($M_A$= 10 TeV). All have positive sign of $\mu,M_1$, and $M_2$, except cyan
         (-,+,+), olive-green (+,+,-), orange (+,-,+) and purple (+,-,-). Open symbols
          are for $|M_2-2 M_1|/|M_1|>0.1$ (i.e. not close to the GUT-unification case).
\label{fig:broadbrush}}
     \end{center}
     \end{figure}
If one is interested in the {\it guaranteed} reach, rather than the {\it possible} reach,
one should not rely on assumptions that are not directly testable.
In essence, this means to concentrate on the {\it  exclusion reach}.
In SUSY, the fundamental principle that sparticles and particles have the same
couplings and the same quantum-numbers (except for spin), sets a scene where such a program is possible.
It implies that cross-sections and decay modes are completely know within SUSY itself.
In particular, if R-parity conservation is assumed, it means that there is
always one completely know process, namely NLSP production, followed by
the decay of the NLSP to it's SM partner and the LSP, if kinematically allowed, with 100 \% branching ratio.
In estimating the exclusion reach, rather than the discovery reach, it is essential to find
the {\it most challenging} situations. Such a scenario is easily found, and we will consider
  \begin{itemize}
  \item  the MSSM with R-parity conservation, since LEP experience shows that the case
    of R-parity violation is always less demanding at e$^+$e$^-$ machines, and likely to also be
    so at hadron machines.
    %\footnote{
    %A caveat is that we also consider CP-conservation. The experimental implication of CP violation needs study}
  \item  The NLSP is not a sfermion, for the same reason. The  $\stau$ is an exception, in that
    the LEP experience indicates that a $\stau$ NLSP might be even more challenging than a bosino one.
    However, the issue is even more pronounced at a hadron collider \cite{Abdughani:2019wss}.
   \end{itemize}
   
  Under these conditions both the LSP and the NLSP are more or less pure Binos, Winos, or Higgsinos,
  and $M_1 , M_2$ and $\mu$ are the MSSM parameters most influencing the experimental signatures.
  We  consider any values, and combinations of signs, of these parameters,
  up to values that makes the bosinos
    kinematically out-of-reach
    for any new facility, i.e. up to a few TeV.
    We also vary other parameters ($\tan {\beta} $= 3 to 30, $M_A$= 0.5 to 10 TeV, $M_{sfermion}$= 5 to 10 TeV), to verify that
    they have only
    a minor impact on the signatures.
    No other assumptions, such as relations between the parameters due to some specific
    SUSY-breaking mechanism, are done.
    No assumption on prior probabilities is implied, and therefore that the density of points in the
    various projections that will be shown is not of great importance.
    The important observation to be made is whether {\it there are} any points outside excluded regions: this
    implies that the model {\it cannot} be excluded.
    
     \begin{figure}[t]
    \begin{center}
      \subfloat[][Full range, all cases]{\includegraphics [scale=0.38]{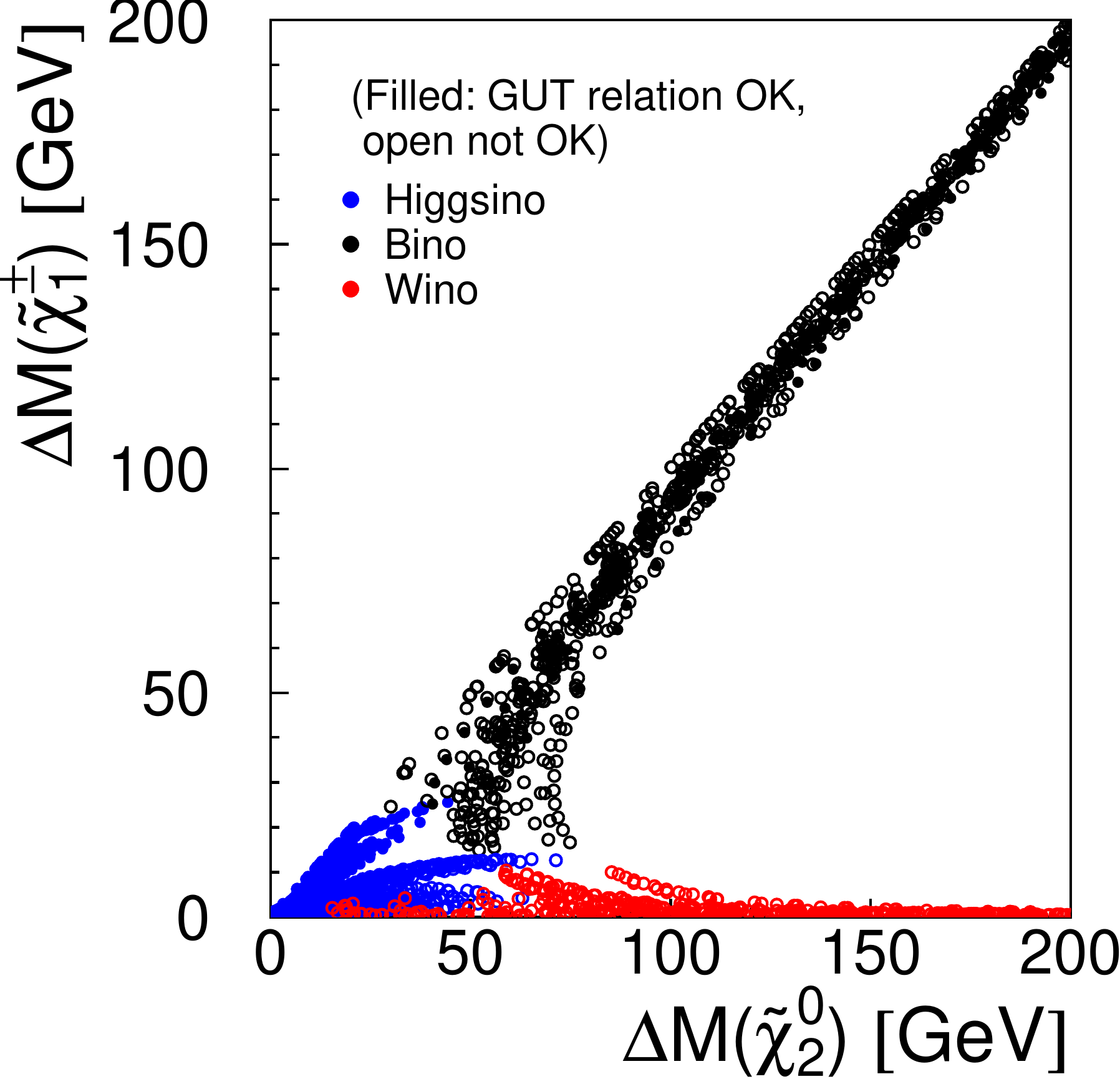}}
      \hskip 0.5cm
      \subfloat[][Blow-up of the Higgsino LSP region]{\includegraphics [scale=0.38]{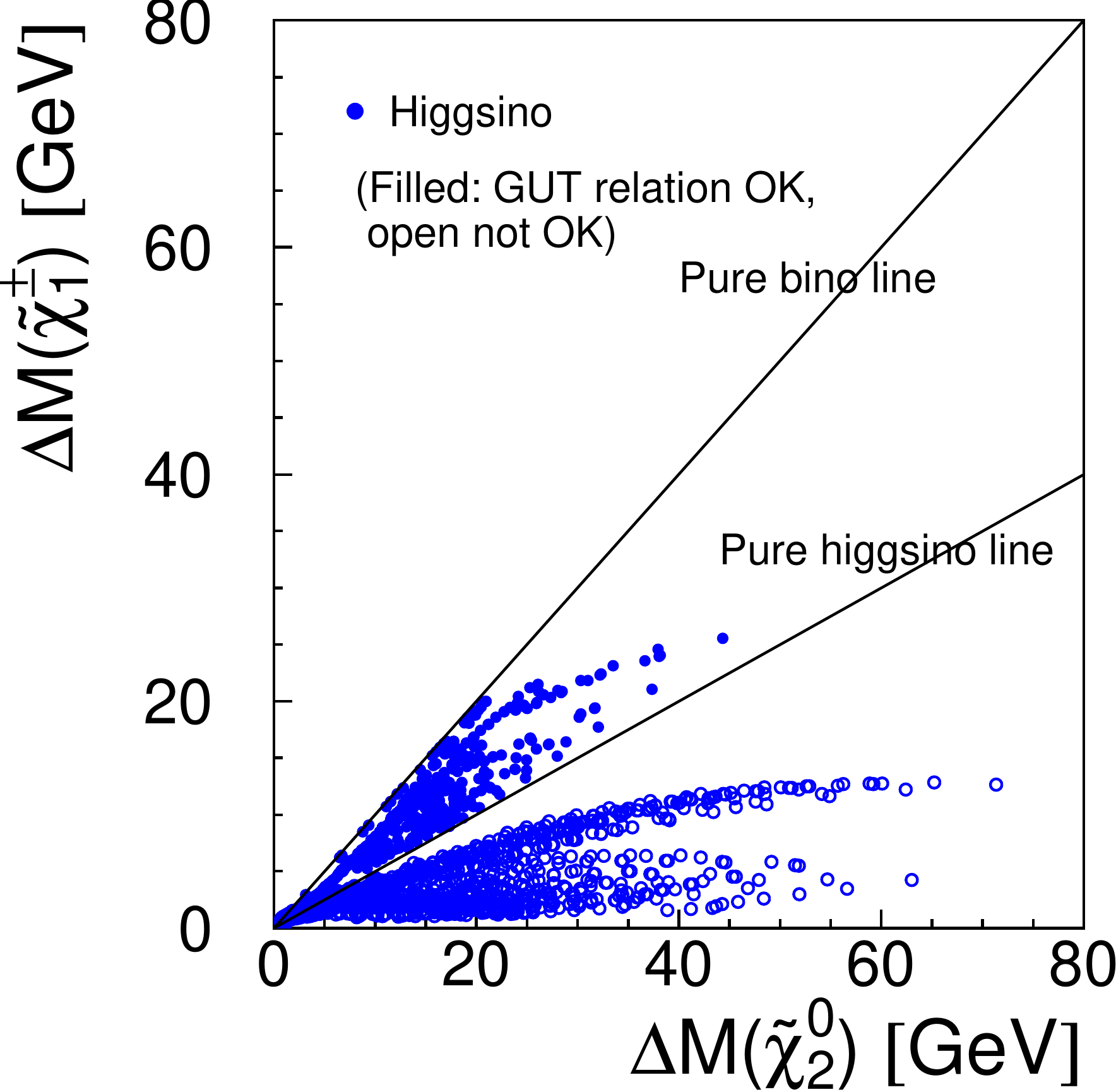}}
           \caption{The mass difference between the LSP and $\XPM{1}$ versus that between the LSP and $\XN{2}$. \label{fig:dmx1dmn2}}
     \end{center}
     \end{figure}
     \begin{figure}[h]
    \begin{center}
      \subfloat[][Higgsino-LSP case]{\includegraphics [scale=0.38]{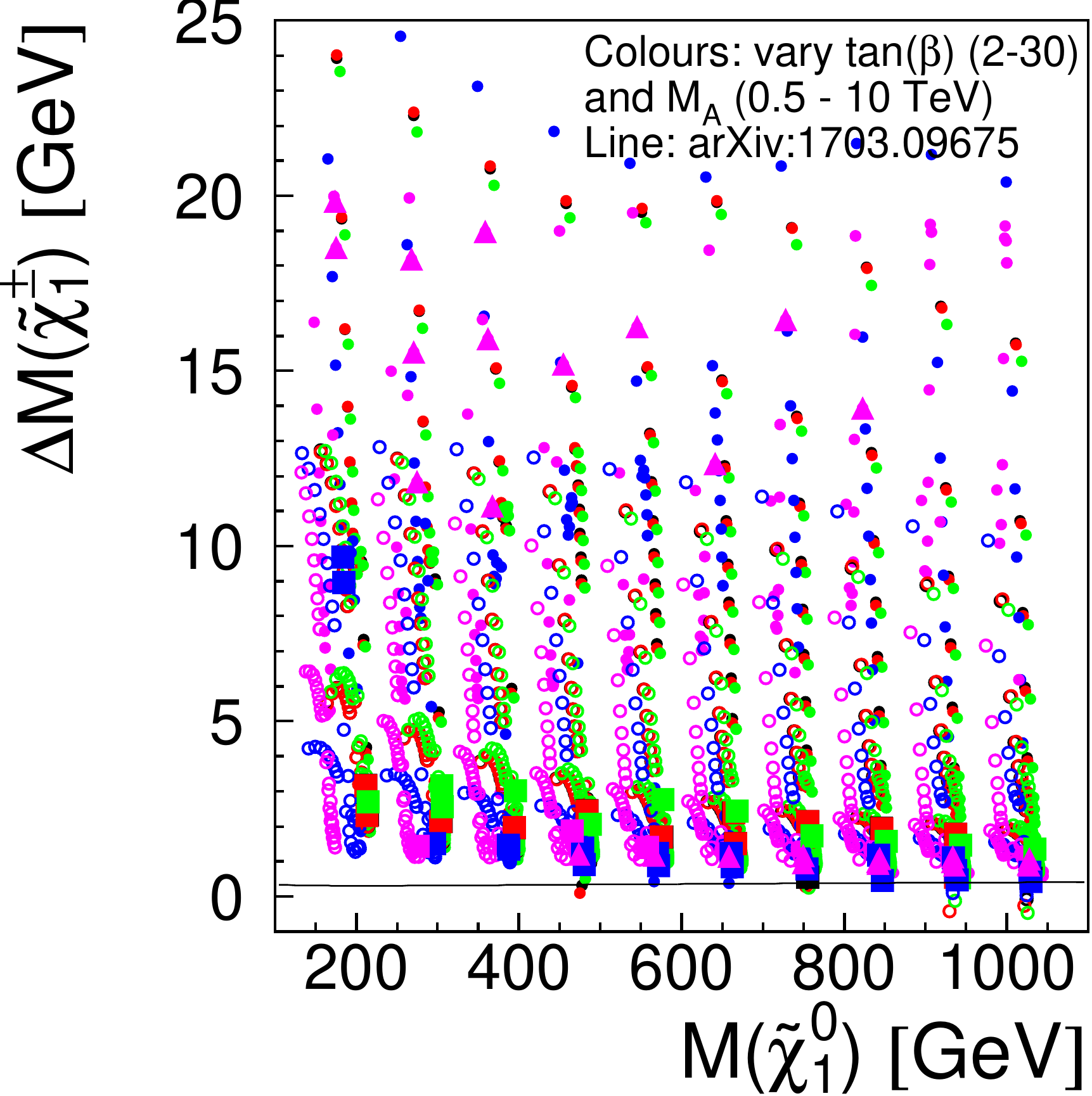}}
      \hskip 1.0cm
      \subfloat[][Wino-LSP case]{\includegraphics [scale=0.38]{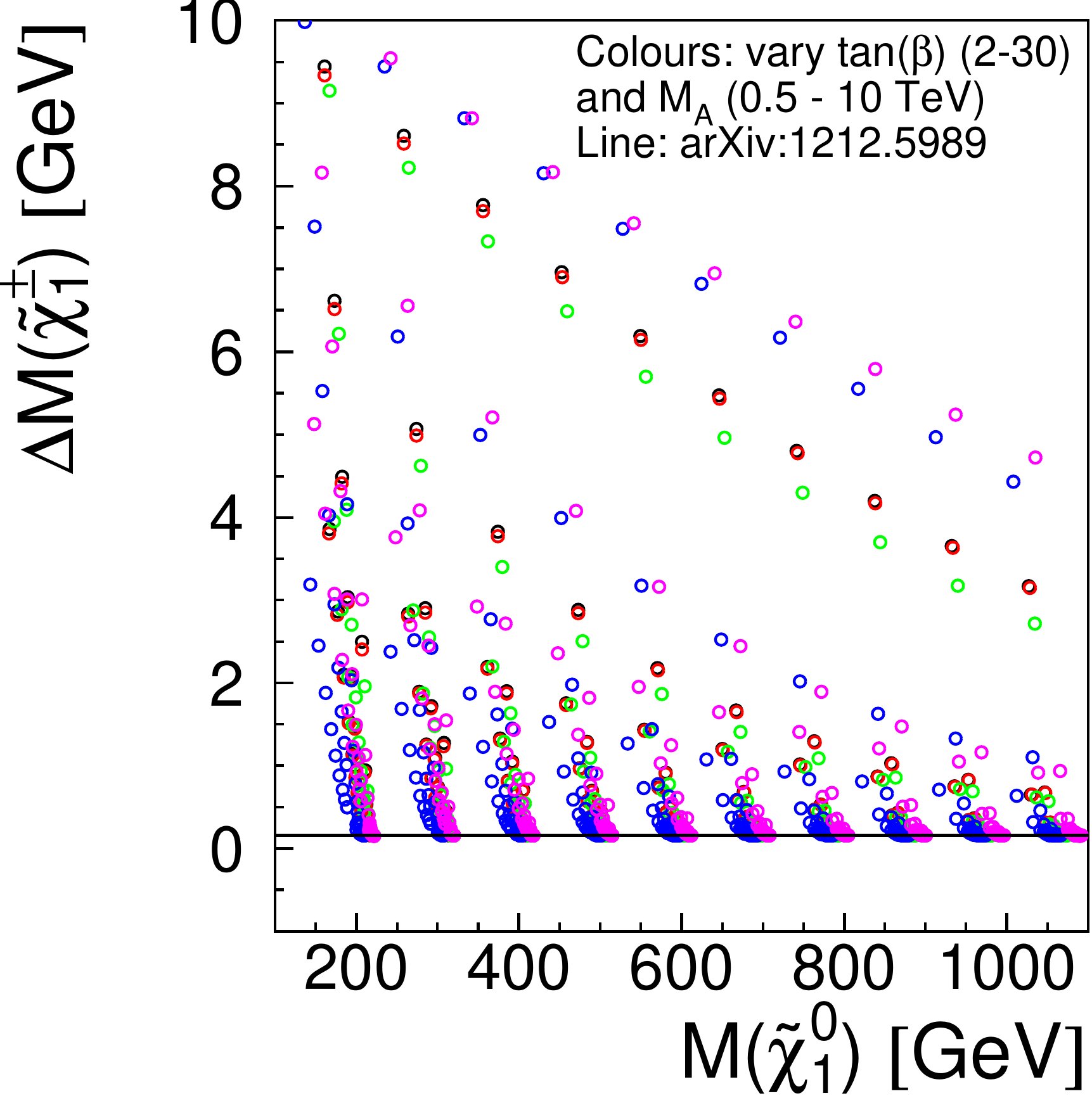}}
      \caption{$\Delta(\MXC{1})$ vs. $M_{LSP}$ in the small $\Delta(M)$ region. In (a), the
        squares are points where  $\Delta(\MXC{1})\approx\Delta(\MXN{2})/2$, and triangles are points where  $\Delta(\MXC{1})\approx\Delta(\MXN{2})$
        Colours as explained in Figure \ref{fig:broadbrush}.\label{fig:dmx1mlsp}}
     \end{center}
     \end{figure}
  Figure \ref{fig:cube} shows the points studied in $M_1 , M_2$ and $\mu$ as three two-dimensional
  projections.
  We proceed to find what  happens with spectra, cross-sections, and decay branching-ratios  when exploiting this ``cube''.
  In order to do so, {\tt SPheno 4.0.5} \mcite{shpeno,*Porod:2003um,*Porod:2011nf} was used to calculate spectra and and decay branching ratios at each point,
  and {\tt Whizard 2.8.0} \mcite{whizard,*Kilian:2007gr,*Moretti:2001zz} was used to find the production cross-sections,
  and to generate parton-level events.
  In addition {\tt FeynHiggs 2.16.0} \mcite{feynhiggs,*Bahl:2018qog,*Bahl:2017aev,*Bahl:2016brp,*Hahn:2013ria,*Frank:2006yh,*Degrassi:2002fi,*Heinemeyer:1998np,*Heinemeyer:1998yj}
  was used to calculate the expected mass of the SM-like higgs boson,  and as a double-check of the sparticle mass-spectrum.
            \begin{figure}
    \begin{center}
      \subfloat[][]{\includegraphics [scale=0.9]{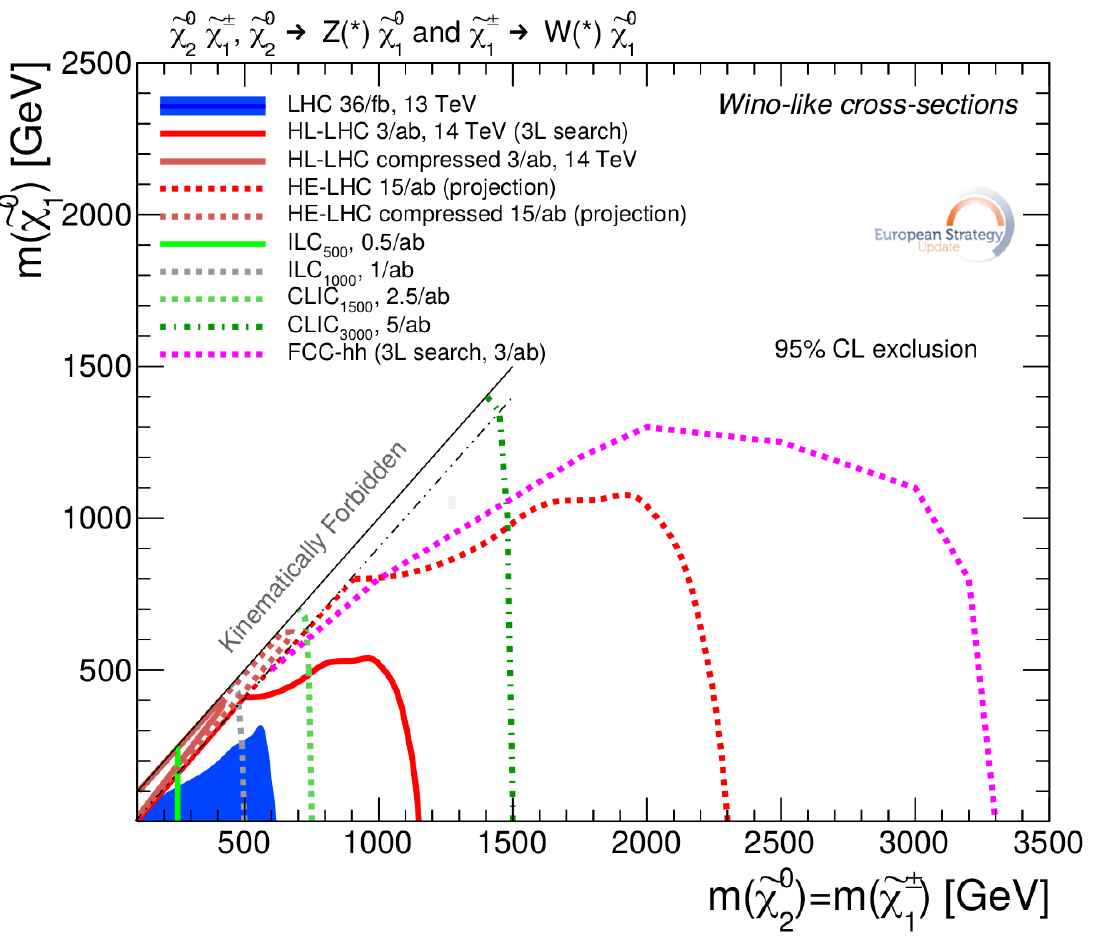}}
      
        \subfloat[][ $\XPM{1} \rightarrow Z \XN{1}$]{\includegraphics [scale=0.9]{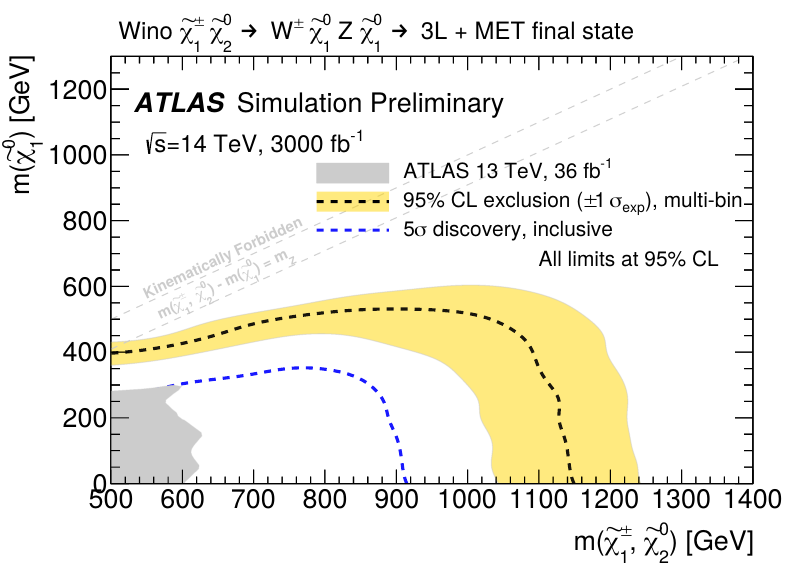}}
      \subfloat[][$\XPM{1} \rightarrow h \XN{1}$]{\includegraphics [scale=0.9]{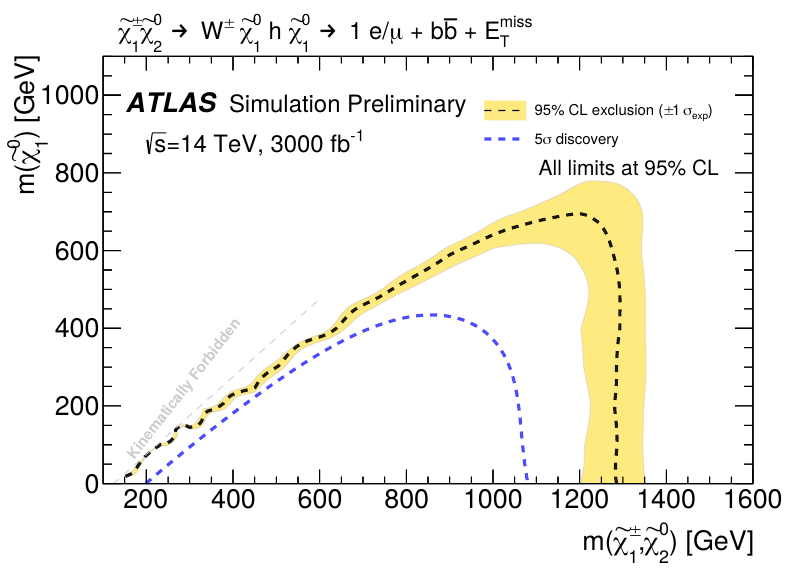}}
      \caption{The reaches in the high $\Delta(M)$ (Bino-LSP) region, as
        reported in \cite{Strategy:2019vxc} (top),
        and the two projections to HL-LHC from ATLAS \cite{ATL-PHYS-PUB-2018-048} (bottom).(b) corresponds to
        the solid red line in (a).\label{fig:bbbino}}
     \end{center}
     \end{figure}
         Around 80 \% of the points had calculated higgs-mass agreeing with the experimental value
  at the 2$\sigma$ level of the theoretical uncertainty, with the exception
  of the points with the highest of the three $\tan{\beta}$ values in our scan,
  namely $\tan{\beta}$=30, where only 7, 9 and 23 \% of the points were in the range for
  Wino-, Bino- and Higgsino-LSP, respectively.
  None of the features shown in the following figures, however, change if demanding that the calculated higgs-mass
  was in the two standard deviation range.
  The main features are shown in Figure \ref{fig:broadbrush}.
     One observes that, except for Bino-LSP, the LSP-NLSP splitting is small.
  The colours indicate different settings of the secondary parameters; the observation is  that they don't matter much.
  In addition, the open circles indicate cases where GUT-scale unification
  of $M_1$ and $M_2$ is not possible\footnote{
    If $M_1$ and $M_2$ are unified at the GUT scale, the different RGE running of the two results
    in the relation $M_2=(3g^2/5g^{\prime 2})M_1 \approx 2M_1$ at the weak scale. The maximally
    stretched difference between the LSP and the NLSP
    occurs when the LSP is pure Bino, and the NLSP a pure Wino. In this case $M_{LSP}=M_1$ and $M_{NLSP}=M_2=2M_1$.
  A Higgsino admixture in these states can only make the difference {\it smaller}.}.

In many models, the next-to-next lightest SUSY particle (the NNLSP) is close in mass to the
NLSP.
     Therefore, another aspect of experimental importance is the mass-differences to the LSP
     of both the lightest chargino and of the second lightest neutralino: either of these
     could be either the NLSP or the NNLSP.
     This aspect is shown in Figure \ref{fig:dmx1dmn2} which shows $\Delta(M)$ for $\XPM{1}$ versus that of $\XN{2}$.
     One notes three distinct regions    
      \begin{itemize}
      \item  Bino LSP: Both mass differences quite similar, but can take any value;
      \item  Wino LSP: $\Delta(\MXC{1})$ will be small, while  $\Delta(\MXN{2})$ can vary largely;
      \item  Higgsino LSP: Both mass differences often small.
      \end{itemize}
    Note, however, that in the Higgsino LSP case, few models are on the ``Higgsino line'',
      i.e. the case where the chargino is {\it exactly} in the middle of
      mass-gap between the first and second neutralino.
 Finally, Figure~\ref{fig:dmx1mlsp} show  $\Delta(M)$ for $\XPM{1}$ vs. $M_{LSP}$
 for a Higgsino LSP or a Wino LSP.
 Here, one can note that in both scenarios, quite a large spread is possible, and
 that some Higgsino models actually have a chargino LSP.
 The last feature is a point of disagreement between the results of {\tt SPheno} and {\tt FeynHiggs} -
 the latter does not find models with a chargino LSP\footnote{There are differences to be expected as in case of
   {\tt SPheno} this is a pure $\overline{\mathrm{DR}}$ calculation whereas in {\tt FeynHiggs} on-shell calculation is preformed.This issue needs further investigation.}.

     \section{SUSY In the Briefing-book}
     In the physics Briefing-book of the update of the European strategy for particle
     physics \cite{Strategy:2019vxc}, the reach of searches for electroweak SUSY particles
     for different proposed future accelerators are
     presented in chapter 8.3.2, and illustrated by two figures. We will discuss these in this section.
     \subsection{Bino LSP\label{sect:bbbino}}
    \begin{wrapfigure}{R}{9cm}
    \begin{center}

      \subfloat[][$M_1 , \mu$>0, $M_2$<0]{\includegraphics [scale=0.25]{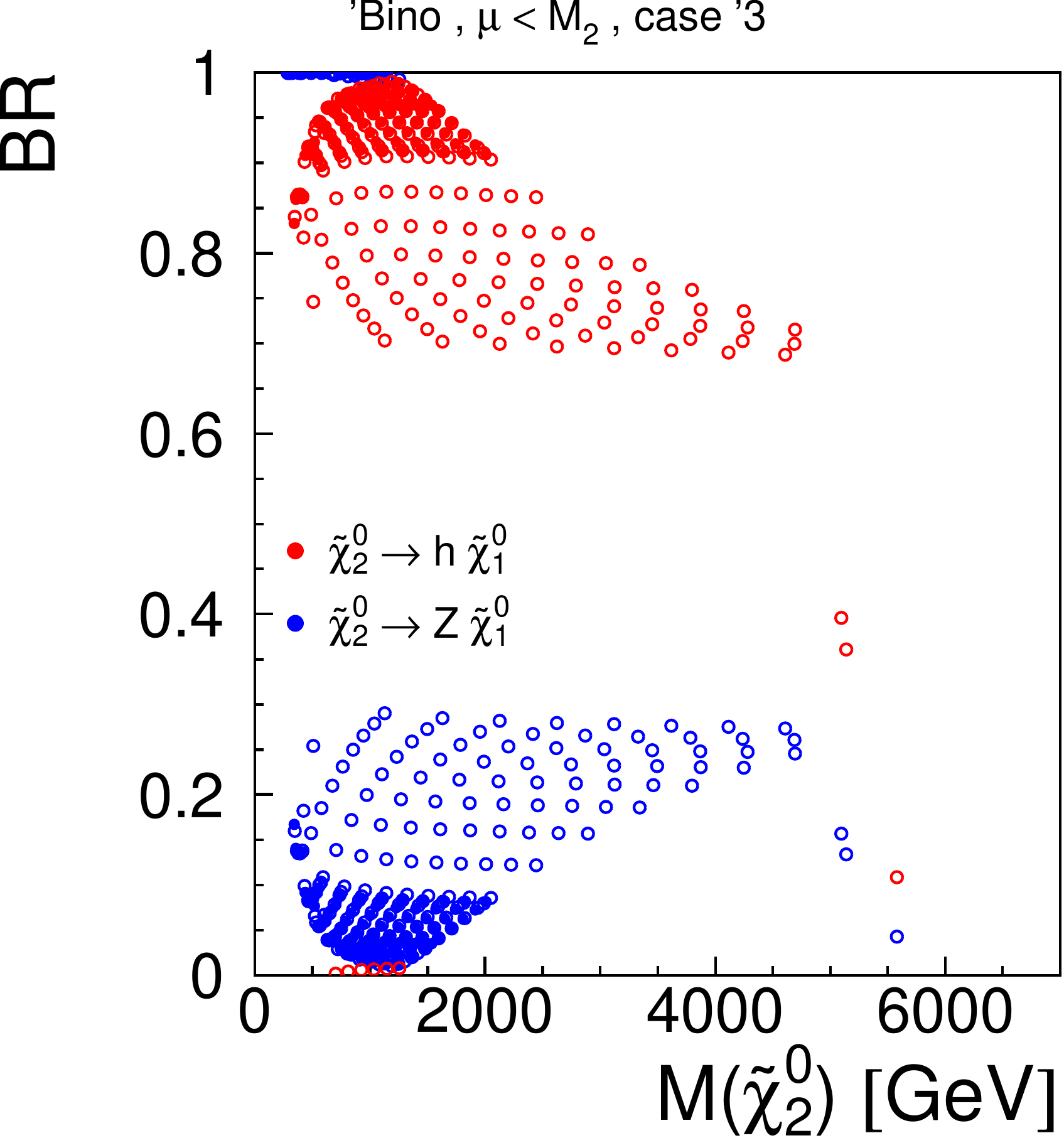}}
      \subfloat[][$M_1 , M_2<0$, $\mu$ >0]{\includegraphics [scale=0.25]{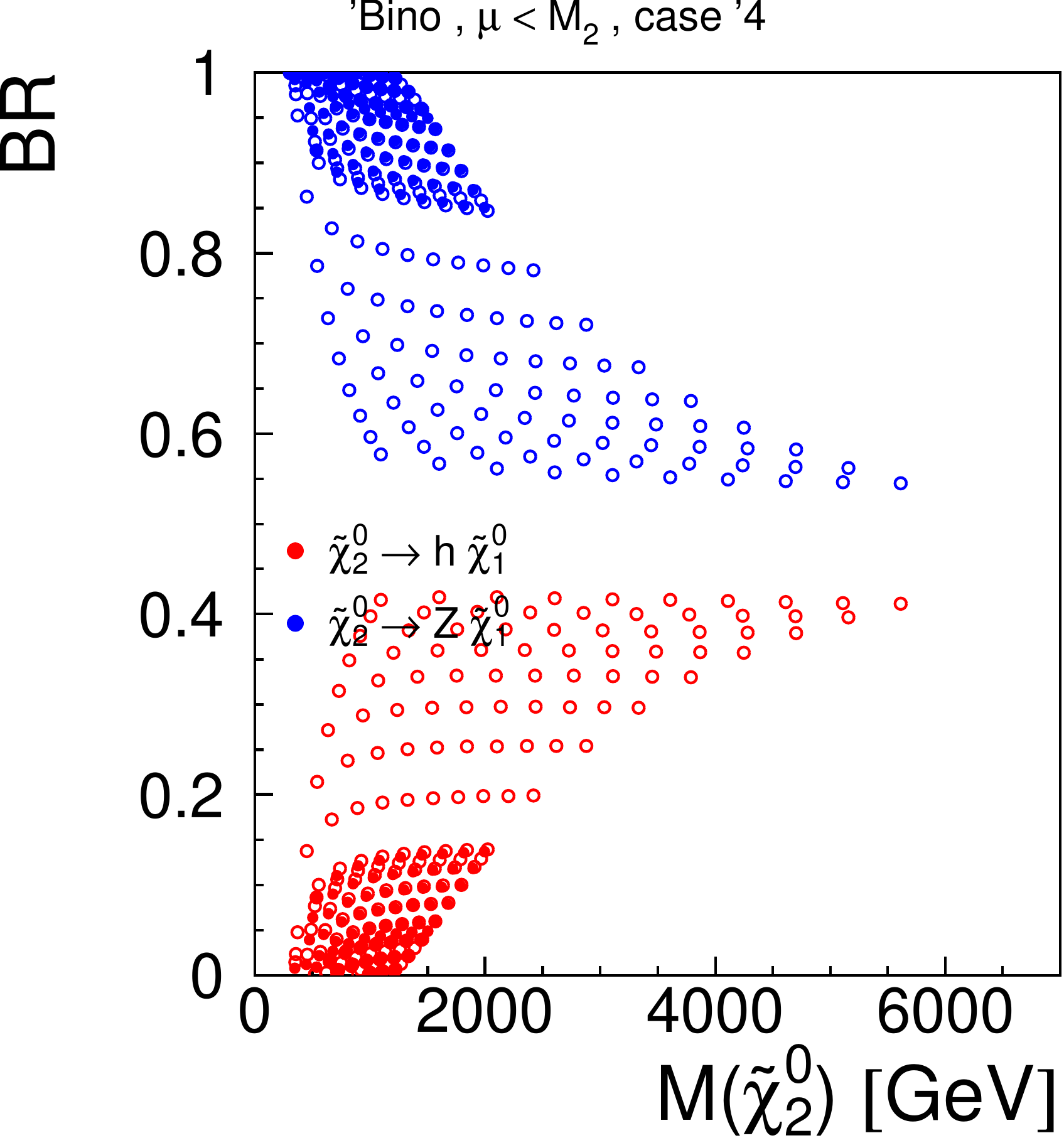}}
      \caption{Branching ratios of $\XPM{1} \rightarrow Z \XN{1}$ (blue)
        and  $\XPM{1} \rightarrow h \XN{1}$ (red) in the Bino LSP case,
        with $|\mu| < |M_2|$, and different signs of $\mu, M_1$, and $M_2$.
        The same grid in absolute values of  $\mu, M_1$, and $M_2$ is used in
        both (a) and (b).\label{fig:brnsigns}}
     \end{center}
     \end{wrapfigure}
    Figure 8.9 in \cite{Strategy:2019vxc} (reproduced here as Figure \ref{fig:bbbino}) shows the
     estimated reaches in the Bino LSP case, i.e. for the large $\Delta(M)$ case.
     The signature  for this scenario  at pp-colliders are events with large
     missing transverse momentum (MET).
     Due to the large mass-difference, the missing momentum originates from the invisible SUSY particles themselves,
     i.e. there is no  need for a system recoiling against the SUSY particles.
     To first order, this makes the analyses robust, since only the mass-difference
     is needed to predict the signal topologies.
     However, there are still a number of model-dependencies, discussed below.
     The sources of the curves shown for the various pp options
     are from the projection to HL-LHC by the ATLAS collaboration \cite{ATL-PHYS-PUB-2018-048}.
     The result presented in that publication is the
     solid red line in Figure \ref{fig:bbbino}a, and the actual plot from the paper is reproduced as Figure \ref{fig:bbbino}b.
     This curve is extrapolated giving the HE-LHC curve (red-dotted).
     Several things should be noted: The curve shown is the exclusion reach,
     not the discovery reach (for the CLIC and ILC curves, the differences between exclusion and discovery reach
     are less than the width of the lines in the figure).
     The ATLAS result is only shown down to $\MXC{1}$= 500 GeV,
     the region below this is just a guide-the-eye straight line.
     The the chosen decay-mode ($\XPM{1} \rightarrow Z \XN{1}$) is the most sensitive at low  $\Delta{M}$.
     The other mode  ($\XPM{1} \rightarrow h \XN{1}$), shown in Figure \ref{fig:bbbino}c is less
     powerful in this region. On the other hand the higgs mode is expected to probe higher $\MXC{1}$ at
     the highest $\Delta{M}$. At CLIC or ILC, one does not need to make such a distinction.
     The issue of the dominant decay mode is important, as illustrated in Figure \ref{fig:brnsigns}.
     In these figures, we show the branching ratios in Bino-LSP models (i.e. models where $M_1$ is the smallest
     of the bosino parameters), when only the relative signs
     of $M_1,M_2$ and $\mu$ are modified.
     %\footnote{For any value of  $M_1,M_2$ and $\mu$ there are four
     %  distinct choices of signs, not eight, since one sign can be rotated away}.
     The observation is that whether the $Z$ or the $h$ mode is dominant depends crucially of the
     choice of the relative sign, and hence that the exclusion-region should be the {\it intersection} of
     Figures  \ref{fig:bbbino}b and c, not the {\it union}.

\begin{figure}[t]
        \begin{center}
      \subfloat[][Higgsino LSP]{\includegraphics [scale=0.27]{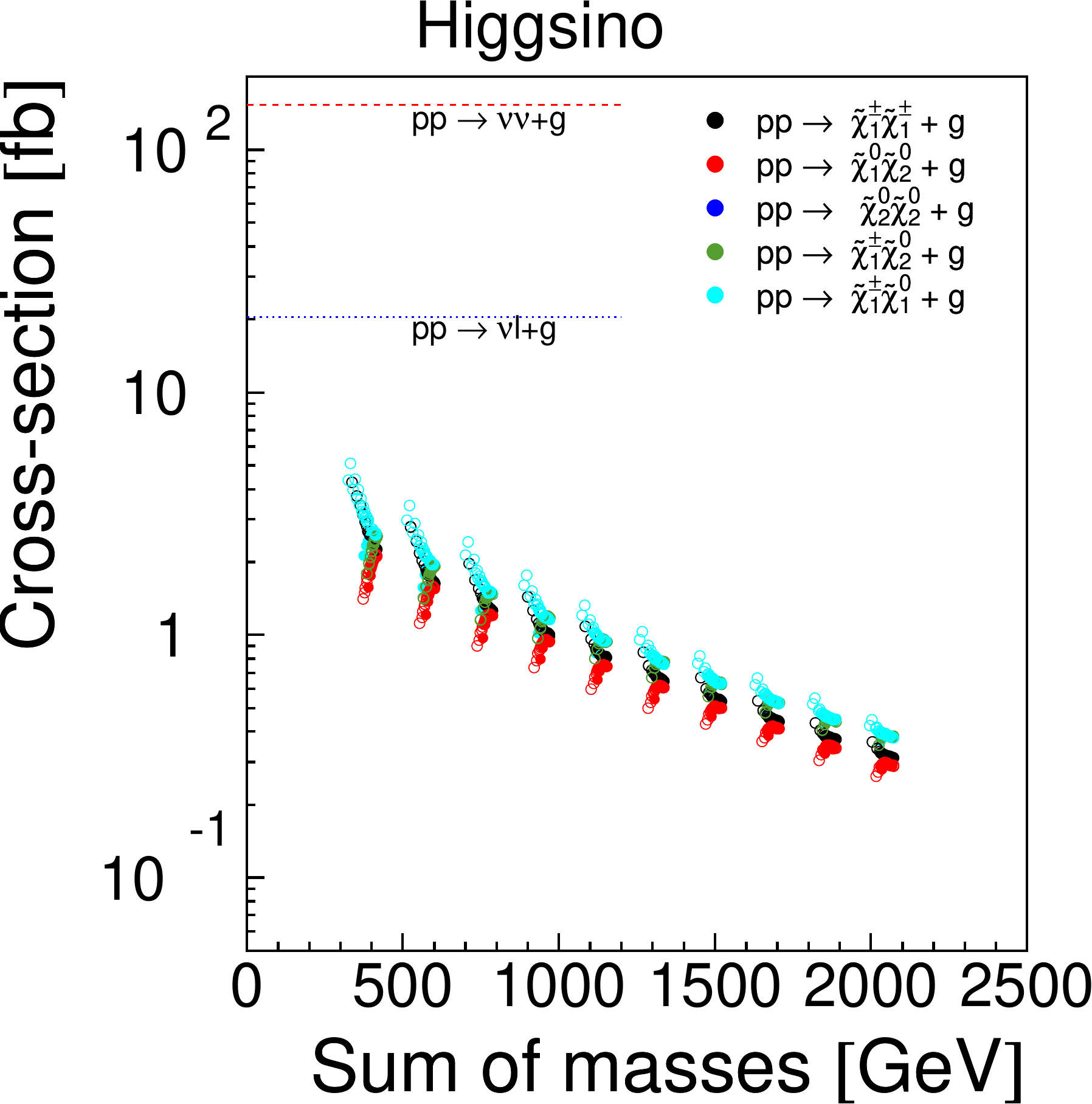}}
      \subfloat[][Wino LSP]{\includegraphics [scale=0.27]{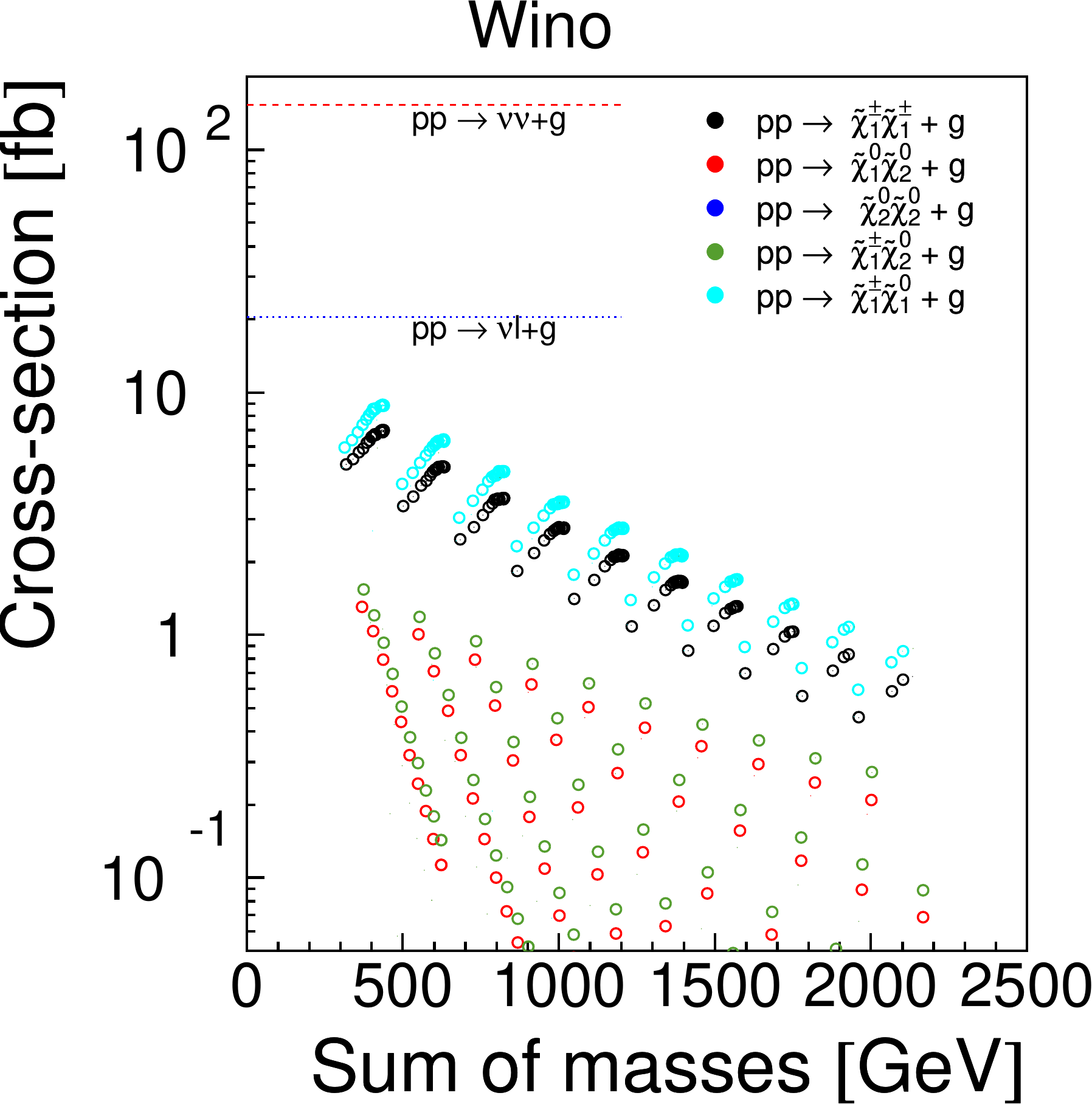}}
      \subfloat[][Bino LSP]{\includegraphics [scale=0.27]{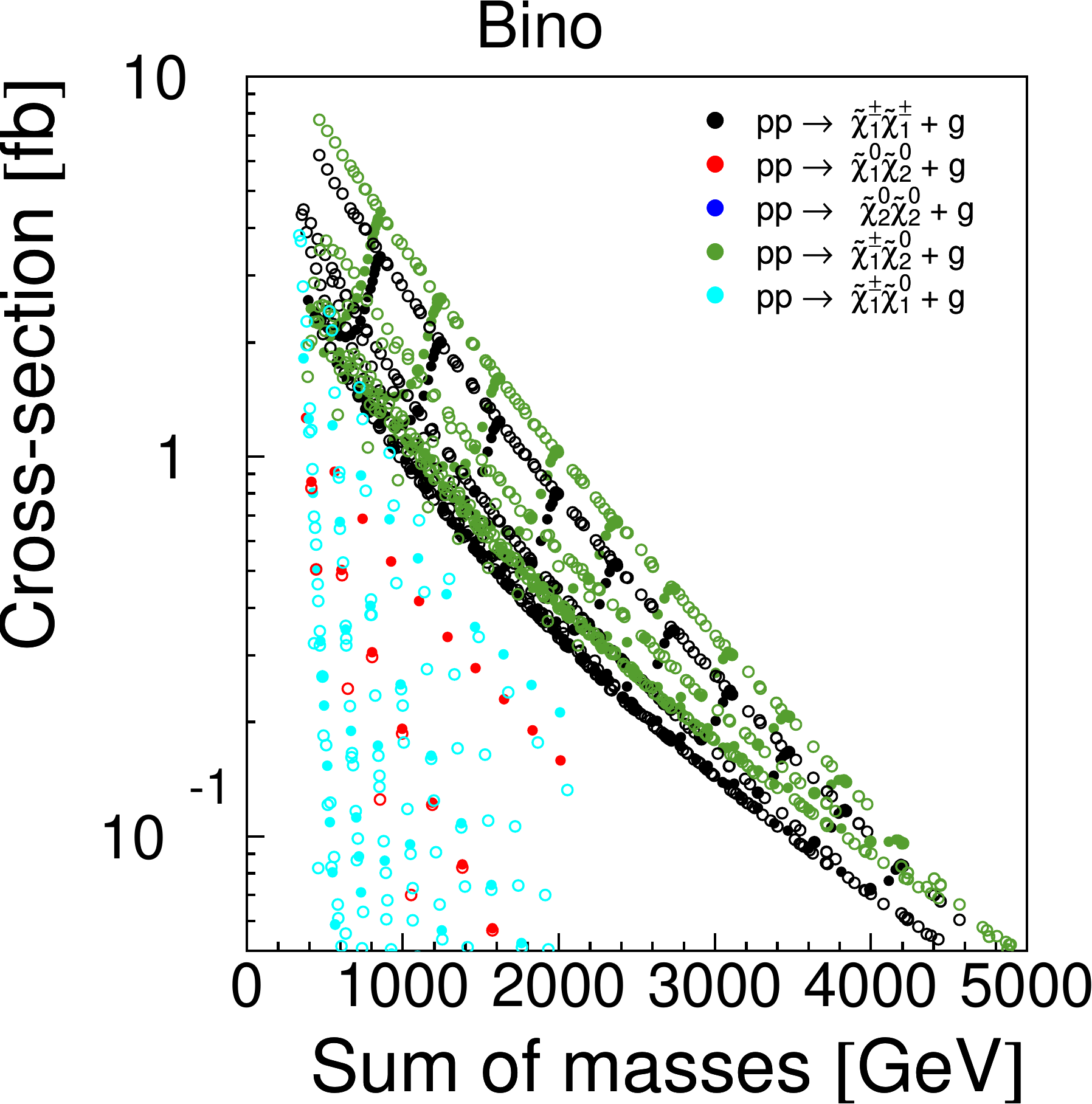}}
      \caption{Cross sections  for $pp \rightarrow$ two uncoloured bosinos + a gluon,
        as a function of the sum of the masses of the two bosinos.
               The five different final states are shown separately, as indicated in the figures.\label{fig:fccxsect}}
     \end{center}
     \end{figure}

\begin{figure}[b]
        \begin{center}
      \subfloat[][]{\includegraphics [scale=0.26]{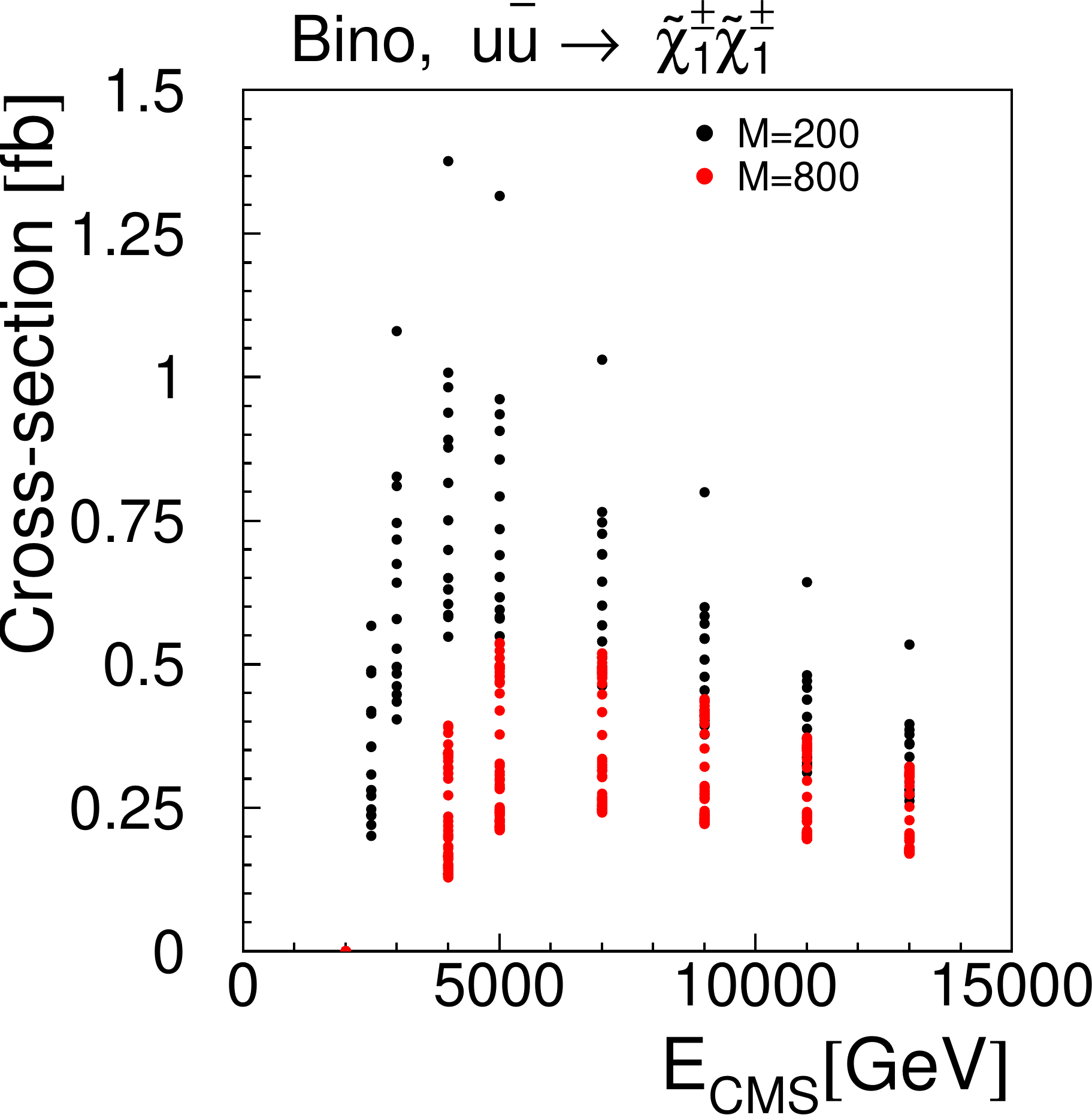}}
       \subfloat[][]{\includegraphics [scale=0.26]{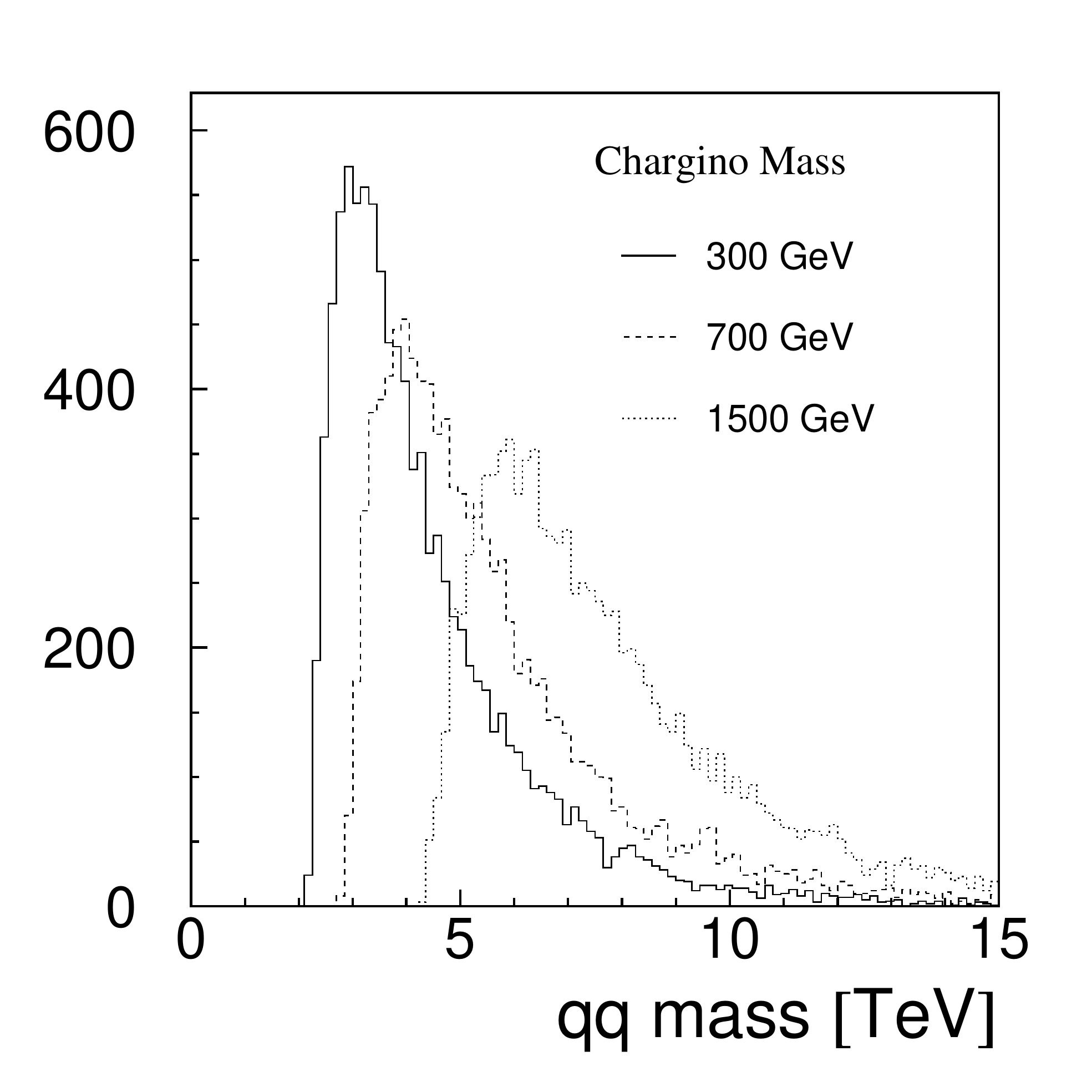}}
      \subfloat[][]{\includegraphics [scale=0.26]{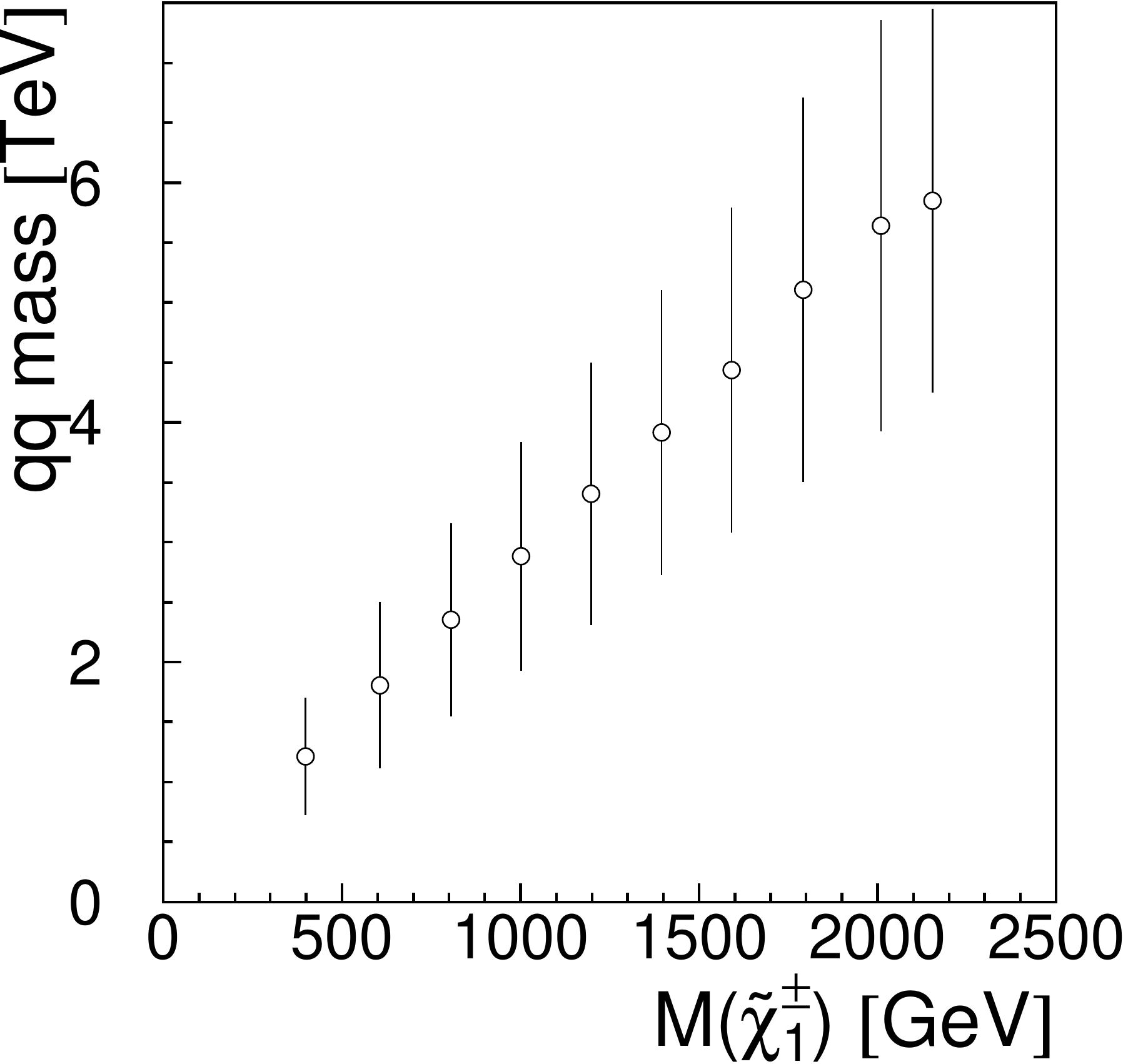}}
      \caption{Properties of $\XPM{1}$ production:
        (a) Cross-section at fixed $m_{qq}$; (b) Distributions of $m_{qq}$ at different $\MXC{1}$ in pp;
        (c) Average $m_{qq}$ vs. $\MXC{1}$ in pp. The error-bars
        represent the r.m.s. of the distribution, not the r.m.s. of the mean.\label{fig:xsectvsmqq}}
     \end{center}
   \end{figure}

     One can note that the exclusion region remains below a line with slope $\sim 1/2$
     when luminosity and/or energy is increased. The reason for this is as follows:
     Figure \ref{fig:fccxsect} shows how the cross-section varies with the sum of the
     two bosino masses at FCChh-conditions. Here a simple setup - which is nevertheless adequate for
     illustrating the scaling behaviour - was used to calculate
     cross-section $\times$ branching-ratios using {\tt Whizard}. The process is
     $pp \rightarrow$ uncoloured bosinos + gluon, with the {\tt Whizard}-default parton
     density function (CTEQ6L1\cite{Pumplin:2002vw}). Two observations can be made. Firstly, there is a
     close to exponential fall of the cross-section with mass. Secondly, the
     cross-section at any given mass can vary by a factor $\sim$ 2, by varying the parameters
     of the model.
    The exponential fall-off with increasing mass comes around for the following reason:
     If the mass of the interacting quark-pair would be fixed (equivalent to the
     situation at a lepton collider, where the invariant mass of the initial state
     is fixed (= 2$\times E_{beam}$)), the cross section versus the mass of
     the produced bosino pair initially rises proportionally to $\beta$ - typical for fermion pair-production - followed
     by a fall-off proportional to $\frac{1}{s}$, see Figure \ref{fig:xsectvsmqq}a.
     Once this is folded with the distribution of $m_{qq}$ given by the rapidly falling
     parton densities\footnote{Note that for the Drell-Yan production of the bosino pair,
       at least one of the partons must come from the sea}, the actual distribution on $m_{qq}$,
     given that a bosino production took place shows a distribution - albeit broad - that correlates with
     the mass of the bosino pair (Figure \ref{fig:xsectvsmqq}b). This correlation is
     close to linear, as can be seen in \ref{fig:xsectvsmqq}c.

   \begin{figure}[t]
        \begin{center}
          \subfloat[][]{\includegraphics [scale=0.38]{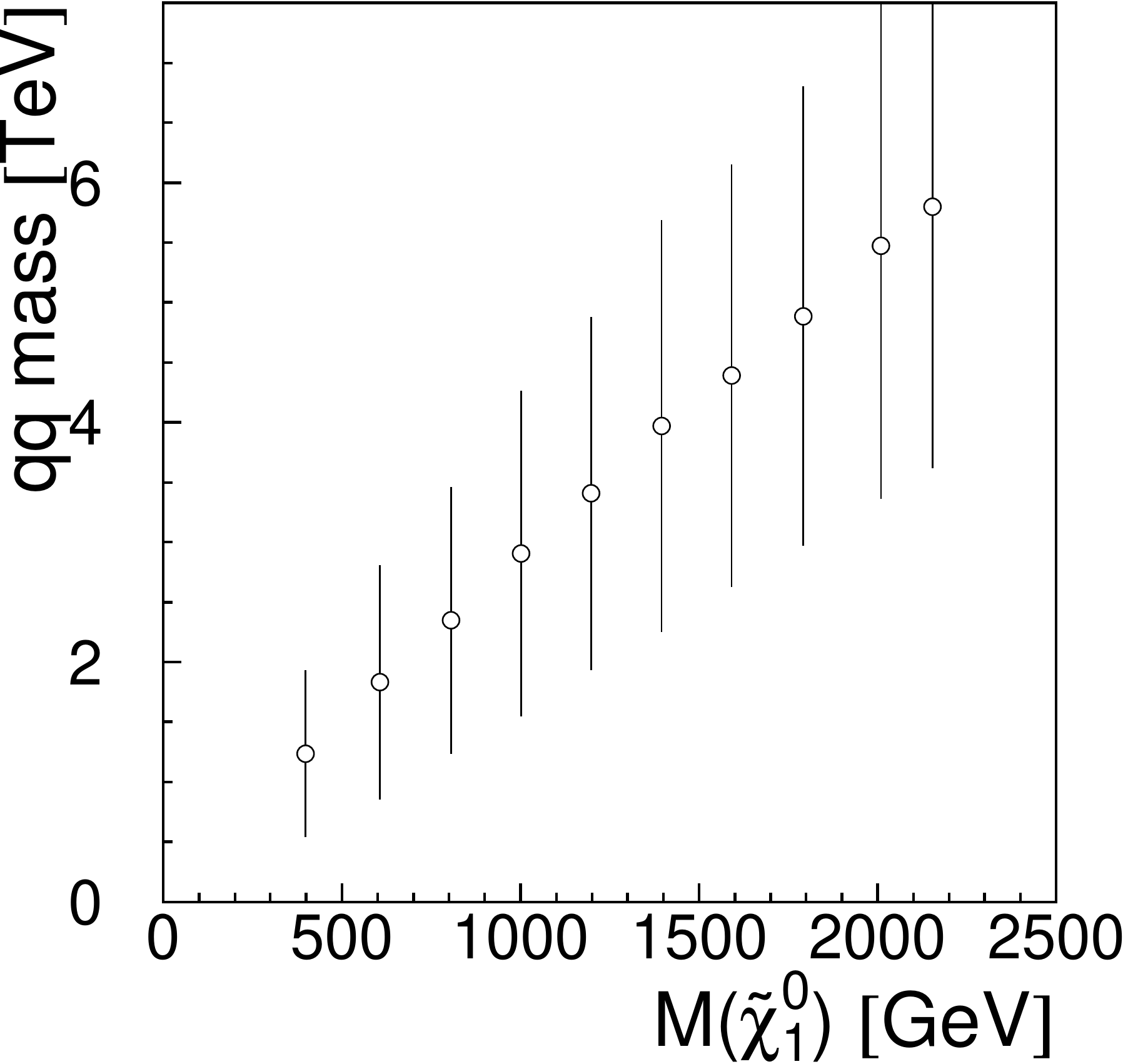}}
       \hskip 0.3cm
      \subfloat[][]{\includegraphics [scale=0.38]{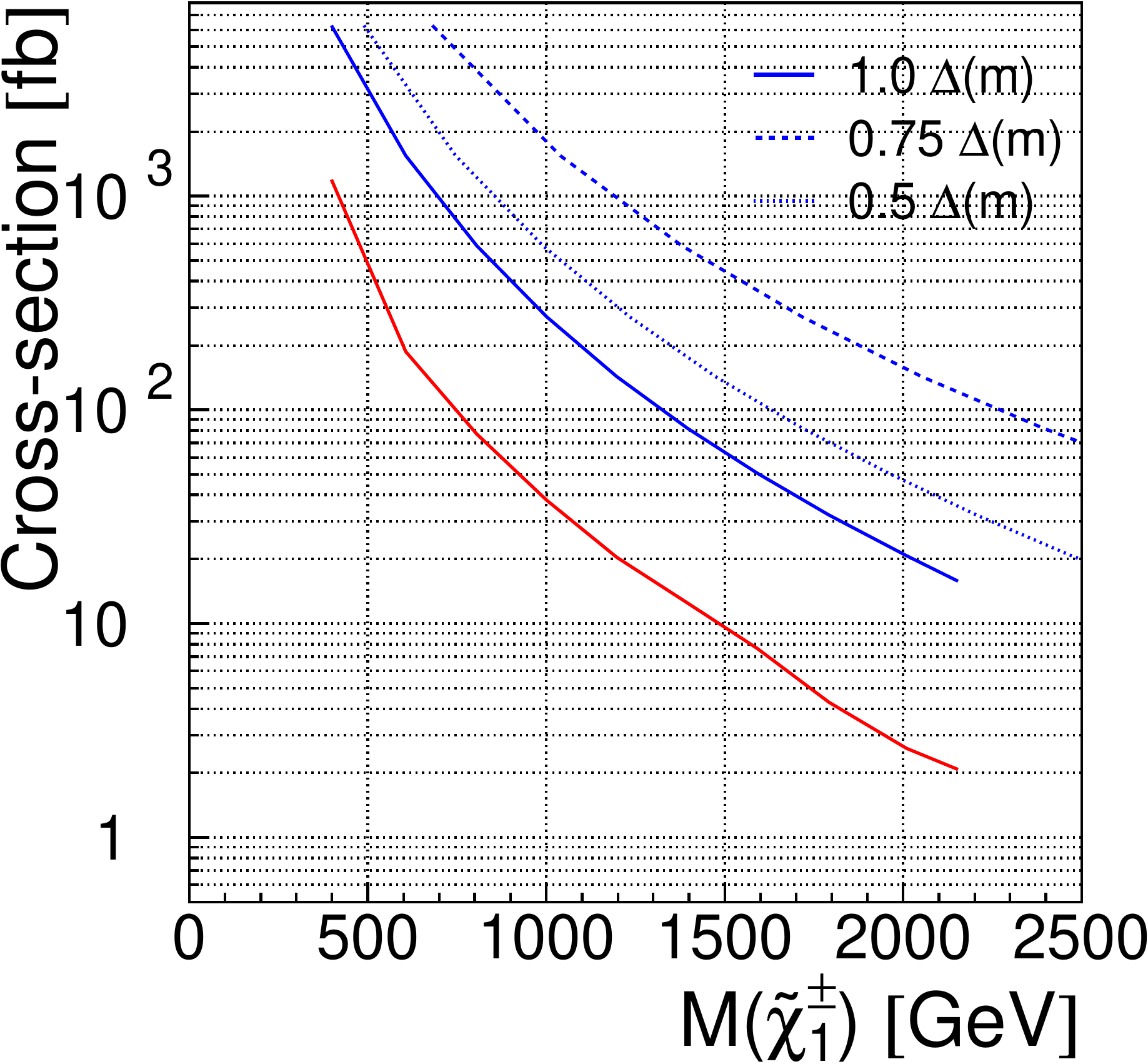}}
       \caption{Properties of signal and background  of $\XPM{1}$ production.
         (a) Average $m_{qq}$ in $pp \rightarrow Zg \rightarrow \nu\bar{\nu}g$
         versus $\MXC{1}$ when the cut on the $\nu\bar{\nu}$ transverse momentum is adjusted according to the
         expected missing transverse momentum of a $\XPM{1}$. It is set to 0.85 $\MXC{1}$. (b)
         Cross-section of $\XPM{1}$-pair production (red-solid),
         together with that of $pp \rightarrow Zg \rightarrow \nu\bar{\nu}g$
         with cuts on the  $\nu\bar{\nu}$ increasing with
         $\MXC{1}$ for three choices of $\Delta(M)$, keeping the cut at 75 \% of the
         highest possible missing $p_T$ for the signal: nominal (blue-solid, cut = 0.85 $\MXC{1}$),
         $\frac{3}{4}\Delta(M)_{nom.}$ (blue-dotted,  cut = 0.7 $\MXC{1}$),
         and  $\frac{1}{2}\Delta(M)_{nom.}$ (blue-dashed,  cut = 0.5 $\MXC{1}$).\label{fig:fccbckxsect}}
     \end{center}
     \end{figure}

Generally, the maximum missing momentum due to the invisible LSPs is
\begin{align}
    \bcancel{p}_{max} =  & 2 \gamma_{f\bar{f}} \beta_{f\bar{f}} \gamma_{\scriptscriptstyle NLSP} E_{\scriptscriptstyle LSP} + 2 \gamma_{f\bar{f}}  \gamma_{\scriptscriptstyle NLSP} p_{\scriptscriptstyle LSP}
\end{align}
In the Bino case, the initial $f\bar{f}$-system need not be boosted, so $\gamma_{f\bar{f}} \beta_{f\bar{f}} \approx 0$, $ \gamma_{f\bar{f}} \approx 1$,
$\gamma_{\scriptscriptstyle NLSP} = E_{\scriptscriptstyle NLSP} / M_{\scriptscriptstyle NLSP} \approx M_{f\bar{f}}/ 2 M_{\scriptscriptstyle NLSP}$ and
\begin{align}
    \bcancel{p}_{max} \approx  & 2 \frac{M_{f\bar{f}}}{2 M_{\scriptscriptstyle NLSP}} p_{\scriptscriptstyle LSP} \approx  2 \frac{M_{f\bar{f}} }{2 M_{N\scriptscriptstyle LSP}} \frac {M^2_{\scriptscriptstyle NLSP}-M^2_{\scriptscriptstyle LSP}}{2 M_{\scriptscriptstyle NLSP}}
\end{align}
where the last step is because 
at interesting points, the SUSY particle mass is much above it's SM partner (even if this is a $W$, $Z$ or $h$).
From Figure \ref{fig:xsectvsmqq}c, we know that $M_{f\bar{f}} \approx 3 M_{\scriptscriptstyle NLSP}$, and
\begin{align}
    \bcancel{p}_{max} \approx  & \frac{3}{2} M_{\scriptscriptstyle NLSP} \left ( 1 - (\frac {M_{\scriptscriptstyle LSP}}{M_{\scriptscriptstyle NLSP}})^2 \right )
\end{align}

Hence, at these LSP-NLSP mass-ratios, the missing $p_T$ due to the invisible LSP is proportional to
the  bosino-mass.
     This means that one can increase the missing $p_T$  cut while conserving a given the signal efficiency as one
     searches for higher bosino masses.
     The missing  $p_T$ from irreducible background (typically Dell-Yan + gluon, with
     $Z\rightarrow \nu\bar{\nu}$) is obviously independent for the bosino masses, but does
     depend on the required missing $p_T$, essentially the $p_T$ of the gluon.
     Figure \ref{fig:fccbckxsect}a shows that if the $p_T$ cut applied is the one
     that keeps the same signal efficiency at any given $\MXC{1}$,
     the $M_{f\bar{f}}$ of the background events passing the cut also follows a linear trend
     quite similar to that of the signal, seen in \ref{fig:xsectvsmqq}c.
     In the figure, the cut has been set to 0.85 $\MXC{1}$, which according to
     Eq. 3 corresponds to 0.75 $\times$ the maximal missing $p_T$ from $\XPM{1}$ pair-production,
     when $M_{NLSP}=2M_{LSP}$, as it is at the border of the currently excluded region.
     \begin{figure}[b]
    \begin{center}
      \subfloat[][Z mode]{\includegraphics [scale=0.9]{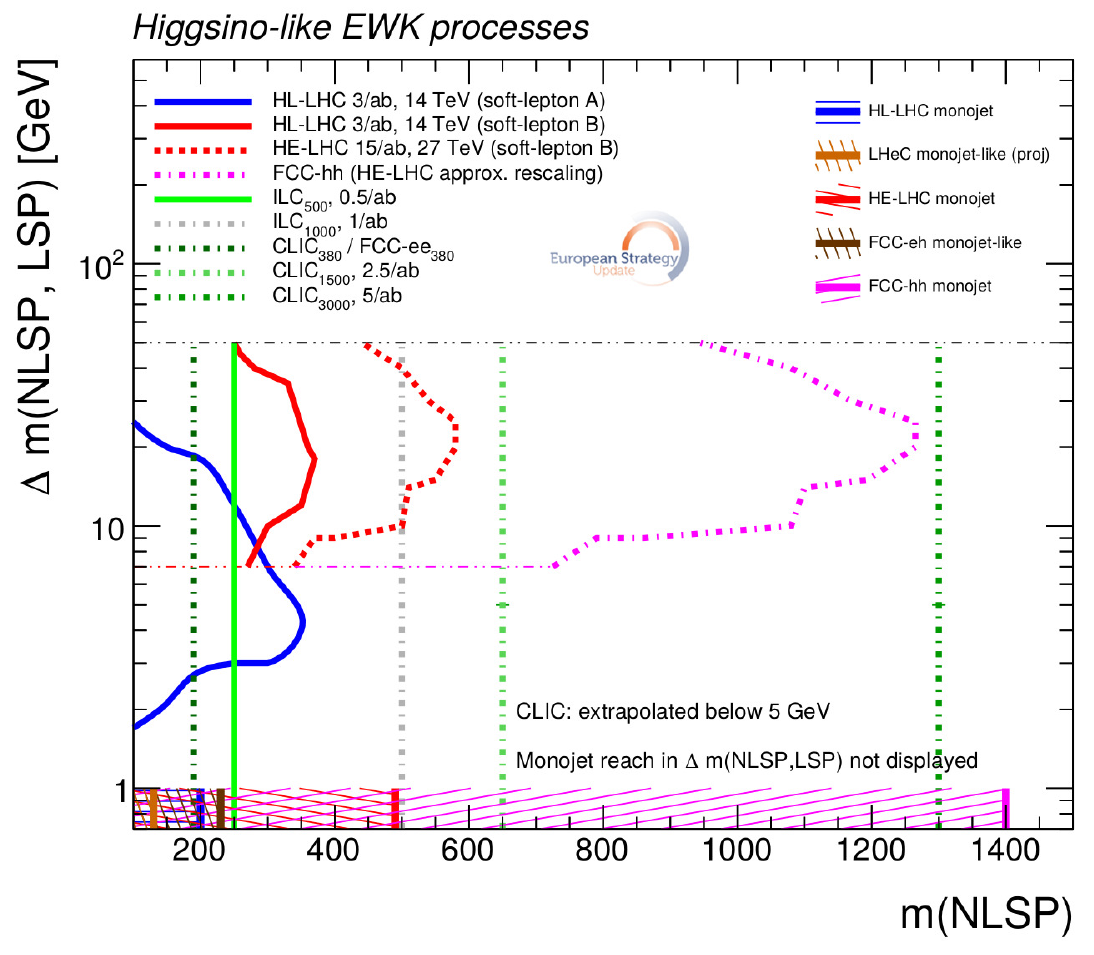}}
      
      \subfloat[][Z mode]{\includegraphics [scale=0.4]{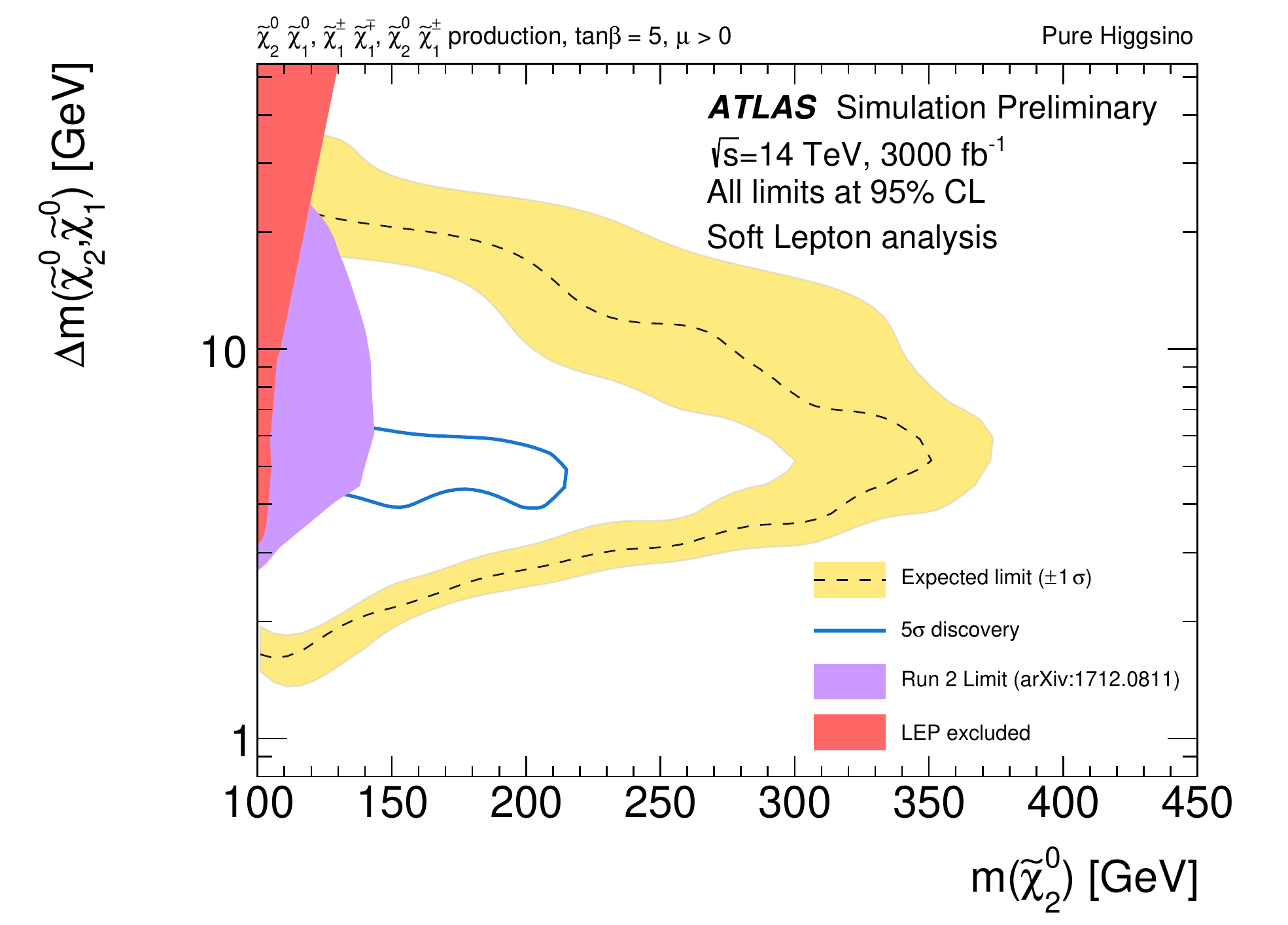}}
      \subfloat[][Z mode]{\includegraphics [scale=0.35]{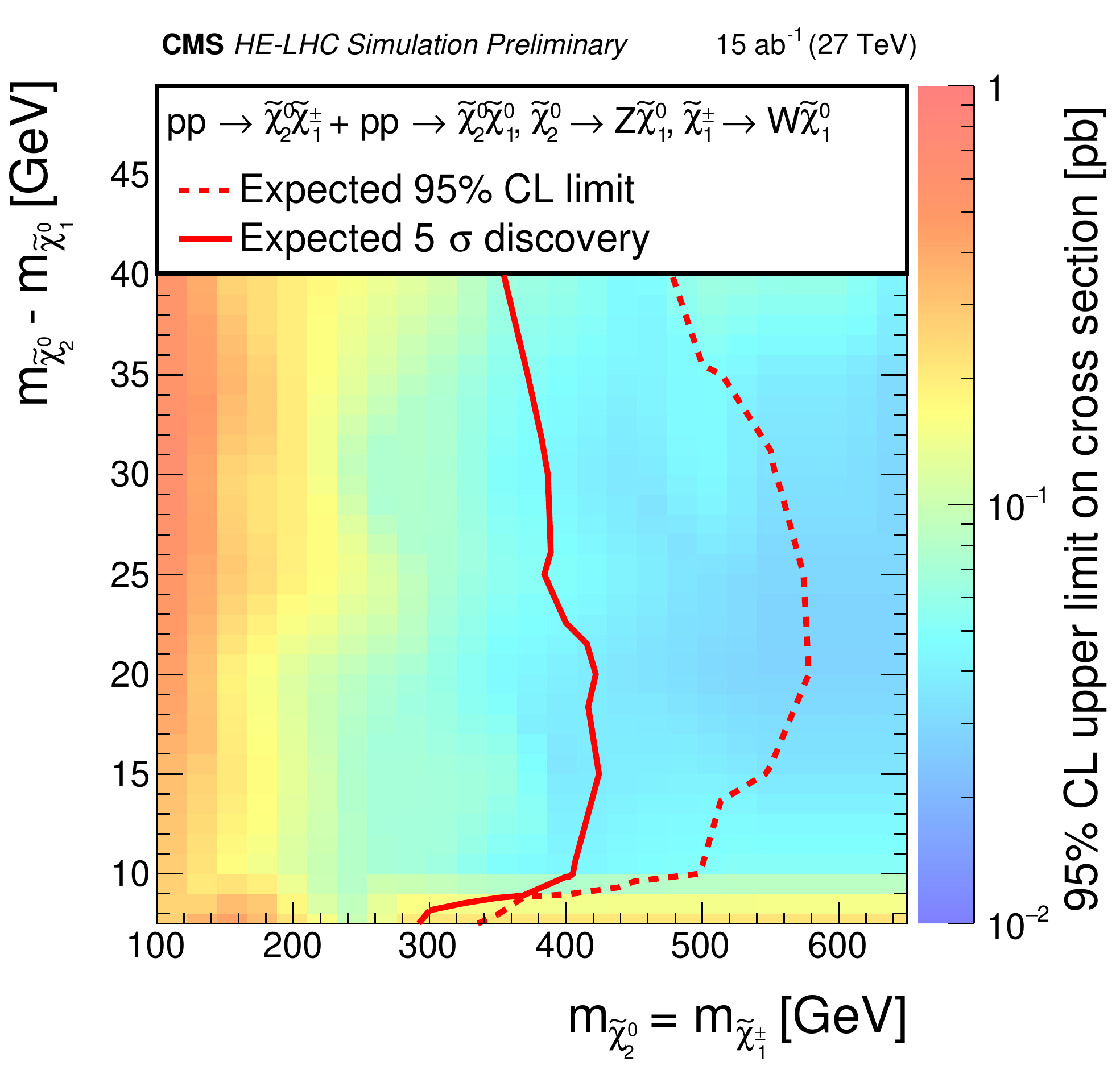}}
      \caption{The reaches in the low $\Delta(M)$ (Higgsino- or Wino-LSP) region, as
        reported in \cite{Strategy:2019vxc} (top),
        and the two projections to HL-LHC from ATLAS \cite{ATL-PHYS-PUB-2018-031} and CMS\cite{Sirunyan:2018iwl,CMS:2018qsc}
        (bottom).(b) corresponds to
        the solid blue line in (a), (c) to the red line\label{fig:bbwinohiggsino}}
     \end{center}
     \end{figure}
     The cross-section for this process therefore also falls exponentially with the required
     $p_T$, (see Figure \ref{fig:fccbckxsect}b) meaning that the 
     signal-to-background ratio will remain constant along lines through the origin in the $\MXC{1}$ vs. $M_{LSP}$ plane.
     On the other hand, Eq. 3 also shows that the  missing  $p_T$ decreases with increasing
     $M_{\scriptscriptstyle LSP}/M_{\scriptscriptstyle NLSP}$,   meaning that to exclude lower $\Delta(M)$ at the same efficiency requires
     a decrease in the cut, leading to a large increase in the background.
%     E.g. the cut that keeps the same efficiency on the line $M_{LSP} = 3 M_{NLSP}/4$ as that on $M_{LSP} = M_{NLSP}/2$ 
%     (i.e. on a line on which the gap is reduced by a factor of two),
%     requires to apply the cut appropriate for half the $M_{\scriptscriptstyle NLSP}$.
     Comparing the solid and dashed blue lines in Figure \ref{fig:fccbckxsect}b shows that to half
     the excluded $\Delta(M)$ at $\MXC{1}$ = 2 TeV would require $\sim$ 10 times
     more luminosity, assuming that no new background sources would start contributing (e.g. jet-energy resolution, jet-energy scale, non-direct
     neutrinos, etc.), which clearly is an unrealistic best-case.

     To conclude this discussion of the Bino LSP case, we note that although the signal is robust,
     there are a number of issues that must be taken into account: The analyses are typically performed
     using a set of processes involving production of different combinations of LSPs, NLSPs and NNLSPs.
     The sensitivity is different for different channels, and Figure \ref{fig:fccxsect}c shows that the cross-sections,
     and their ratios, can vary substantially between models.
     Furthermore, we also noted that the dominating decay mode of the second neutralino is strongly dependent
     on the relative signs of $\mu, M_1$ and $M_2$, and that the sensitivity of the analyses depends also on this.
     To claim that an $M_{NLSP}$-$M_{LSP}$ is excluded, the analysis must be done assuming the least favourable
     production-process and least favourable decay-mode.
     Finally, we pointed out that to extend the coverage to higher NLSP masses at constant LSP mass,
     while retaining the same
     signal efficiency can be done by making the cut on MET {\it stronger}, and that the
     signal-to-background ratio will remain constant when doing so.
     In contrast,
     to
     extend the coverage to higher LSP masses at constant NLSP mass (i.e. to lower $\Delta(M)$) at constant
     signal efficiency, one must
     make the MET-cut {\it weaker}, and thus making the signal-to-background ratio lower.
     A lower MET-cut also implies that proportionally more background originates from fake MET due
     to detector effects, or from non-prompt neutrinos.
     The conclusion is that while progress with increased (parton) luminosity in the $M_{NLSP}$ direction
     is substantial, the progress into the region of lower $\Delta(M)$ will be much less pronounced.
  \subsection{Wino/Higgsino LSP}

     Figure 8.10 in \cite{Strategy:2019vxc} (reproduced here as Figure \ref{fig:bbwinohiggsino}) shows the
     estimated reaches in the Wino or Higgsino LSP case, i.e. for the small $\Delta(M)$ case.
     The two curves come from the HL-LHC projections from ATLAS \cite{ATL-PHYS-PUB-2018-031} (solid blue) and CMS \cite{Sirunyan:2018iwl,CMS:2018qsc}
     %CidVidal:2018eel}
     % ,(=1801.01846,1812.07831)
     (solid red).
     In the CMS case, energy and luminosity
     scaling extrapolation to HE-LHC (dashed red) and FCChh (dashed magenta) are also done.
     Both collaborations make assumptions on the spectrum, however different ones:
     ATLAS assumes $\Delta(\MXC{1})= \Delta(\MXN{2})/2$, while CMS assumes  $\Delta(\MXC{1})= \Delta(\MXN{2})$.
     In Figure \ref{fig:dmx1mlsp}a, the points in our scan that fulfil the ATLAS condition are marked with squares
     and those that fulfil CMS one are marked with triangles.
    \begin{figure}[b]
    \begin{center}
      \subfloat[][Higgsino-LSP, from  \cite{Han:2018wus}]{\includegraphics [scale=1.0]{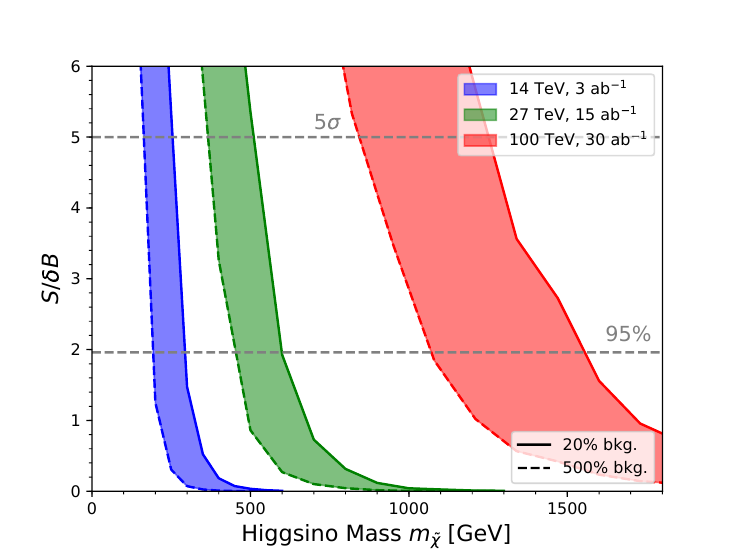}}
      \subfloat[][Wino-LSP,  from \cite{Han:2018wus}]{\includegraphics [scale=1.0]{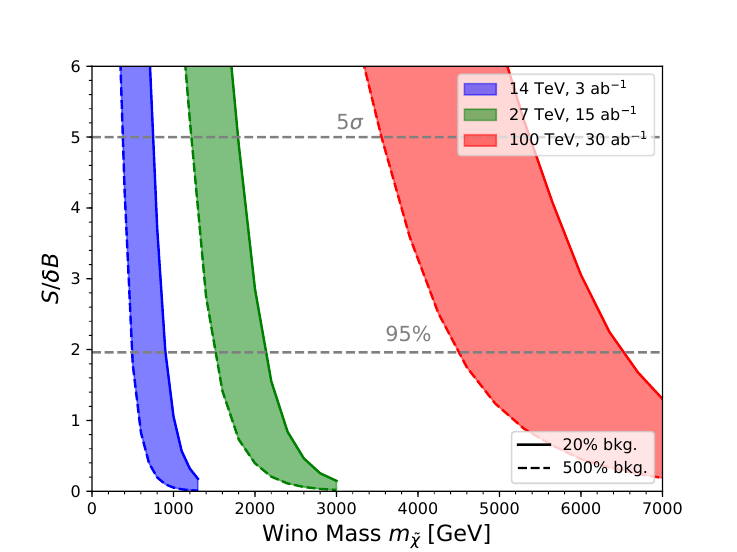}}
      
      \subfloat[][Higgsino-LSP, from \cite{Benedikt:2018csr} ]{\includegraphics [scale=1.05]{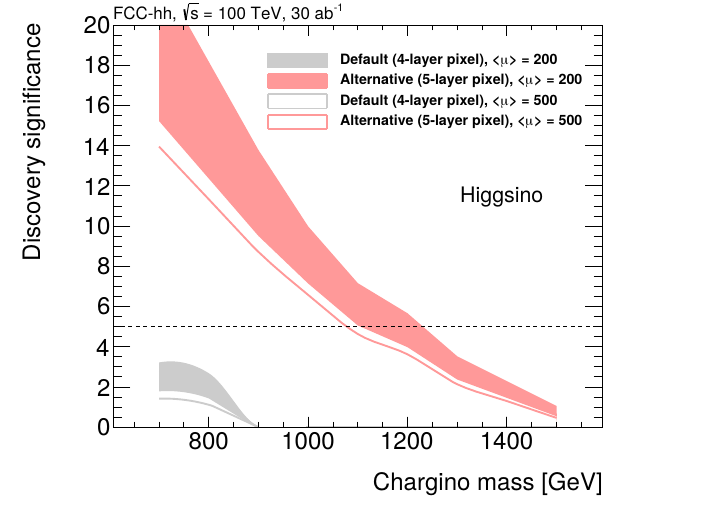}}
      \subfloat[][Wino-LSP,  from \cite{Benedikt:2018csr}]{\includegraphics [scale=1.05]{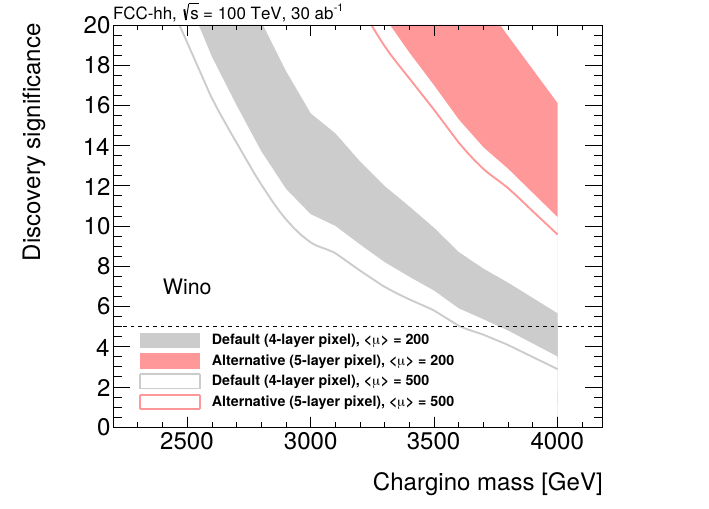}}
      \caption{The sensitivities for the ``Disappearing tracks'' at future pp-colliders.
        In (c) and (d), the grey bands corresponds to the actual FCChh conceptual detector.\label{fig:delphesdis}}
     \end{center}
   \end{figure}
     
     The reason for the sharp cut-off at low mass-differences, seen in \ref{fig:bbwinohiggsino}a,
     is that these searches require leptonic
     decays of the NLSP to be possible to extract a signal from a huge, mainly hadronic, QCD background.
     Lepton identification is therefore essential, and this requires that the particle reaches the
     barrel calorimeters. The cut-off then appears because below this mass-difference, the decay products
     are so soft that they are bent back by the detector B-field inside the radius of the calorimeters.
     This cut-off would be at higher $\Delta(M)$ at FCChh, since the reference detector design \cite{Benedikt:2018csr} foresees
     a considerably larger inner radius of the barrel calorimeter system ($\sim$ 2 m, while ATLAS and
     CMS have an inner radius of 1.5 m and 1.3 m, respectively), and a stronger B-field
     ( 4 T vs. 2(3.8)  T for ATLAS(CMS)).
     From this, one also sees that the CMS properties are closer to the FCChh detector in this respect:
     cut-off transverse momenta are 1.2, 0.7 and 0.4 GeV for FCChh, CMS and ATLAS, respectively.
 In addition, the analyses use the combination of NLSP-NLSP, NLSP-NNLSP and NNLSP-NNLSP production, and assume
 certain relations between the mass difference to the LSP of the NLSP or NNLSP,
 different for ATLAS and CMS, as mentioned above.
Our scan shows that
these relations are quite particular cases.
Whether these assumptions are
essential or not is an open question, but we do conclude that the soft-lepton analysis will progress
to higher NLSP masses, but not to lower $\Delta(M)$, and remains model-dependent.

     \begin{figure}[b]
    \begin{center}
      \subfloat[][Higgsino LSP, SPheno]{\includegraphics [scale=0.27]{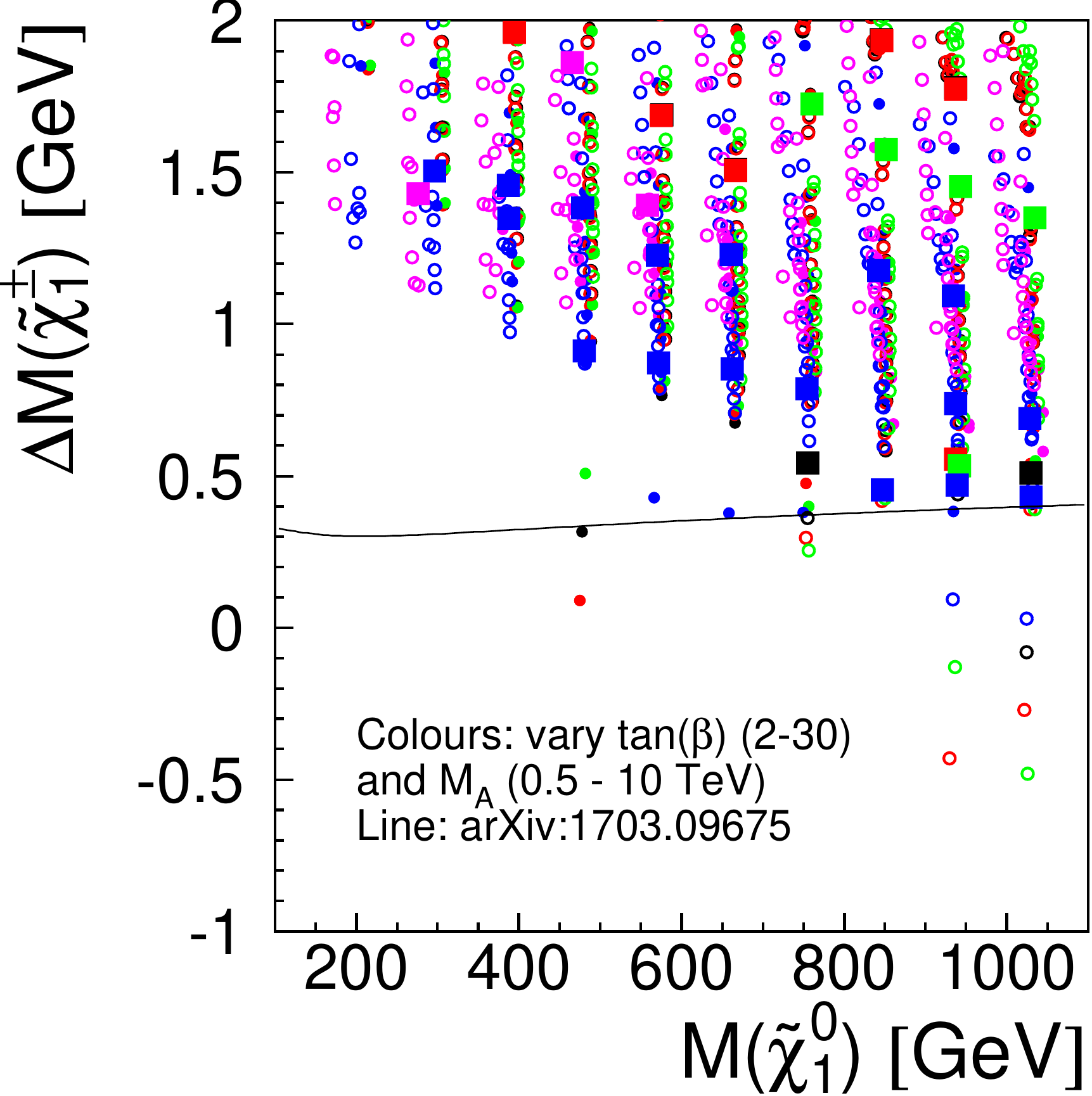}}
      \hskip 1.0cm
      \subfloat[][Higgsino LSP, SPheno]{\includegraphics [scale=0.27]{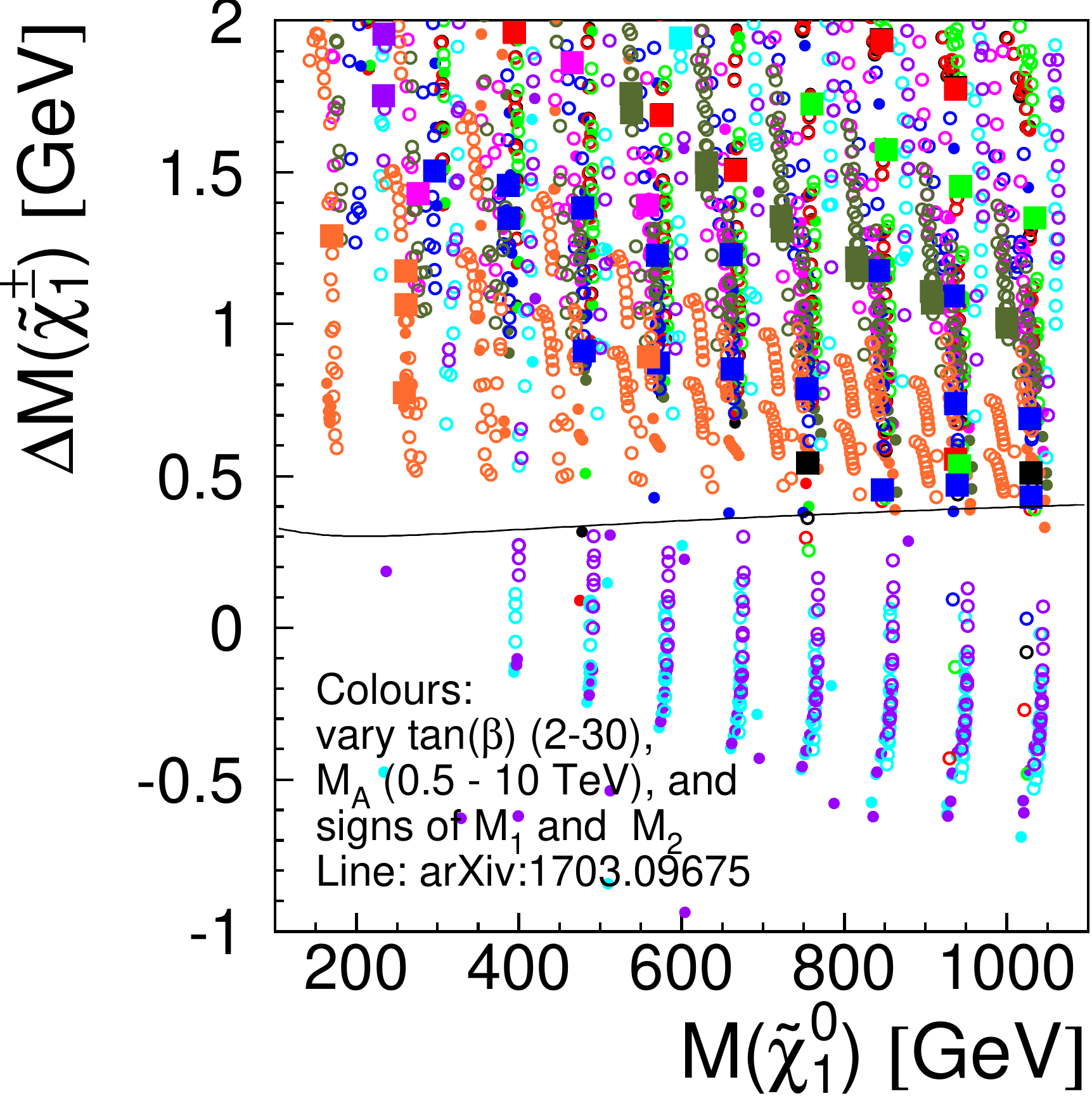}}
      
      \subfloat[][Higgsino LSP, FeynHiggs]{\includegraphics [scale=0.27]{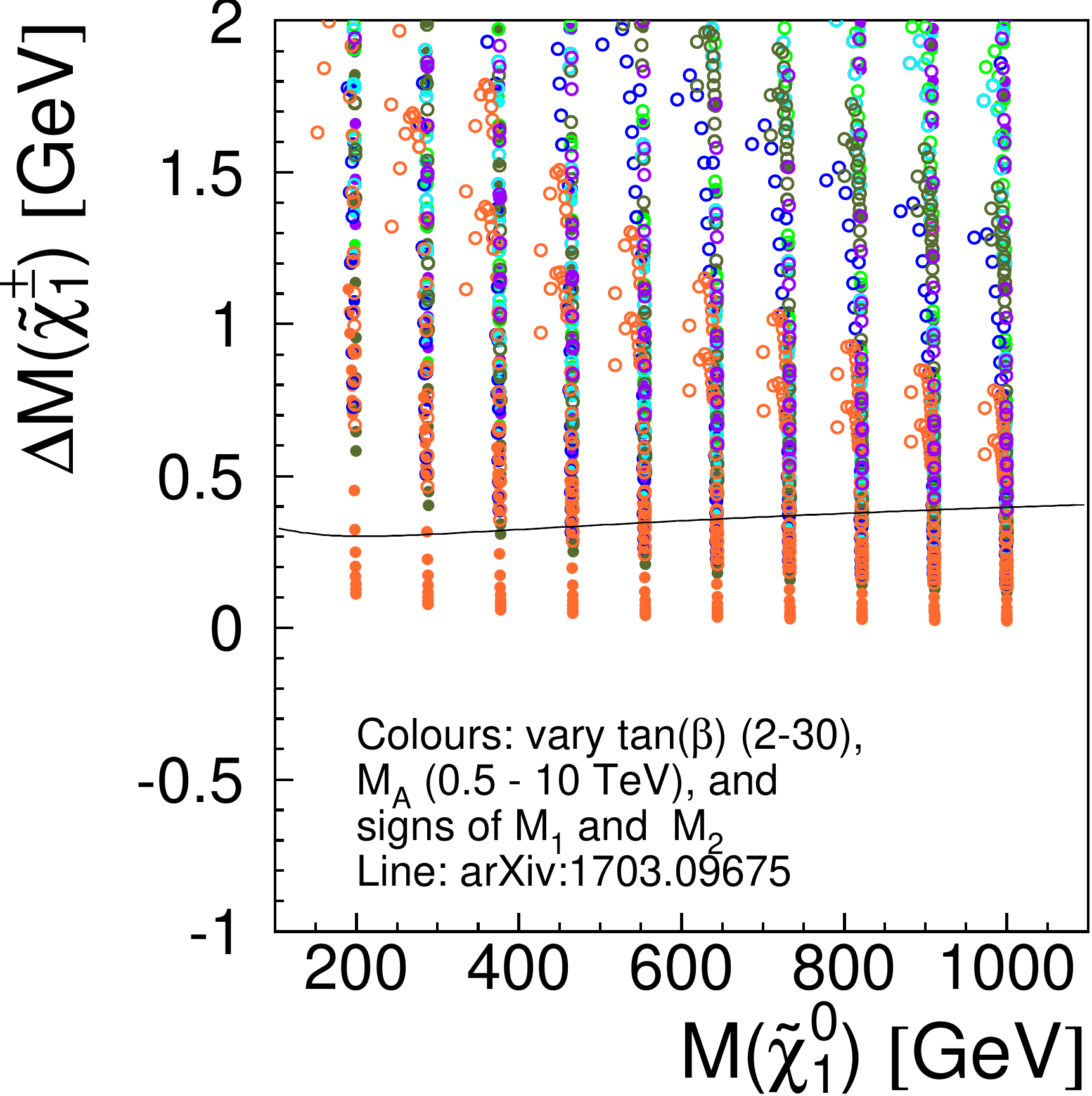}}
      \hskip 1.0cm
      \subfloat[][Wino LSP, SPheno]{\includegraphics [scale=0.27]{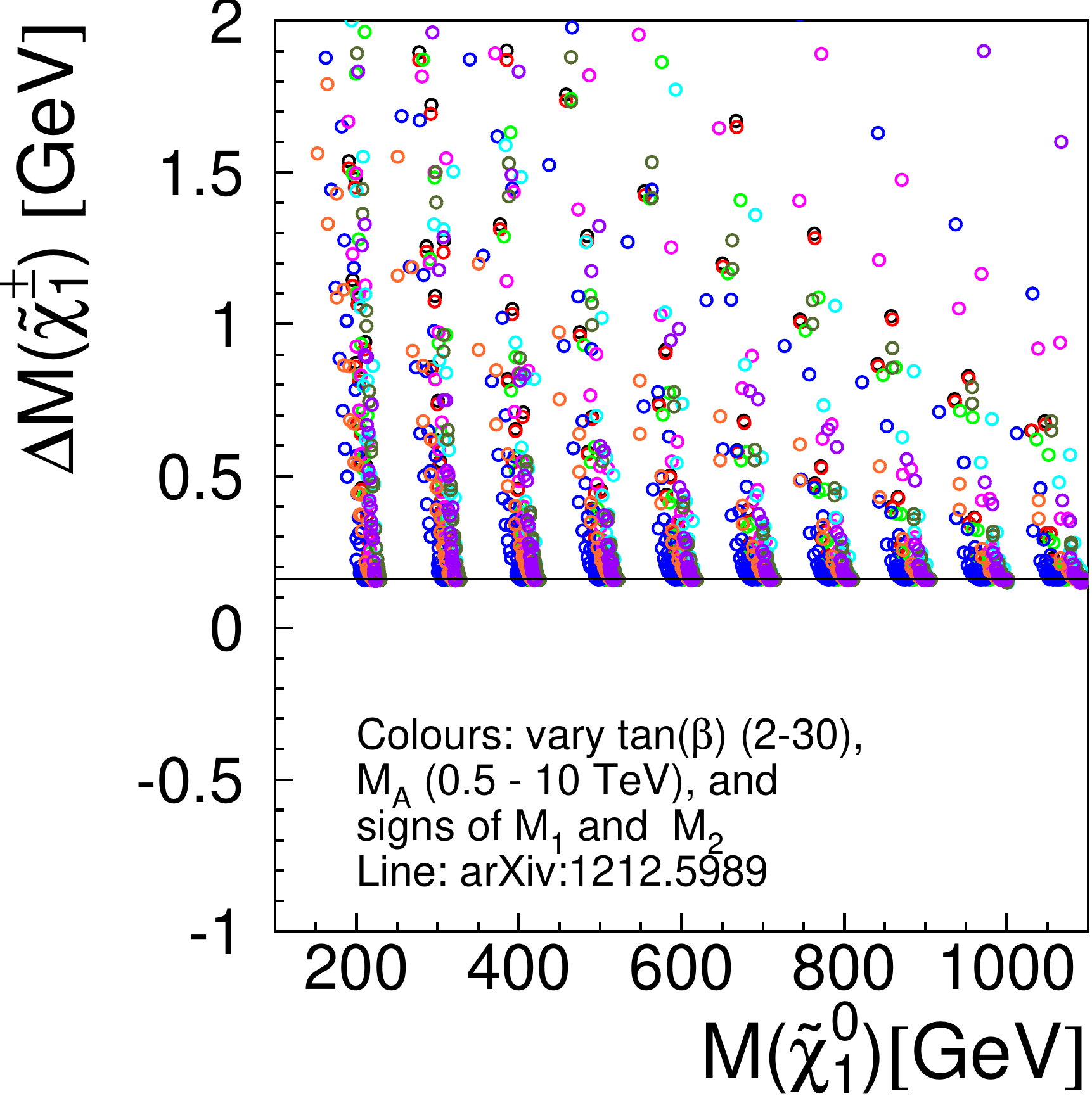}}
      \caption{Zoom-in of Figure \ref{fig:dmx1mlsp}, showing $\Delta(\MXC{1})$ vs. $M_{LSP}$ in the small $\Delta(M)$ region.
        In (a) only models with $\mu, M_1$, and $M_2$ all positive
        are shown.
        In (b) and (c) all models are shown, for both spectrum calculation codes we used.
        The lines are from \cite{Fukuda:2017jmk} (a,b,c) and \cite{Ibe:2012sx}(d),
        and are the mass-differences used in the calculation of the reaches shown in Figures \ref{fig:delphesdis}
        and \ref{fig:delphesmonox}. In (a) and (b), the
        squares are points where  $\Delta(\MXC{1})=\Delta(\MXN{2})/2$, i.e. the ``deep Higgsino'' region;
        the colours are explained in Figure \ref{fig:broadbrush}.\label{fig:bosinodmzoom}}
     \end{center}
   \end{figure}
    The hatched band at the bottom of Figure \ref{fig:bbwinohiggsino}a shows the reach at very low $\Delta(M)$. The upper edge of
       the band at  $\Delta(M)$=1 GeV should not be taken literally; only the reach in $M_{LSP}$ is relevant.
     Two methods are used to estimate the reach at very low $\Delta(M)$: ``Disappearing tracks'' and ``Mono-X''.
     The ``Disappearing tracks'' signature, which consists of a topology where a reconstructed trajectory
     terminates inside the tracking volume, indicating that a decay took place where the decay product
     had a momentum below the threshold of detectability. This signature is effective for cases with low mass-differences,
     since this  potentially implies both a long lifetime of the primary NLSP, and little energy release in it's
     decay.
     The  ``Mono-X'' technique is effective if the decay products of the produced bosinos are so soft that
     they are invisible, or if only LSP-pairs are produced. One then searches for the a large un-balanced
     $p_{T}$ from an initial state radiation, which could be a gluon, photon, $Z$, $W$ or $h$. In pp-collisions,
     the gluon would be the prime candidate.
     
    \begin{figure}[t]
    \begin{center}
       \subfloat[][Higgsino LSP]{\includegraphics [scale=0.27]{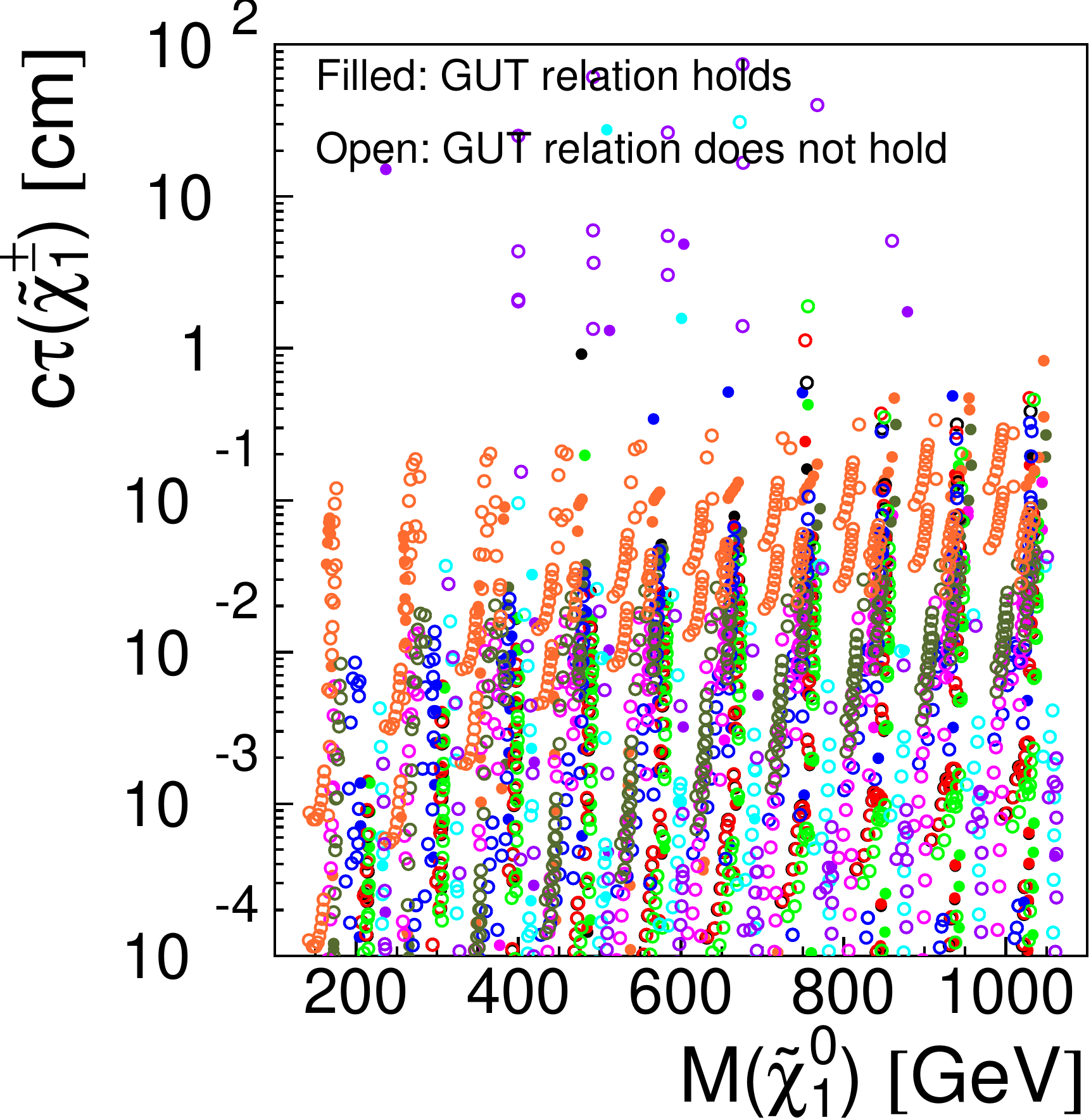}}
       \subfloat[][Wino LSP]{\includegraphics [scale=0.27]{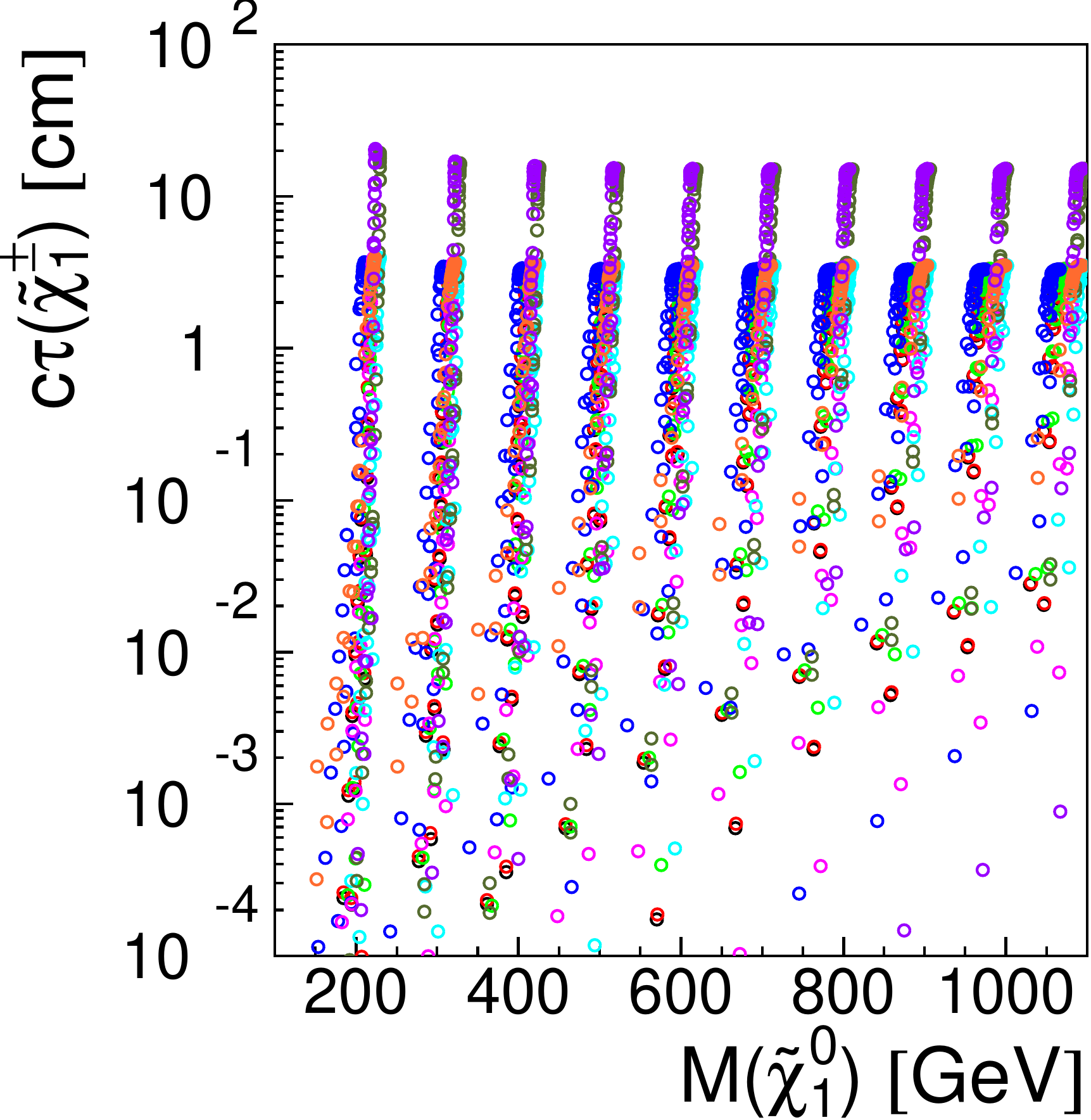}}
         \subfloat[][]{\includegraphics [scale=0.27]{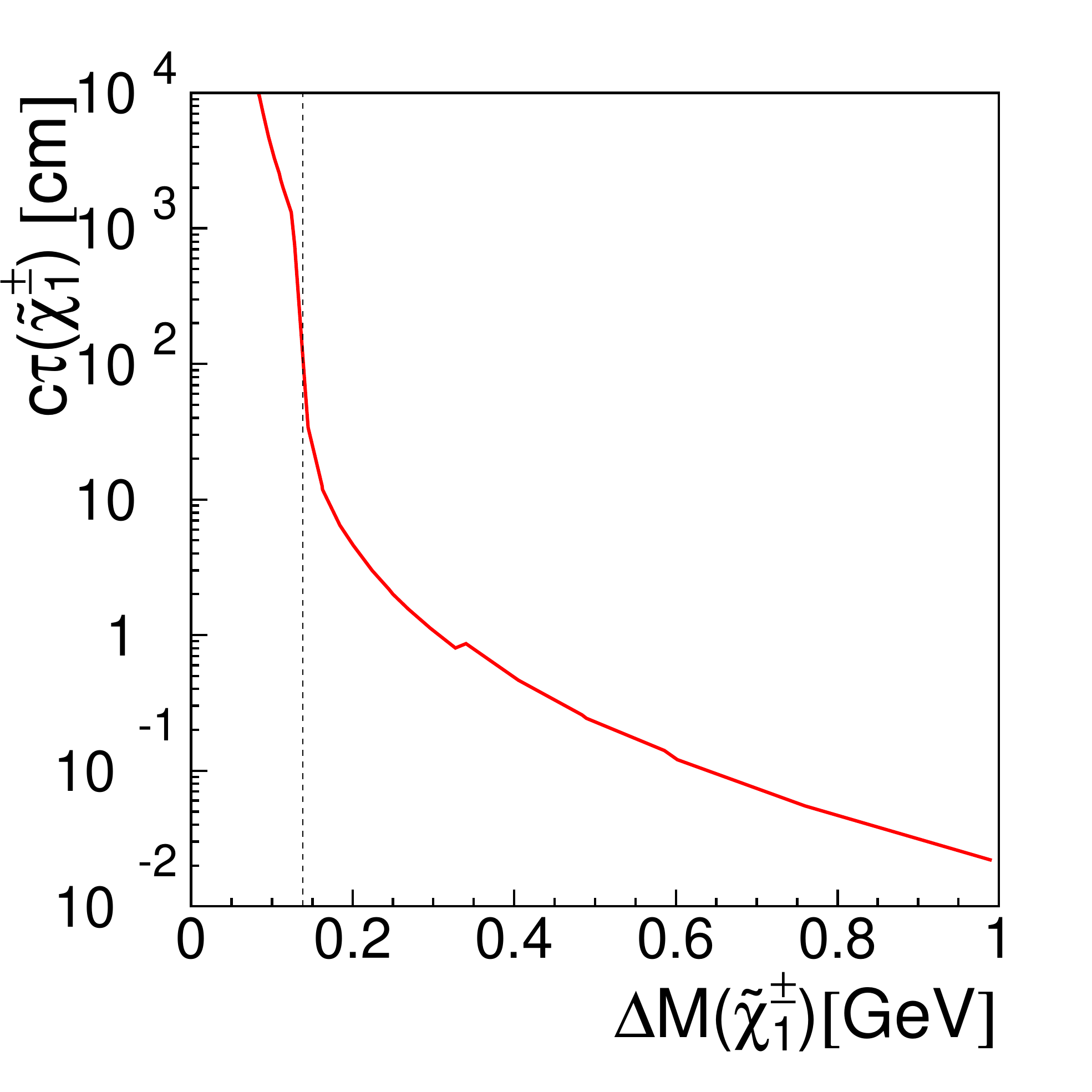}}
         \caption{$c\tau(\XPM{1})$ vs. $M_{LSP}$ for Higgsino (a) and Wino (b) LSP, and  $c\tau(\XPM{1})$ vs. $\MXC{1}$ for
           Higgsino LSP (c).
           Colours as explained in Figure \ref{fig:broadbrush}.\label{fig:ctau}}
     \end{center}
   \end{figure}
     The ``Disappearing tracks'' method was used by FCChh (in the CDR\cite{Benedikt:2018csr}), as well as in
     \cite{Han:2018wus}. The two results
     are shown in Figure \ref{fig:delphesdis}. The upper row is from \cite{Han:2018wus}, while the lower row is
     from \cite{Benedikt:2018csr}. In the latter, the grey curves should be considered: the pink ones shows what could be
     obtained if the innermost layer of the vertex detector would be placed much closer to the beam than what is
     assumed to be the closest conceivable radius, given the radiation levels expected \cite{Benedikt:2018csr}.
     One can observe a large discrepancy between the results in the upper and lower row in the figure.
     Both are based on Delphes parameterised fast simulation \cite{deFavereau:2013fsa}, but
     the FCChh analysis is more realistic, in that it assumes
     a detector as assumed in the CDR, and with a more FCChh-like number of pile-up events (even though it only assumes
     a pile-up of 500, rather than the number 955 that is stated elsewhere in the CDR).
     The analysis of \cite{Han:2018wus} simply uses the current ATLAS setup of Delphes, presumably meaning a pile-up level
     of LHC, which is some 20 times lower than that foreseen at FCChh.
     It should be noted that the CDR detector (the grey bands) has its closest layer closer than that of the current ATLAS,
     and should actually be more powerful than ATLAS, in stark contrast with what is seen.
   One observes that for Higgsinos, the significance of a signal only barely reaches two sigma.
   This is quite different from what is found in \cite{Han:2018wus}, and reflects more realistic simulation would yield.

   The key element for the ``Disappearing tracks'' analysis is the magnitude of $\Delta(M)$.
   Figure \ref{fig:bosinodmzoom}a,b is a zoom in of Figure  \ref{fig:dmx1mlsp}a, showing the
   Higgsino LSP case. In the figure the
   absolute lower limit of $\Delta(M)$ mentioned in the Briefing-book, which was given in
   \cite{Fukuda:2017jmk}, is also shown. Figure \ref{fig:bosinodmzoom}a shows models where
   $\mu$, $M_1$ and $M_2$ are all
   positive. In this case, the lower limit is respected, but reached only for few models.
   Figure \ref{fig:bosinodmzoom}b shows the situation after the full scan
   where any combinations of signs of  $\mu$, $M_1$ and $M_2$ is allowed.
   Clearly the limit is violated, and this is because the calculation in \cite{Fukuda:2017jmk}
   only refers to {\it SM} effects on the mass-splitting, assuming that mixing effects between the
   SUSY fields are negligible. This situation occurs in the ``deep Higgsino'' region where  $M_1$ and $M_2 > > \mu$.
    In Figure \ref{fig:dmx1dmn2}b, the models that are in this region are those that lies on the line
   labelled ``Pure Higgsino line''.
   One also notes that many models, in particular those where $\mu$ is negative, features a
   chargino LSP.
    %the and are thus already excluded from cosmological arguments, if
   %R-parity is conserved.
   As already mentioned, {\tt SPheno} and {\tt FeynHiggs} give different results in this case, and
   Figure \ref{fig:bosinodmzoom}c shows the spectrum under the same conditions as in Figure \ref{fig:bosinodmzoom}b,
   but calculated with   {\tt FeynHiggs} rather than {\tt SPheno}. One sees that
   {\tt FeynHiggs} does not yield chargino LSPs, but does not seem to respect the limiting mass-difference.
   This observation is interesting, but not essential for the question of guaranteed exclusion: The important feature
   in this respect is that there {\it are} models with $\Delta(M)$ = 1 GeV and above at all $M_{LSP}$, i.e.
   far above what is reachable with the ``Disappearing tracks'' method,
   and that the two codes agree on this.
   In Figure \ref{fig:bosinodmzoom}d, the corresponding zoom of Figure  \ref{fig:dmx1mlsp}b is shown,
   and illustrates the Wino LSP case.
   The line in \ref{fig:bosinodmzoom}d is the lower limit given in \cite{Ibe:2012sx},
   which as can be seen is respected, but is by no means attained by all models.
   It is also worth mentioning that the Wino LSP scenario, by it's very construction,
   does not allow for GUT-scale $M_1$-$M_2$ unification.

    \begin{figure}[b]
    \begin{center}
      \subfloat[][Higgsino LSP]{\includegraphics [scale=1.]{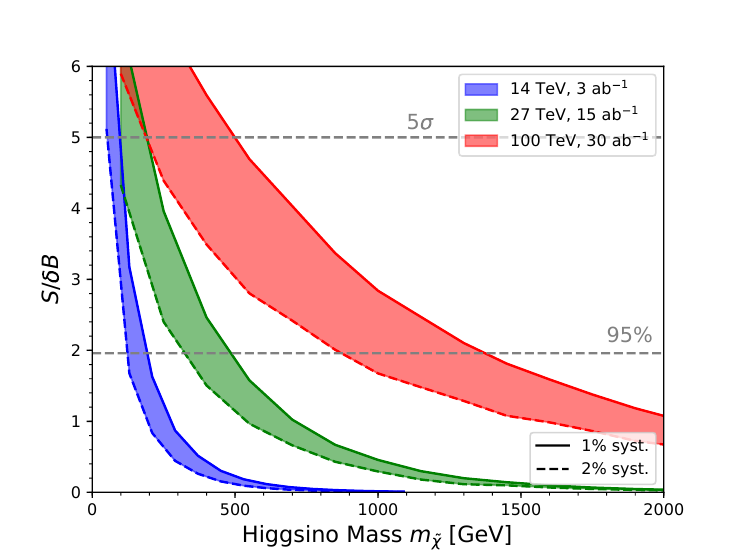}}
      \subfloat[][Wino LSP Z]{\includegraphics [scale=1.]{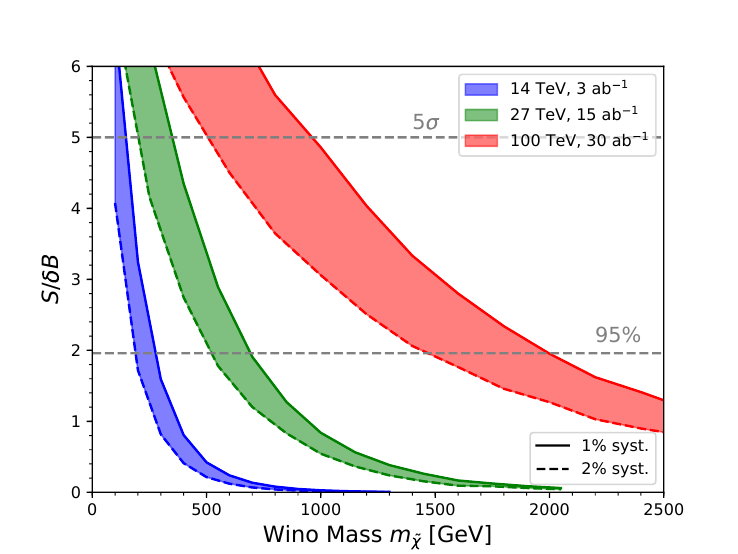}}
      \caption{The sensitivities for the ``Mono-X'' technique at future pp-colliders. From \cite{Han:2018wus}.
        \label{fig:delphesmonox}}
     \end{center}
     \end{figure}
   For the ``Disappearing track'' analysis, the decay-length needs to be macroscopic.
   In \cite{Aaboud:2017mpt}, the ATLAS collaboration reported their results
   on this type of search at 13 TeV.
   They found that the search is effective for lifetimes of about 200 ps,
   corresponding to a $c\tau$ of about 6 cm.
   Figure \ref{fig:ctau} shows $c\tau$ for the different considered models.
   One can see that in the Higgsino case, hardly any points in our scan would yield a decay-length
   long enough - in fact most of the models have a $c\tau$ below 1 mm.
   In the Wino LSP case, on the other hand, there are good chances that $c\tau$ would
   be 1 cm or more. There are, however, even in this case models where $c\tau$ is below 1 mm,
   so a non-observation of a signal cannot be used to infer that the Wino LSP hypothesis is
   excluded.
   In Figure \ref{fig:ctau}c, the dependence of $c\tau$ on $\Delta(M)$ for a Higgsino LSP
   is shown. One can note that $c\tau$ becomes above 1 mm only for $\Delta(M)$ less than
   600 MeV, so in fact the excluded region from disappearing tracks is off the vertical
   scale in Briefing-book Figure \ref{fig:bbwinohiggsino}.

   For the ``Mono-X'' signature, the source of the limits is \cite{Han:2018wus}, and the
   key figures from that publication are shown in
     Figure \ref{fig:delphesmonox}. It should be noted that these figures were included in the  HE/HL-LHC
     input to ESU \cite{esuhehl},
     not the FCChh one \cite{esufcchh}, nor in the FCChh CDR \cite{Benedikt:2018csr}.
     As mentioned above, the analysis is based on Delphes fast simulation using the ATLAS-card,
     and we saw that when applied to the  ``Disappearing tracks'' it gave results far better than
     those of the more realistic analysis in the FCChh CDR.
     %\footnote{This presumably means that LHC pile-up
     %  conditions were assumed. The authors do not comment on this}.
     Furthermore, by scrutinising the dependence of the significance of a signal versus the
     mass and assumed systematic errors, one
     can conclude that the results are systematics limited, with systematics assumed to be between 1 and 2 \%.
     This can be contrasted to existing ``Mono-X'' analyses from both ATLAS \cite{Aaboud:2017phn} and CMS \mcite{cmsmomox,*Chatrchyan:2012me,*Sirunyan:2017jix}, which
     both estimate systematic errors at the level of 10 \%, with a pile-up 20 times lower than that expected at FCChh.
     It is also noteworthy that there to date are no results from ATLAS nor CMS where their ``Mono-X'' searches have
     been used to infer any conclusions about SUSY.

     \section{Summary of the ILC projections\label{sec:ilcproj}}
  \begin{figure}[b]
    \begin{center}
  \subfloat[][Higgsino LSP]{\includegraphics [scale=0.37]{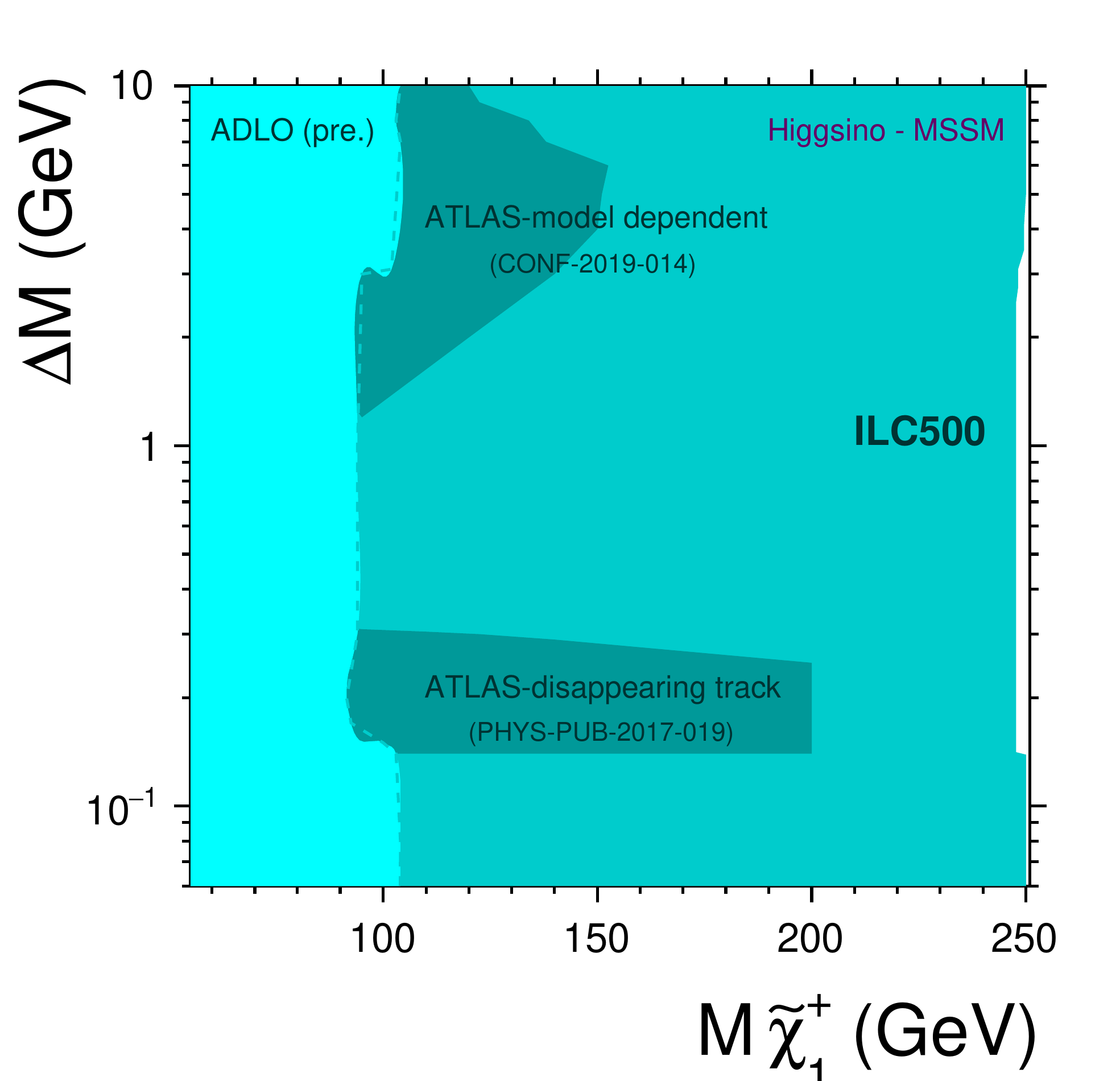}}
  \subfloat[][Wino LSP]{\includegraphics [scale=0.37]{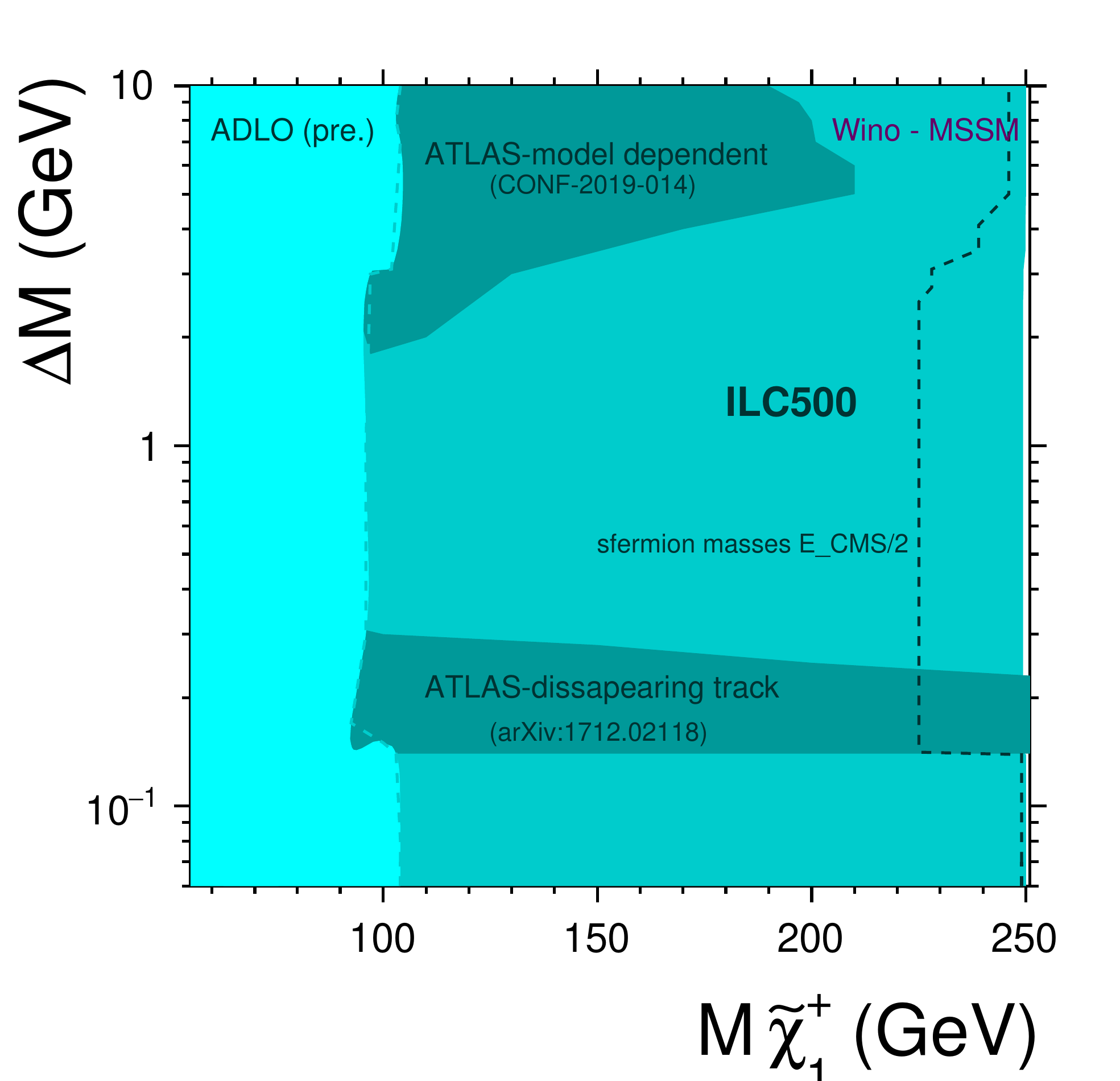}}
  \caption{Exclusion reaches for $\XPM{1}$ at LEP II~\cite{lepsusywg}, ILC-500 \cite{PardodeVera:2020zlr},
    and LHC \cite{Aad:2019qnd,ATL-PHYS-PUB-2017-019,Aaboud:2017mpt,ATL-PHYS-PUB-2017-019}.
    The LEP II, ILC, and ATLAS ``disappearing tracks'' analyses are valid without any assumption, while the 
    ATLAS ``soft leptons'' assumes $M_1$ and $M_2 > > \mu$ \label{fig:ilchinowino}}
\end{center}
\end{figure}
     In this section, we make a brief summary of the expected performance of the SUSY searches at the ILC.
      \begin{figure}[t]
    \begin{center}
  \subfloat[][Full range]{\includegraphics [scale=0.37]{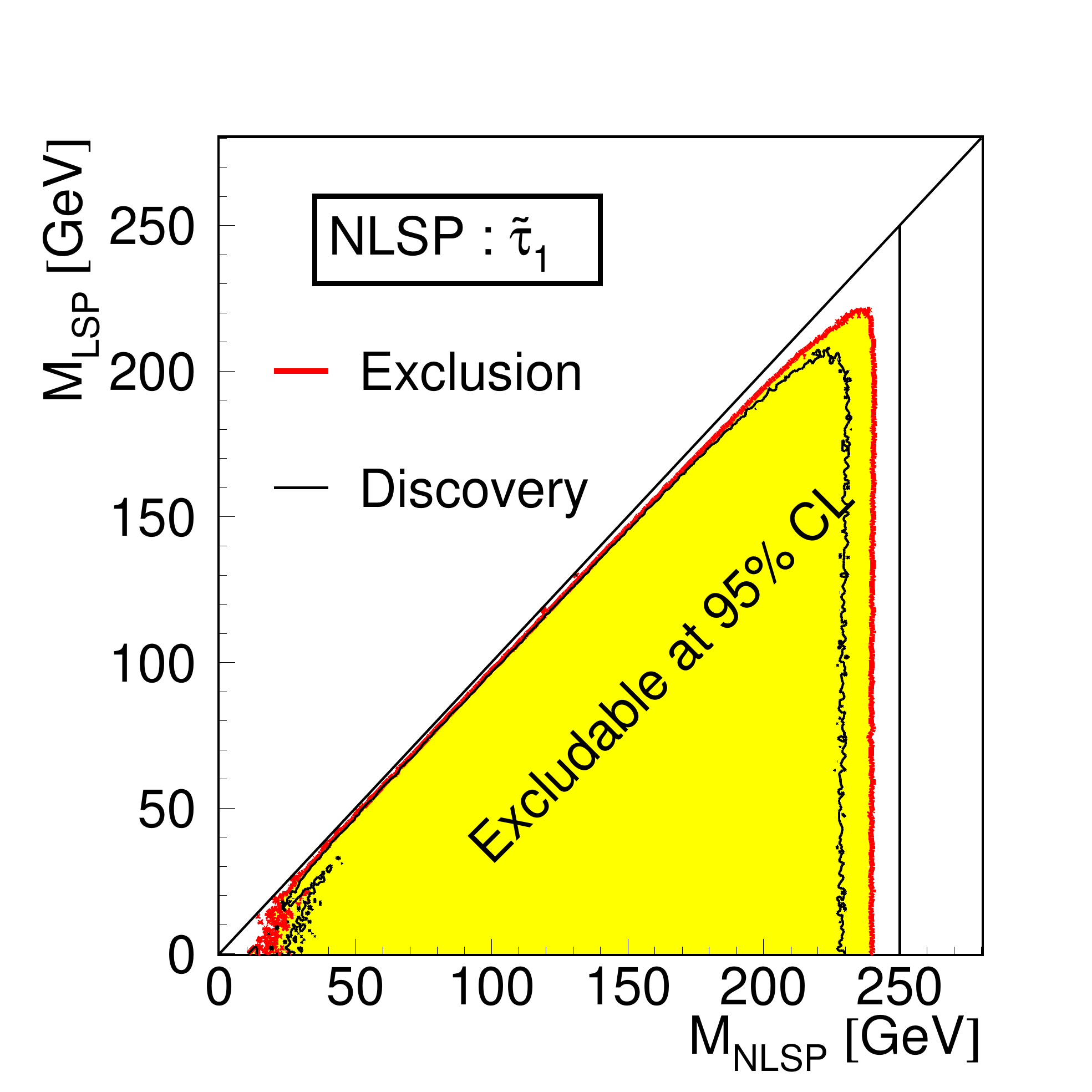}}
  \subfloat[][Zoom to last 50 GeV]{\includegraphics [scale=0.37]{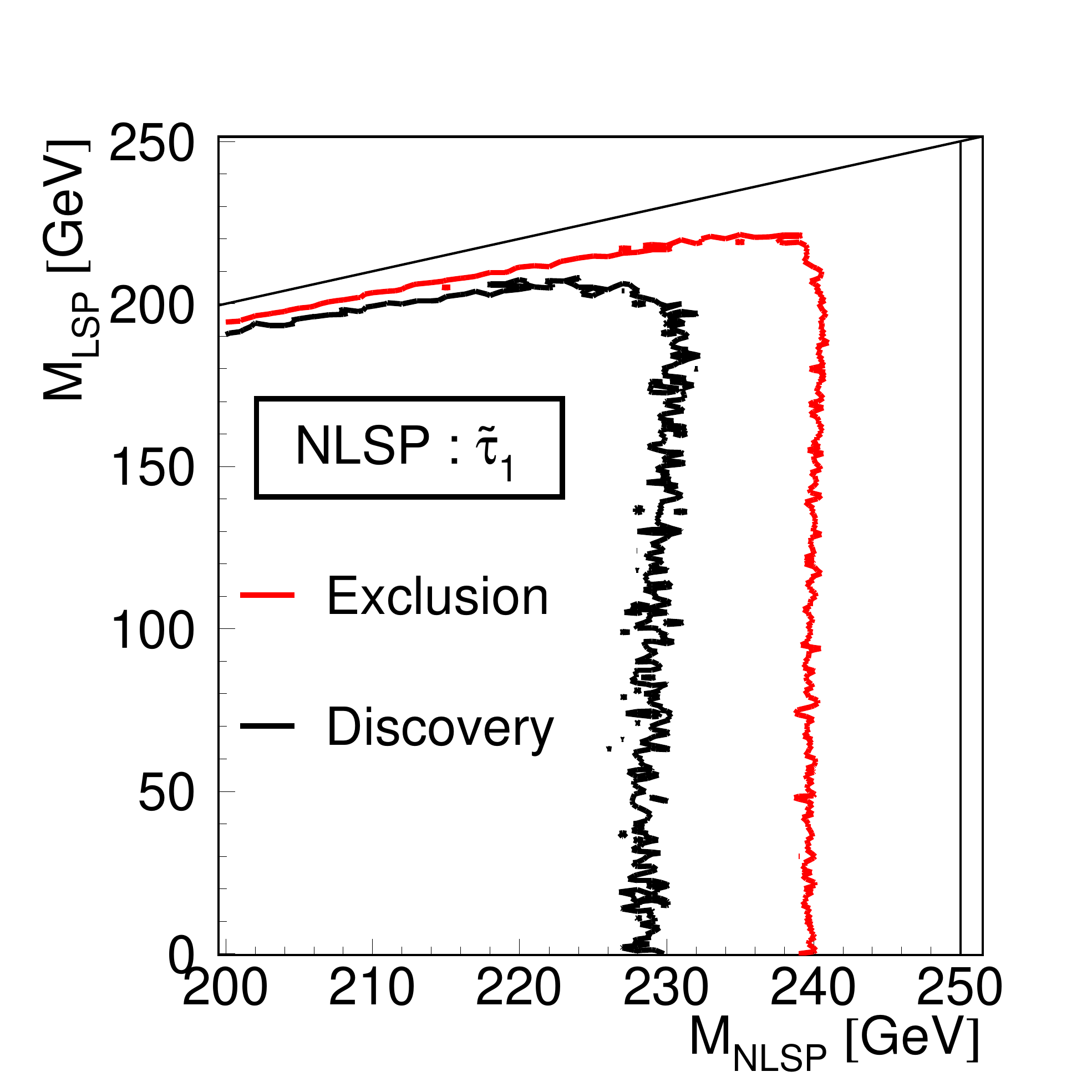}}
        \caption{Reach at ILC-500 for the $\stau$ NLSP case. From \cite{Berggren:2013vna} \label{fig:ilcstau}}
\end{center}
\end{figure}
    A more comprehensive account can be found in \cite{Bambade:2019fyw,Fujii:2017ekh,Berggren:2013bua}.
     Figure \ref{fig:ilchinowino}, from \cite{PardodeVera:2020zlr}, shows the reach of an 500 GeV ILC
     in the search for $\XPM{1}$ in the Higgsino- and Wino-LSP cases.
     These projections were obtained by extrapolation of the LEP II results \cite{lepsusywg}, using 
     background-levels and signal-efficiencies as reported in \cite{lepsusywg}, assuming no other
     ameliorations over LEP II than increased beam energy, beam-polarisation, and data set size.
     This is clearly a very conservative assumption as it neglects the progress in detector
     technology, reported in volume 4 of the ILC TDR \cite{Behnke:2013lya}.
     In \cite{PardodeVera:2020zlr}, it is shown that if $\Delta(M) > 3$ GeV, exclusion and
     discovery-reach are only a few 100 MeV apart, and if $\Delta(M)$ is between 3 GeV and $m_\pi$,
     \begin{figure}[b]
    \begin{center}
  \subfloat[][$\Delta{M}$=1.6 GeV. From \cite{Berggren:2013vfa}]{\includegraphics [scale=0.25]{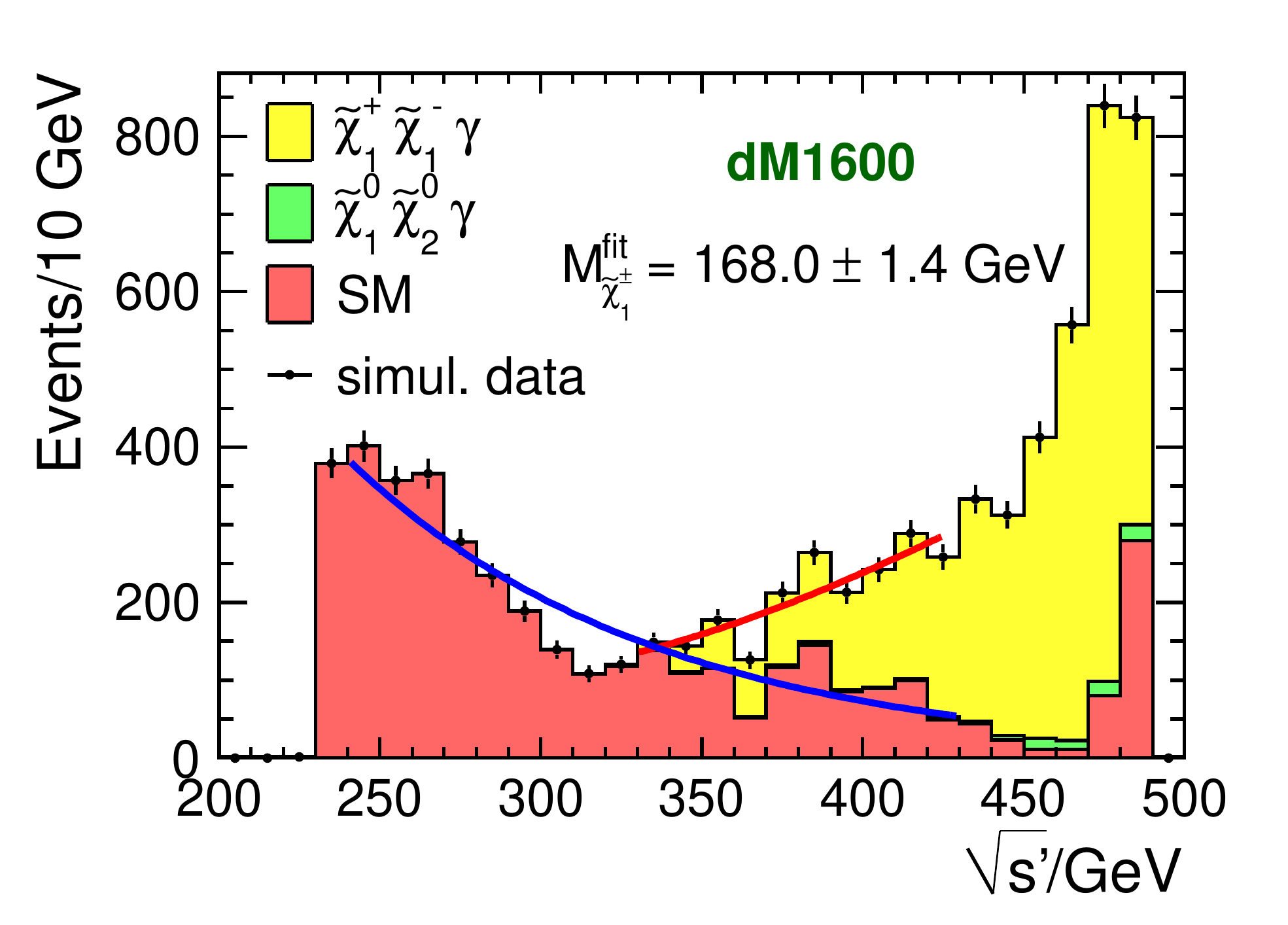}}
  \subfloat[][$\Delta{M}$=4.4 GeV. From \cite{Baer:2019gvu}]{\includegraphics [scale=0.24]{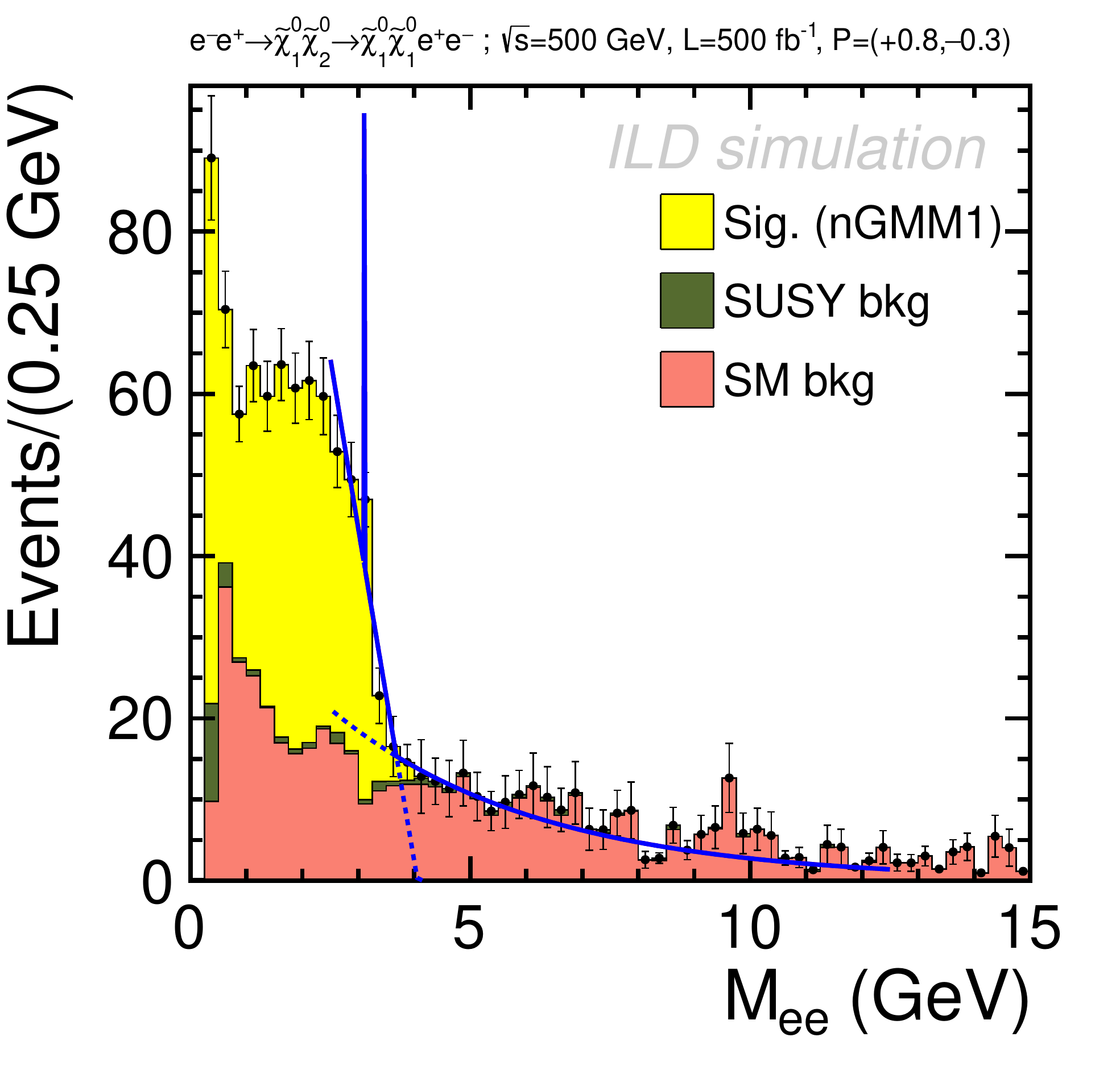}}
  \subfloat[][$\Delta{M}$=9.7 GeV. From \cite{Baer:2019gvu}]{\includegraphics [scale=0.24]{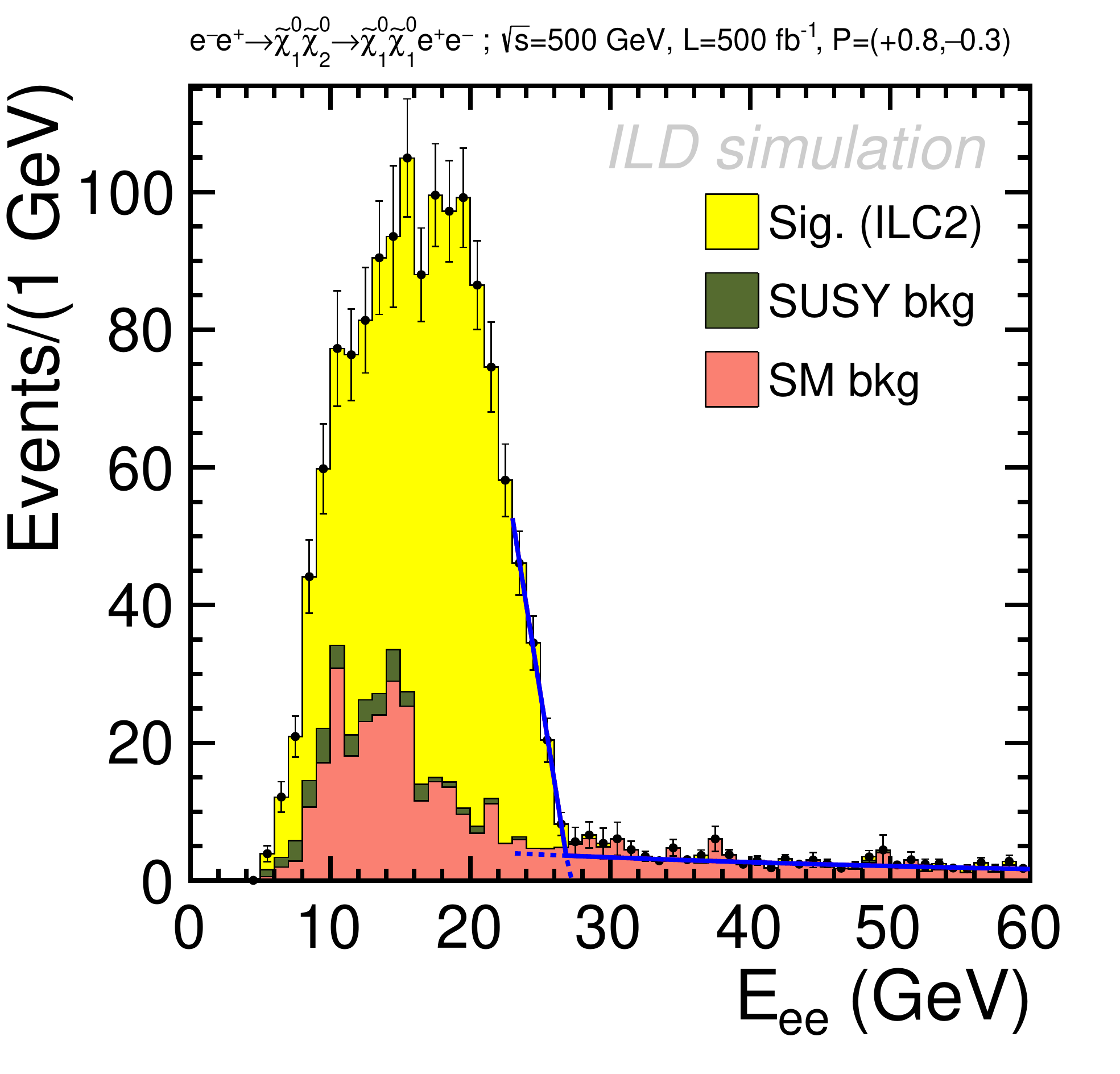}}
  \caption{Examples of Higgsino signals at ILC-500, and various models and mass-differences.
    The assumed integrated luminosity
    is 500 fb$^{-1}$ in all cases. In (a) $\XPM{1}$ is the NLSP, in (b) and (c) it is
    the $\XN{2}$. The spike in (b) is the $J/\Psi$.
     \label{fig:ilchinopoints}}
\end{center}
\end{figure}
     they are at most 5 GeV from each other. The only caveat is in a very particular situation
     where destructive interference between the s-channel and the sneutrino-induced t-channel
     could reduce the production cross-section drastically for the Wino-LSP case \cite{Choi:1998ut}. This only
     happens when the sneutrino mass is close to the beam-energy, and in most of that
     parameter-space, the sneutrino, not the $\XPM{1}$, is the NLSP.
     The experimental implications of such a low mass sneutrino
     were not studied\footnote{In \cite{MoortgatPick:1999ck}, a theoretical study of this
       situation was undertaken. The experimental issues will be a topic for future studies.
      }.

     Even in this case,
     exclusion is guaranteed up to $\MXC{1}$=246 GeV, discovery to 243 GeV,
     if $\Delta(M) > 3$ GeV.
     There is a substantial  loss of reach only in the region where $\Delta(M)$ is between 3 GeV and $m_\pi$, where
     exclusion is guaranteed up to $\MXC{1}$=225 GeV, discovery to 205 GeV.
     However, one should keep in mind that the main reason for the drop in efficiency at LEP II in this
     $\Delta(M)$ region was trigger-efficiency, and the ILC detectors are to be run trigger-less.
In \cite{Berggren:2013vna}, a SUSY parameter scan using detailed fast simulation of the ILD at ILC was done,
to establish the reach in the experimentally worst case NLSP, namely the $\stone$ with the mixing angle
in the $\stau$ sector that minimises the production cross-section.
The results are shown in Figure \ref{fig:ilcstau}, showing that also in this case,
exclusion (discovery) is possible to 10 (20) GeV below the kinematic limit, already a modest integrated
luminosity of 500 fb$^{-1}$, less than a third of what is expected at the favourable beam-polarisation settings
({\it viz.} right-handed electron, left-handed positron).
One can note that the limits are valid only if $\Delta(M)$ > 3-4 GeV, 
however no dedicated low $\Delta(M)$ analysis was done in \cite{Berggren:2013vna} - it is the
subject of an ongoing study.
The same publication also studied the arguably most favourable NLSP candidate, the smuon, and found
that exclusion (discovery) would be assured at 2(4) GeV below the kinematic limit.
Several full simulation studies have been done at particular Higgsino-LSP model-points, typically with modest to very low
$\Delta(M)$ \cite{Berggren:2013vfa,Baer:2019gvu}.
A few examples are shown in Figure \ref{fig:ilchinopoints}, and illustrate how clean the signal is expected to be.

%  \subsection*{\it SUSY: All-in-one}
%xx     \begin{itemize}
%xx     \item  Figures in book
%xx     \item  Figures from refs
%xx     \item  What refs
%xx     \end{itemize}
% From \cite{Bambade:2019fyw}
%>>>       \begin{figure}[h]
%>>>    \begin{center}
%>>>      \subfloat[][Z mode]{\includegraphics [scale=0.38]{plots/fig_08d_w_lep_w_lhc2019_lodm_w_bmark_w_hlproj_w500_w1000_rad_nat_only_C1}}
%>>>          \caption{ILC: {\color{red}No assumptions}. ATLAS:  Assumes $M_1$ and $M_2 > > \mu$ \label{fig:reachsum}}
%>>>%Higgsino_ilc_lep_lhc}
%>>>\end{center}
%>>>     \end{figure}
%>>>     Figure \ref{fig:reachsum}, from \cite{Bambade:2019fyw}, summarises the situation.
%>>>     It shows the mentioned $\XPM{1}$ reach for the ILC, the current ATLAS results, the extrapolated
%>>>     discovery reach of HL-LHC (for the $\XPM{1} \rightarrow Z \XN{1}$ mode, it coincides
%>>>     with the exclusion reach when the unknown branching ratio is taken into account,
%>>>     as discussed in Sect. \ref{sect:bbbino}), and the positions of models studied in full simulation at the ILC
%>>>     in one figure.

\section{Summary and Conclusions}
We have discussed the landscape of possible MSSM models that could
have a next-to-lightest SUSY particle (an NLSP) in reach of future
HEP facilities. We have concentrated on the case of an electroweak bosino
NLSP, as this in almost all cases is the most challenging one, in addition to
being quite likely.
In doing so, we scanned over a grid of values of $\mu$, $M_1$, and $M_2$,
with the only constraint that the NLSP should not be heavier than a few TeV.
We did {\it not} require that the models contained a viable dark matter
candidate, solved the naturalness problem, gave an explanation to
the g-2 anomaly, etc, nor that there were any particular relations
between the parameters. In this way, the study carries no prejudice
on any SUSY breaking scheme, nor the possible existence of other
beyond the standard model phenomena.
We confronted our findings on possible spectra, cross-sections and decay
branching-ratios to projections done for the various options for
future facilities, taking into consideration the detail and maturity
of both the projects and the individual analyses.
We concentrated on future pp-colliders, in particular FCChh,
only briefly touching upon the e$^+$e$^-$-colliders (mainly because
the conclusion about the latter are very simple: either discovery or
exclusion is guaranteed up a few GeV below the kinematic limit,
under all circumstances).

For the high $\Delta(M)$ {\it Bino-region},
the signal at pp-colliders is unambiguous, in the sense that it consists of missing
transverse momentum (MET), originating from the invisible SUSY particles themselves,
without need for a system recoiling against the SUSY particles.
We note, however, that the relative contributions from different possible processes
can vary over a large range, as can the decay branching ratios. Since the sensitivity
of the analyses depends on this, to claim exclusion one must establish which of these
yields the lowest sensitivity. This is usually not done, but rather a single
representative model is assumed.
A further observation is that
there is a simple scaling of the reach to be expected (Sect. \ref{sect:bbbino}),
which is corroborated by the data at LHC at 7 and 13 TeV respectively, and the
thorough HL-LHC projections.
This leads to the expectation that the reach will be extended far at the highest mass-differences,
while only modest progress can be hoped for to lower mass-differences.
Models currently excluded
are only such where  unification of $M_1$ and $M_2$ at the
GUT-scale does not occur - only little progress can be expected into the region were
GUT-scale unification is possible.

For the low $\Delta(M)$ region, the {\it Wino-} and {\it Higgsino-} regions,
the MET from SUSY itself is too small to consist a signal-signature, and
more channel-specific searches are needed, in conjunction with the presence of a sizeable
system recoiling against the SUSY particles.
At mass-differences down to a few GeV, leptonic decays can be searched for.
Due to the need for lepton-identification, this method will not be able to reach
as low mass differences at FCChh as what is attained at LHC.
At mass differences an order of magnitude lower,
the lifetime of the NLSP might become big enough that its decay in the
detector would be observable (the ``Disappearing track'' technique).
FCChh prospects for Higgsinos with this technique are not promising,
while they are for Winos. This is due to the expected lower mass-differences
in the latter case.
In existing analyses from ATLAS and CMS of both these techniques, as well as in the
projections, very specific model-points have been assumed, usually corresponding to
situations where the mass-splitting is only due to SM loop-effects, ignoring the
effects of mixing in the SUSY sector. Our parameter-scan shows that these
assumptions are quite aggressive, and completely different mass-spectra are common-place.
In fact, with one of the spectrum calculators we used, some models actually acquire a $\XPM{1}$ LSP.

A second technique to probe $\Delta(M)$ below the cut-off of the soft-lepton technique is
the ``Mono-X'' one, where the decay of the NLSP is assumed to be undetectable, and the
signal would be the presence of a high $P_T$ mono-jet (or photon, Z, W or higgs), recoiling
against an invisible system.
The power of this technique has not been evaluated to a level that allows for any conclusions,
nor has it been used by ATLAS or CMS in a SUSY context.

In conclusion,
future pp-colliders do have a large {\it discovery reach}, where it is
permissible to assume that the model realised in nature is {\it favourable}.
However, the {\it exclusion reach}, where one must assume that the model
realised is the {\it least favourable}, is quite modest, and has not been evaluated
in large detail.
Notwithstanding the gaps in finding the least favourable model,
one can already note that the regions where the mass-differences are considerable, but
still small enough to allow models with GUT-scale $M_1$-$M_2$ unification
will to a large extent remain uncovered.
Furthermore, the low mass-difference regions leaves gaps both
above and below the region that the soft-lepton method can cover,
regions where both Higgsino- and Wino-LSP models thrive.
A window of opportunity exists at very small differences,
but only for very specific models.

None of these shortcomings are present for future TeV-scale e$^+$e$^-$ colliders.
At these facilities, SUSY will be excluded or discovered up to the kinematic
limit, under all circumstances, and remain {\it the} option to exhaustively test the
hypothesis of weak-scale SUSY.

\printbibliography[title=References]

@article{Wess:1974tw,
      author         = "Wess, J. and Zumino, B.",
      title          = "{Supergauge Transformations in Four-Dimensions}",
      journal        = "Nucl. Phys.",
      volume         = "B70",
      year           = "1974",
      pages          = "39-50",
      doi            = "10.1016/0550-3213(74)90355-1",
      note           = "[,24(1974)]",
      SLACcitation   = "%%CITATION = NUPHA,B70,39;%%"
}

@article{Nilles:1983ge,
      author         = "Nilles, Hans Peter",
      title          = "{Supersymmetry, Supergravity and Particle Physics}",
      journal        = "Phys. Rept.",
      volume         = "110",
      year           = "1984",
      pages          = "1-162",
      doi            = "10.1016/0370-1573(84)90008-5",
      reportNumber   = "UGVA-DPT-1983-12-412",
      SLACcitation   = "%%CITATION = PRPLC,110,1;%%"
}

@article{Haber:1984rc,
      author         = "Haber, Howard E. and Kane, Gordon L.",
      title          = "{The Search for Supersymmetry: Probing Physics Beyond the
                        Standard Model}",
      journal        = "Phys. Rept.",
      volume         = "117",
      year           = "1985",
      pages          = "75-263",
      doi            = "10.1016/0370-1573(85)90051-1",
      reportNumber   = "UM-HE-TH-83-17, SCIPP-85-47",
      SLACcitation   = "%%CITATION = PRPLC,117,75;%%"
}

@article{Barbieri:1982eh,
      author         = "Barbieri, Riccardo and Ferrara, S. and Savoy, Carlos A.",
      title          = "{Gauge Models with Spontaneously Broken Local
                        Supersymmetry}",
      journal        = "Phys. Lett.",
      volume         = "119B",
      year           = "1982",
      pages          = "343",
      doi            = "10.1016/0370-2693(82)90685-2",
      reportNumber   = "CERN-TH-3365",
      SLACcitation   = "%%CITATION = PHLTA,119B,343;%%"
}

@article{Heister:2001nk,
      author         = "Heister, A. and others",
      title          = "{Search for scalar leptons in e+ e- collisions at
                        center-of-mass energies up to 209-GeV}",
      collaboration  = "ALEPH",
      journal        = "Phys. Lett.",
      volume         = "B526",
      year           = "2002",
      pages          = "206-220",
      doi            = "10.1016/S0370-2693(01)01494-0",
      eprint         = "hep-ex/0112011",
      archivePrefix  = "arXiv",
      primaryClass   = "hep-ex",
      reportNumber   = "CERN-EP-2001-086",
      SLACcitation   = "%%CITATION = HEP-EX/0112011;%%"
}

@article{Heister:2003zk,
      author         = "Heister, A. and others",
      title          = "{Absolute mass lower limit for the lightest neutralino of
                        the MSSM from e+ e- data at s**(1/2) up to 209-GeV}",
      collaboration  = "ALEPH",
      journal        = "Phys. Lett.",
      volume         = "B583",
      year           = "2004",
      pages          = "247-263",
      doi            = "10.1016/j.physletb.2003.12.066",
      reportNumber   = "CERN-EP-2003-077",
      SLACcitation   = "%%CITATION = PHLTA,B583,247;%%"
}

@article{Heister:2002mn,
      author         = "Heister, A. and others",
      title          = "{Search for charginos nearly mass degenerate with the
                        lightest neutralino in e+ e- collisions at center-of-mass
                        energies up to 209-GeV}",
      collaboration  = "ALEPH",
      journal        = "Phys. Lett.",
      volume         = "B533",
      year           = "2002",
      pages          = "223-236",
      doi            = "10.1016/S0370-2693(02)01584-8",
      eprint         = "hep-ex/0203020",
      archivePrefix  = "arXiv",
      primaryClass   = "hep-ex",
      reportNumber   = "CERN-EP-2002-020",
      SLACcitation   = "%%CITATION = HEP-EX/0203020;%%"
}

@article{Abdallah:2003xe,
      author         = "Abdallah, J. and others",
      title          = "{Searches for supersymmetric particles in e+ e-
                        collisions up to 208-GeV and interpretation of the results
                        within the MSSM}",
      collaboration  = "DELPHI",
      journal        = "Eur. Phys. J.",
      volume         = "C31",
      year           = "2003",
      pages          = "421-479",
      doi            = "10.1140/epjc/s2003-01355-5",
      eprint         = "hep-ex/0311019",
      archivePrefix  = "arXiv",
      primaryClass   = "hep-ex",
      reportNumber   = "CERN-EP-2003-007",
      SLACcitation   = "%%CITATION = HEP-EX/0311019;%%"
}

@article{Achard:2003ge,
      author         = "Achard, P. and others",
      title          = "{Search for scalar leptons and scalar quarks at LEP}",
      collaboration  = "L3",
      journal        = "Phys. Lett.",
      volume         = "B580",
      year           = "2004",
      pages          = "37-49",
      doi            = "10.1016/j.physletb.2003.10.010",
      eprint         = "hep-ex/0310007",
      archivePrefix  = "arXiv",
      primaryClass   = "hep-ex",
      reportNumber   = "CERN-EP-2003-059",
      SLACcitation   = "%%CITATION = HEP-EX/0310007;%%"
}

@article{Abbiendi:2003ji,
      author         = "Abbiendi, G. and others",
      title          = "{Search for anomalous production of dilepton events with
                        missing transverse momentum in e+ e- collisions at
                        s**(1/2) = 183-Gev to 209-GeV}",
      collaboration  = "OPAL",
      journal        = "Eur. Phys. J.",
      volume         = "C32",
      year           = "2004",
      pages          = "453-473",
      doi            = "10.1140/epjc/s2003-01466-y",
      eprint         = "hep-ex/0309014",
      archivePrefix  = "arXiv",
      primaryClass   = "hep-ex",
      reportNumber   = "CERN-EP-2003-040",
      SLACcitation   = "%%CITATION = HEP-EX/0309014;%%"
}

@online{lepsusywg,
  author = "{ALEPH, DELPHI, L3 and OPAL experiments}",
  title = {LEPSUSYWG, Reports  04-01.1, 04-02.1, 01-03.1, and 02.-04.1},
  year = {2004},
  url ={http://lepsusy.web.cern.ch/lepsusy/Welcome.html},
%%  urldate = {2004-06-22}
}

@article{Bagnaschi:2017tru,
      author         = "Bagnaschi, E. and others",
      title          = "{Likelihood Analysis of the pMSSM11 in Light of LHC
                        13-TeV Data}",
      journal        = "Eur. Phys. J.",
      volume         = "C78",
      year           = "2018",
      number         = "3",
      pages          = "256",
      doi            = "10.1140/epjc/s10052-018-5697-0",
      eprint         = "1710.11091",
      archivePrefix  = "arXiv",
      primaryClass   = "hep-ph",
      reportNumber   = "KCL-PH-TH-2017-22, CERN-PH-TH-2017-087, CERN-TH-2017-087,
                        DESY-17-059, FTPI-MINN-17-17, UMN-TH-3701-17,
                        IFT-UAM-CSIC-17-035",
      SLACcitation   = "%%CITATION = ARXIV:1710.11091;%%"
}

@incollection{Bagnaschi:2018zwg,
      author         = "Bagnaschi, Emanuele and Bechtle, Philip and Haller,
                        Johannes and Kogler, Roman and Peiffer, Thomas and
                        Stefaniak, Tim and Weiglein, Georg",
      title          = "{Global SM and BSM Fits using Results from LHC and other
                        Experiments}",
      editor         = "Haller, Johannes and Grefe, Michael",
      booktitle      = "Particles, Strings and the Early Universe: The Structure
                        of Matter and Space-Time",
      year           = "2018",
      pages          = "203-230",
      doi            = "10.3204/PUBDB-2018-00782/B8",
      SLACcitation   = "%%CITATION = INSPIRE-1699702;%%"
}

@article{Caron:2016hib,
      author         = "Caron, Sascha and Kim, Jong Soo and Rolbiecki, Krzysztof
                        and Ruiz de Austri, Roberto and Stienen, Bob",
      title          = "{The BSM-AI project: SUSY-AI-generalizing LHC limits on
                        supersymmetry with machine learning}",
      journal        = "Eur. Phys. J.",
      volume         = "C77",
      year           = "2017",
      number         = "4",
      pages          = "257",
      doi            = "10.1140/epjc/s10052-017-4814-9",
      eprint         = "1605.02797",
      archivePrefix  = "arXiv",
      primaryClass   = "hep-ph",
      SLACcitation   = "%%CITATION = ARXIV:1605.02797;%%"
}

@article{Aad:2015baa,
      author         = "Aad, Georges and others",
      title          = "{Summary of the ATLAS experiment's sensitivity to
                        supersymmetry after LHC Run 1 - interpreted in the
                        phenomenological MSSM}",
      collaboration  = "ATLAS",
      journal        = "JHEP",
      volume         = "10",
      year           = "2015",
      pages          = "134",
      doi            = "10.1007/JHEP10(2015)134",
      eprint         = "1508.06608",
      archivePrefix  = "arXiv",
      primaryClass   = "hep-ex",
      reportNumber   = "CERN-PH-EP-2015-214",
      SLACcitation   = "%%CITATION = ARXIV:1508.06608;%%"
}

@article{Porod:2003um,
      author         = "Porod, Werner",
      title          = "{SPheno, a program for calculating supersymmetric
                        spectra, SUSY particle decays and SUSY particle production
                        at e+ e- colliders}",
      journal        = "Comput. Phys. Commun.",
      volume         = "153",
      year           = "2003",
      pages          = "275-315",
      doi            = "10.1016/S0010-4655(03)00222-4",
      eprint         = "hep-ph/0301101",
      archivePrefix  = "arXiv",
      primaryClass   = "hep-ph",
      reportNumber   = "ZU-TH-01-03",
      SLACcitation   = "%%CITATION = HEP-PH/0301101;%%"
}

@article{Porod:2011nf,
      author         = "Porod, W. and Staub, F.",
      title          = "{SPheno 3.1: Extensions including flavour, CP-phases and
                        models beyond the MSSM}",
      journal        = "Comput. Phys. Commun.",
      volume         = "183",
      year           = "2012",
      pages          = "2458-2469",
      doi            = "10.1016/j.cpc.2012.05.021",
      eprint         = "1104.1573",
      archivePrefix  = "arXiv",
      primaryClass   = "hep-ph",
      SLACcitation   = "%%CITATION = ARXIV:1104.1573;%%"
}

@article{Kilian:2007gr,
      author         = "Kilian, Wolfgang and Ohl, Thorsten and Reuter, Jurgen",
      title          = "{WHIZARD: Simulating Multi-Particle Processes at LHC and
                        ILC}",
      journal        = "Eur. Phys. J.",
      volume         = "C71",
      year           = "2011",
      pages          = "1742",
      doi            = "10.1140/epjc/s10052-011-1742-y",
      eprint         = "0708.4233",
      archivePrefix  = "arXiv",
      primaryClass   = "hep-ph",
      reportNumber   = "DESY-11-126, EDINBURGH-2010-36, FR-PHENO-2010-037,
                        SI-HEP-2010-18",
      SLACcitation   = "%%CITATION = ARXIV:0708.4233;%%"
}

@article{Moretti:2001zz,
      author         = "Moretti, Mauro and Ohl, Thorsten and Reuter, Jurgen",
      title          = "{O'Mega: An Optimizing matrix element generator}",
      year           = "2001",
      pages          = "1981-2009",
      eprint         = "hep-ph/0102195",
      archivePrefix  = "arXiv",
      primaryClass   = "hep-ph",
      reportNumber   = "IKDA-2001-06, LC-TOOL-2001-040",
      SLACcitation   = "%%CITATION = HEP-PH/0102195;%%"
}

@article{Strategy:2019vxc,
      author         = "Richard Keith Ellis and others",
      title          = "{Physics Briefing Book}",
      year           = "2019",
      eprint         = "1910.11775",
      archivePrefix  = "arXiv",
      primaryClass   = "hep-ex",
      reportNumber   = "CERN-ESU-004",
      SLACcitation   = "%%CITATION = ARXIV:1910.11775;%%"
}

@techreport{ATL-PHYS-PUB-2018-048,
      title         = "{Prospects for searches for staus, charginos and
                       neutralinos at the high luminosity LHC with the ATLAS
                       Detector}",
      institution   = "CERN",
      collaboration = "ATLAS Collaboration",
      address       = "Geneva",
      number        = "ATL-PHYS-PUB-2018-048",
      month         = "Dec",
      year          = "2018",
      reportNumber  = "ATL-PHYS-PUB-2018-048",
      url           = "https://cds.cern.ch/record/2651927",
}

@article{Pumplin:2002vw,
      author         = "Pumplin, J. and Stump, D. R. and Huston, J. and Lai, H.
                        L. and Nadolsky, Pavel M. and Tung, W. K.",
      title          = "{New generation of parton distributions with
                        uncertainties from global QCD analysis}",
      journal        = "JHEP",
      volume         = "07",
      year           = "2002",
      pages          = "012",
      doi            = "10.1088/1126-6708/2002/07/012",
      eprint         = "hep-ph/0201195",
      archivePrefix  = "arXiv",
      primaryClass   = "hep-ph",
      reportNumber   = "MSU-HEP-011101",
      SLACcitation   = "%%CITATION = HEP-PH/0201195;%%"
}

@techreport{ATL-PHYS-PUB-2018-031,
      title         = "{ATLAS sensitivity to winos and higgsinos with a highly
                       compressed mass spectrum at the HL-LHC}",
      institution   = "CERN",
      collaboration = "ATLAS Collaboration",
      address       = "Geneva",
      number        = "ATL-PHYS-PUB-2018-031",
      month         = "Nov",
      year          = "2018",
      reportNumber  = "ATL-PHYS-PUB-2018-031",
      url           = "https://cds.cern.ch/record/2647294",
}

@article{Sirunyan:2018iwl,
      author         = "Sirunyan, Albert M and others",
      title          = "{Search for new physics in events with two soft
                        oppositely charged leptons and missing transverse momentum
                        in proton-proton collisions at $\sqrt{s}=$ 13 TeV}",
      collaboration  = "CMS",
      journal        = "Phys. Lett.",
      volume         = "B782",
      year           = "2018",
      pages          = "440-467",
      doi            = "10.1016/j.physletb.2018.05.062",
      eprint         = "1801.01846",
      archivePrefix  = "arXiv",
      primaryClass   = "hep-ex",
      reportNumber   = "CERN-EP-2017-336, CMS-SUS-16-048",
      SLACcitation   = "%%CITATION = ARXIV:1801.01846;%%"
}

@article{Benedikt:2018csr,
      author         = "Abada, A. and others",
      title          = "{FCC-hh: The Hadron Collider}",
      collaboration  = "FCC",
      journal        = "Eur. Phys. J. ST",
      volume         = "228",
      year           = "2019",
      number         = "4",
      pages          = "755-1107",
      doi            = "10.1140/epjst/e2019-900087-0",
      reportNumber   = "CERN-ACC-2018-0058",
      SLACcitation   = "%%CITATION = 00619,228,755;%%"
}

@article{Han:2018wus,
      author         = "Han, Tao and Mukhopadhyay, Satyanarayan and Wang, Xing",
      title          = "{Electroweak Dark Matter at Future Hadron Colliders}",
      journal        = "Phys. Rev.",
      volume         = "D98",
      year           = "2018",
      number         = "3",
      pages          = "035026",
      doi            = "10.1103/PhysRevD.98.035026",
      eprint         = "1805.00015",
      archivePrefix  = "arXiv",
      primaryClass   = "hep-ph",
      reportNumber   = "PITT-PACC-1806",
      SLACcitation   = "%%CITATION = ARXIV:1805.00015;%%"
}

@article{Fukuda:2017jmk,
      author         = "Fukuda, Hajime and Nagata, Natsumi and Otono, Hidetoshi
                        and Shirai, Satoshi",
      title          = "{Higgsino Dark Matter or Not: Role of Disappearing Track
                        Searches at the LHC and Future Colliders}",
      journal        = "Phys. Lett.",
      volume         = "B781",
      year           = "2018",
      pages          = "306-311",
      doi            = "10.1016/j.physletb.2018.03.088",
      eprint         = "1703.09675",
      archivePrefix  = "arXiv",
      primaryClass   = "hep-ph",
      reportNumber   = "IPMU-17-0050, UT-17-11, KYUSHU-RCAPP-2017-03",
      SLACcitation   = "%%CITATION = ARXIV:1703.09675;%%"
}

@article{Ibe:2012sx,
      author         = "Ibe, Masahiro and Matsumoto, Shigeki and Sato, Ryosuke",
      title          = "{Mass Splitting between Charged and Neutral Winos at
                        Two-Loop Level}",
      journal        = "Phys. Lett.",
      volume         = "B721",
      year           = "2013",
      pages          = "252-260",
      doi            = "10.1016/j.physletb.2013.03.015",
      eprint         = "1212.5989",
      archivePrefix  = "arXiv",
      primaryClass   = "hep-ph",
      reportNumber   = "ICRR-REPORT-641-2012-30, IPMU-12-0231, UT-12-44",
      SLACcitation   = "%%CITATION = ARXIV:1212.5989;%%"
}

@article{Aaboud:2017phn,
      author         = "Aaboud, Morad and others",
      title          = "{Search for dark matter and other new phenomena in events
                        with an energetic jet and large missing transverse
                        momentum using the ATLAS detector}",
      collaboration  = "ATLAS",
      journal        = "JHEP",
      volume         = "01",
      year           = "2018",
      pages          = "126",
      doi            = "10.1007/JHEP01(2018)126",
      eprint         = "1711.03301",
      archivePrefix  = "arXiv",
      primaryClass   = "hep-ex",
      reportNumber   = "CERN-EP-2017-230",
      SLACcitation   = "%%CITATION = ARXIV:1711.03301;%%"
}

@article{Chatrchyan:2012me,
      author         = "Chatrchyan, Serguei and others",
      title          = "{Search for Dark Matter and Large Extra Dimensions in
                        Monojet Events in $pp$ Collisions at $\sqrt{s}=7$ TeV}",
      collaboration  = "CMS",
      journal        = "JHEP",
      volume         = "09",
      year           = "2012",
      pages          = "094",
      doi            = "10.1007/JHEP09(2012)094",
      eprint         = "1206.5663",
      archivePrefix  = "arXiv",
      primaryClass   = "hep-ex",
      reportNumber   = "CMS-EXO-11-059, CERN-PH-EP-2012-168",
      SLACcitation   = "%%CITATION = ARXIV:1206.5663;%%"
}

@article{Sirunyan:2017jix,
      author         = "Sirunyan, A. M. and others",
      title          = "{Search for new physics in final states with an energetic
                        jet or a hadronically decaying $W$ or $Z$ boson and
                        transverse momentum imbalance at $\sqrt{s}=13\text{
                        }\text{ }\mathrm{TeV}$}",
      collaboration  = "CMS",
      journal        = "Phys. Rev.",
      volume         = "D97",
      year           = "2018",
      number         = "9",
      pages          = "092005",
      doi            = "10.1103/PhysRevD.97.092005",
      eprint         = "1712.02345",
      archivePrefix  = "arXiv",
      primaryClass   = "hep-ex",
      reportNumber   = "CMS-EXO-16-048, CERN-EP-2017-294",
      SLACcitation   = "%%CITATION = ARXIV:1712.02345;%%"
}

@online{esuhehl,
  author = {Dainese, A.  and others},
  title = {The physics potential of HE-LHC},
  year = {2019},
  url = {https://indico.cern.ch/event/765096/contributions/3296016/attachments/1785350/2906423/HELHC.pdf}
%%  urldate = {},
}

@online{esufcchh,
author = {Benedikt, M. ,  and others},
title = {Future Circular Collider The Hadron Collider (FCC-hh)},
  year = {2019},
url={https://indico.cern.ch/event/765096/contributions/3298184/attachments/1786069/2907901/133_ESPP18_FCChh_181115-FCC_V0600_MainText.pdf},
%%urldate={},
}

@inproceedings{PardodeVera:2020zlr,
      author         = " N{\'u}{\~n}ez  Pardo de Vera, Mar{\'i}a Teresa and Berggren, Mikael
                        and List, Jenny",
      title          = "{Chargino production at the ILC}",
      booktitle      = "{International Workshop on Future Linear Colliders (LCWS
                        2019) Sendai, Miyagi, Japan, October 28-November 1, 2019}",
      year           = "2020",
      eprint         = "2002.01239",
      archivePrefix  = "arXiv",
      primaryClass   = "hep-ph",
      SLACcitation   = "%%CITATION = ARXIV:2002.01239;%%"
}

@article{Baer:2019gvu,
      author         = "Baer, Howard and Berggren, Mikael and Fujii, Keisuke and
                        List, Jenny and Lehtinen, Suvi-Leena and Tanabe, Tomohiko
                        and Yan, Jacqueline",
      title          = "{The ILC as a natural SUSY discovery machine and
                        precision microscope: from light higgsinos to tests of
                        unification}",
      year           = "2019",
      eprint         = "1912.06643",
      archivePrefix  = "arXiv",
      primaryClass   = "hep-ex",
      reportNumber   = "ILD-PHYS-2019-001, DESY-19-227, KEK Preprint 2019-53",
      SLACcitation   = "%%CITATION = ARXIV:1912.06643;%%"
}

@article{Bambade:2019fyw,
      author         = "Bambade, Philip and others",
      title          = "{The International Linear Collider: A Global Project}",
      year           = "2019",
      eprint         = "1903.01629",
      archivePrefix  = "arXiv",
      primaryClass   = "hep-ex",
      reportNumber   = "DESY 19-037, DESY-19-037, FERMILAB-FN-1067-PPD,
                        IFIC/19-10, IRFU-19-10,
  JLAB-PHY-19-2854, KEK Preprint
                        2018-92, JLAB-PHY-19-2854, KEK
  Preprint 2018-92, LAL/RT
                        19-001, PNNL-SA-142168,
  SLAC-PUB-17412, SLAC-PUB-17412",
      SLACcitation   = "%%CITATION = ARXIV:1903.01629;%%"
}

@article{Fujii:2017ekh,
      author         = "Fujii, Keisuke and others",
      title          = "{The Potential of the ILC for Discovering New Particles}",
      year           = "2017",
      eprint         = "1702.05333",
      archivePrefix  = "arXiv",
      primaryClass   = "hep-ph",
      reportNumber   = "DESY-17-012, KEK-PREPRINT-2016-60, SLAC-PUB-16916,
                        LAL-17-017, MPP-2017-5, IFT-UAM-CSIC-17-008",
      SLACcitation   = "%%CITATION = ARXIV:1702.05333;%%"
}

@inproceedings{Berggren:2013bua,
      author         = "Berggren, Mikael and Han, Tao and List, Jenny and Padhi,
                        Sanjay and Su, Shufang and Tanabe, Tomohiko",
      title          = "{Electroweakino Searches: A Comparative Study for LHC and
                        ILC (A Snowmass White Paper)}",
      booktitle      = "{Proceedings, 2013 Community Summer Study on the Future
                        of U.S. Particle Physics: Snowmass on the Mississippi
                        (CSS2013): Minneapolis, MN, USA, July 29-August 6, 2013}",
 %%     url            = "http://www.slac.stanford.edu/econf/C1307292/docs/submittedArxivFiles/1309.7342.pdf",
      year           = "2013",
      eprint         = "1309.7342",
      archivePrefix  = "arXiv",
      primaryClass   = "hep-ph",
      SLACcitation   = "%%CITATION = ARXIV:1309.7342;%%"
}

@inproceedings{Berggren:2013vna,
      author         = "Berggren, Mikael",
      title          = "{Simplified SUSY at the ILC}",
      booktitle      = "{Proceedings, 2013 Community Summer Study on the Future
                        of U.S. Particle Physics: Snowmass on the Mississippi
                        (CSS2013): Minneapolis, MN, USA, July 29-August 6, 2013}",
      year           = "2013",
      eprint         = "1308.1461",
      archivePrefix  = "arXiv",
      primaryClass   = "hep-ph",
      reportNumber   = "DESY-13-136",
      SLACcitation   = "%%CITATION = ARXIV:1308.1461;%%"
}

@article{Berggren:2013vfa,
      author         = "Berggren, Mikael and Br{\"u}mmer, Felix and List, Jenny and
                        Moortgat-Pick, Gudrid and Robens, Tania and Rolbiecki,
                        Krzysztof and Sert, Hale",
      title          = "{Tackling light higgsinos at the ILC}",
      journal        = "Eur. Phys. J.",
      volume         = "C73",
      year           = "2013",
      number         = "12",
      pages          = "2660",
      doi            = "10.1140/epjc/s10052-013-2660-y",
      eprint         = "1307.3566",
      archivePrefix  = "arXiv",
      primaryClass   = "hep-ph",
      reportNumber   = "DESY-13-098",
      SLACcitation   = "%%CITATION = ARXIV:1307.3566;%%"
}

@article{deFavereau:2013fsa,
      author         = "de Favereau, J. and Delaere, C. and Demin, P. and
                        Giammanco, A. and Lemaître, V. and Mertens, A. and
                        Selvaggi, M.",
      title          = "{DELPHES 3, A modular framework for fast simulation of a
                        generic collider experiment}",
      collaboration  = "DELPHES 3",
      journal        = "JHEP",
      volume         = "02",
      year           = "2014",
      pages          = "057",
      doi            = "10.1007/JHEP02(2014)057",
      eprint         = "1307.6346",
      archivePrefix  = "arXiv",
      primaryClass   = "hep-ex",
      SLACcitation   = "%%CITATION = ARXIV:1307.6346;%%"
}

@article{Bahl:2018qog,
      author         = "Bahl, H. and Hahn, T. and Heinemeyer, S. and Hollik, W.
                        and Paßehr, S. and Rzehak, H. and Weiglein, G.",
      title          = "{Precision calculations in the MSSM Higgs-boson sector
                        with FeynHiggs 2.14}",
      journal        = "Comput. Phys. Commun.",
      volume         = "249",
      year           = "2020",
      pages          = "107099",
      doi            = "10.1016/j.cpc.2019.107099",
      eprint         = "1811.09073",
      archivePrefix  = "arXiv",
      primaryClass   = "hep-ph",
      reportNumber   = "DESY-18-179",
      SLACcitation   = "%%CITATION = ARXIV:1811.09073;%%"
}

@article{Bahl:2017aev,
      author         = "Bahl, Henning and Heinemeyer, Sven and Hollik, Wolfgang
                        and Weiglein, Georg",
      title          = "{Reconciling EFT and hybrid calculations of the light
                        MSSM Higgs-boson mass}",
      journal        = "Eur. Phys. J.",
      volume         = "C78",
      year           = "2018",
      number         = "1",
      pages          = "57",
      doi            = "10.1140/epjc/s10052-018-5544-3",
      eprint         = "1706.00346",
      archivePrefix  = "arXiv",
      primaryClass   = "hep-ph",
      reportNumber   = "DESY-17-072, IFT-UAM-CSIC-17-047, MPP-2017-108",
      SLACcitation   = "%%CITATION = ARXIV:1706.00346;%%"
}

@article{Bahl:2016brp,
      author         = "Bahl, Henning and Hollik, Wolfgang",
      title          = "{Precise prediction for the light MSSM Higgs boson mass
                        combining effective field theory and fixed-order
                        calculations}",
      journal        = "Eur. Phys. J.",
      volume         = "C76",
      year           = "2016",
      number         = "9",
      pages          = "499",
      doi            = "10.1140/epjc/s10052-016-4354-8",
      eprint         = "1608.01880",
      archivePrefix  = "arXiv",
      primaryClass   = "hep-ph",
      SLACcitation   = "%%CITATION = ARXIV:1608.01880;%%"
}

@article{Hahn:2013ria,
      author         = "Hahn, T. and Heinemeyer, S. and Hollik, W. and Rzehak, H.
                        and Weiglein, G.",
      title          = "{High-Precision Predictions for the Light CP -Even Higgs
                        Boson Mass of the Minimal Supersymmetric Standard Model}",
      journal        = "Phys. Rev. Lett.",
      volume         = "112",
      year           = "2014",
      number         = "14",
      pages          = "141801",
      doi            = "10.1103/PhysRevLett.112.141801",
      eprint         = "1312.4937",
      archivePrefix  = "arXiv",
      primaryClass   = "hep-ph",
      reportNumber   = "DESY-13-248, FR-PHENO-2013-018, MPP-2013-317",
      SLACcitation   = "%%CITATION = ARXIV:1312.4937;%%"
}

@article{Frank:2006yh,
      author         = "Frank, M. and Hahn, T. and Heinemeyer, S. and Hollik, W.
                        and Rzehak, H. and Weiglein, G.",
      title          = "{The Higgs Boson Masses and Mixings of the Complex MSSM
                        in the Feynman-Diagrammatic Approach}",
      journal        = "JHEP",
      volume         = "02",
      year           = "2007",
      pages          = "047",
      doi            = "10.1088/1126-6708/2007/02/047",
      eprint         = "hep-ph/0611326",
      archivePrefix  = "arXiv",
      primaryClass   = "hep-ph",
      reportNumber   = "DCPT-06-160, IPPP-06-80, MPP-2006-158, PSI-PR-06-14",
      SLACcitation   = "%%CITATION = HEP-PH/0611326;%%"
}

@article{Degrassi:2002fi,
      author         = "Degrassi, G. and Heinemeyer, S. and Hollik, W. and
                        Slavich, P. and Weiglein, G.",
      title          = "{Towards high precision predictions for the MSSM Higgs
                        sector}",
      journal        = "Eur. Phys. J.",
      volume         = "C28",
      year           = "2003",
      pages          = "133-143",
      doi            = "10.1140/epjc/s2003-01152-2",
      eprint         = "hep-ph/0212020",
      archivePrefix  = "arXiv",
      primaryClass   = "hep-ph",
      reportNumber   = "DCPT-02-126, IPPP-02-63, LMU-11-02, MPI-PHT-2002-73,
                        RM3-TH-02-19",
      SLACcitation   = "%%CITATION = HEP-PH/0212020;%%"
}

@article{Heinemeyer:1998np,
      author         = "Heinemeyer, S. and Hollik, W. and Weiglein, G.",
      title          = "{The Masses of the neutral CP - even Higgs bosons in the
                        MSSM: Accurate analysis at the two loop level}",
      journal        = "Eur. Phys. J.",
      volume         = "C9",
      year           = "1999",
      pages          = "343-366",
      doi            = "10.1007/s100529900006",
      eprint         = "hep-ph/9812472",
      archivePrefix  = "arXiv",
      primaryClass   = "hep-ph",
      reportNumber   = "KA-TP-17-1998, DESY-98-194, CERN-TH-98-405",
      SLACcitation   = "%%CITATION = HEP-PH/9812472;%%"
}

@article{Heinemeyer:1998yj,
      author         = "Heinemeyer, S. and Hollik, W. and Weiglein, G.",
      title          = "{FeynHiggs: A Program for the calculation of the masses
                        of the neutral CP even Higgs bosons in the MSSM}",
      journal        = "Comput. Phys. Commun.",
      volume         = "124",
      year           = "2000",
      pages          = "76-89",
      doi            = "10.1016/S0010-4655(99)00364-1",
      eprint         = "hep-ph/9812320",
      archivePrefix  = "arXiv",
      primaryClass   = "hep-ph",
      reportNumber   = "KA-TP-16-1998, DESY-98-193, CERN-TH-98-389",
      SLACcitation   = "%%CITATION = HEP-PH/9812320;%%"
}

@techreport{CMS:2018qsc,
%      author         = "CMS Collaboration",
      title          = "{Searches for light higgsino-like charginos and
                        neutralinos at the HL-LHC with the Phase-2 CMS detector}",
      collaboration  = "CMS",
      year           = "2018",
      number   = "CMS-PAS-FTR-18-001",
      SLACcitation   = "%%CITATION = CMS-PAS-FTR-18-001;%%"
}

@article{Aad:2019qnd,
      author         = "Aad, Georges and others",
      title          = "{Searches for electroweak production of supersymmetric
                        particles with compressed mass spectra in $\sqrt{s}=13$
                        TeV $pp$ collisions with the ATLAS detector}",
      collaboration  = "ATLAS",
      year           = "2019",
      eprint         = "1911.12606",
      archivePrefix  = "arXiv",
      primaryClass   = "hep-ex",
      reportNumber   = "CERN-EP-2019-242",
      SLACcitation   = "%%CITATION = ARXIV:1911.12606;%%"
}

@article{Aaboud:2017mpt,
      author         = "Aaboud, Morad and others",
      title          = "{Search for long-lived charginos based on a
                        disappearing-track signature in pp collisions at $
                        \sqrt{s}=13 $ TeV with the ATLAS detector}",
      collaboration  = "ATLAS",
      journal        = "JHEP",
      volume         = "06",
      year           = "2018",
      pages          = "022",
      doi            = "10.1007/JHEP06(2018)022",
      eprint         = "1712.02118",
      archivePrefix  = "arXiv",
      primaryClass   = "hep-ex",
      reportNumber   = "CERN-EP-2017-179",
      SLACcitation   = "%%CITATION = ARXIV:1712.02118;%%"
}

@techreport{ATL-PHYS-PUB-2017-019,
      title         = "{Search for direct pair production of higgsinos by the
                       reinterpretation of the disappearing track analysis with
                       36.1 fb$^{-1}$ of $\sqrt{s}=13$ TeV data collected with the
                       ATLAS experiment}",
      institution   = "CERN",
      collaboration = "ATLAS Collaboration",
      address       = "Geneva",
      number        = "ATL-PHYS-PUB-2017-019",
      month         = "Dec",
      year          = "2017",
      reportNumber  = "ATL-PHYS-PUB-2017-019",
      url           = "https://cds.cern.ch/record/2297480",
}

@article{Behnke:2013lya,
      author         = "Abramowicz, Halina and others",
      editor         = "Behnke, Ties and Brau, James E. and Burrows, Philip N.
                        and Fuster, Juan and Peskin, Michael and Stanitzki, Marcel
                        and Sugimoto, Yasuhiro and Yamada, Sakue and Yamamoto,
                        Hitoshi",
      title          = "{The International Linear Collider Technical Design
                        Report - Volume 4: Detectors}",
      year           = "2013",
      eprint         = "1306.6329",
      archivePrefix  = "arXiv",
      primaryClass   = "physics.ins-det",
      reportNumber   = "ILC-REPORT-2013-040, ANL-HEP-TR-13-20,
                        BNL-100603-2013-IR, IRFU-13-59, CERN-ATS-2013-037,
                        COCKCROFT-13-10, CLNS-13-2085, DESY-13-062,
                        FERMILAB-TM-2554, IHEP-AC-ILC-2013-001, INFN-13-04-LNF,
                        JAI-2013-001, JINR-E9-2013-35, JLAB-R-2013-01,
                        KEK-REPORT-2013-1, KNU-CHEP-ILC-2013-1, LLNL-TR-635539,
                        SLAC-R-1004, ILC-HIGRADE-REPORT-2013-003",
      SLACcitation   = "%%CITATION = ARXIV:1306.6329;%%"
}

@article{MoortgatPick:1999ck,
      author         = "Moortgat-Pick, Gudrid A. and Fraas, H.",
      title          = "{Constraining the sneutrino mass in chargino production
                        and decay with polarized beams}",
      booktitle      = "{Electron positron colliders. Proceedings, Epiphany
                        Conference, Cracow, Poland, January 5-10, 1999}",
      journal        = "Acta Phys. Polon.",
      volume         = "B30",
      year           = "1999",
      pages          = "1999-2011",
      eprint         = "hep-ph/9904209",
      archivePrefix  = "arXiv",
      primaryClass   = "hep-ph",
      reportNumber   = "WUE-ITP-99-007",
      SLACcitation   = "%%CITATION = HEP-PH/9904209;%%"
}

@article{Marsh:2015xka,
      author         = "Marsh, David J. E.",
      title          = "{Axion Cosmology}",
      journal        = "Phys. Rept.",
      volume         = "643",
      year           = "2016",
      pages          = "1-79",
      doi            = "10.1016/j.physrep.2016.06.005",
      eprint         = "1510.07633",
      archivePrefix  = "arXiv",
      primaryClass   = "astro-ph.CO",
      reportNumber   = "KCL-PH-TH-2015-50",
      SLACcitation   = "%%CITATION = ARXIV:1510.07633;%%"
}

@article{Choi:1998ut,
      author         = "Choi, S. Y. and Djouadi, A. and Dreiner, Herbert K. and
                        Kalinowski, J. and Zerwas, P. M.",
      title          = "{Chargino pair production in e+ e- collisions}",
      journal        = "Eur. Phys. J.",
      volume         = "C7",
      year           = "1999",
      pages          = "123-134",
      doi            = "10.1007/s100529800957, 10.1007/s100520050392",
      eprint         = "hep-ph/9806279",
      archivePrefix  = "arXiv",
      primaryClass   = "hep-ph",
      reportNumber   = "DESY-98-077, YUMS-98-10, SNUTP-98-52, PM-98-14, IFT-9-98,
                        YUMS-98--10, SNUTP-98--52, DESY-98--77, PM-98--14",
      SLACcitation   = "%%CITATION = HEP-PH/9806279;%%"
}

@article{Abdughani:2019wss,
      author         = "Abdughani, Murat and Wu, Lei",
      title          = "{On the coverage of neutralino dark matter in
                        coannihilations at the upgraded LHC}",
      journal        = "Eur. Phys. J.",
      volume         = "C80",
      year           = "2020",
      number         = "3",
      pages          = "233",
      doi            = "10.1140/epjc/s10052-020-7793-1",
      eprint         = "1908.11350",
      archivePrefix  = "arXiv",
      primaryClass   = "hep-ph",
      SLACcitation   = "%%CITATION = ARXIV:1908.11350;%%"
}

\end{document}